\documentclass[aps,prx,twocolumn,amssymb,longbibliography,amsmath,showpacs]{revtex4-2}

\usepackage[utf8]{inputenc}
\usepackage[colorlinks=true]{hyperref}
\usepackage{amsmath, nccmath}
\usepackage{amssymb}
\usepackage{amsfonts}
\usepackage{graphicx}
\usepackage{verbatim}
\usepackage{bm}
\usepackage{mathbbol}
\usepackage{color}
 \usepackage{mathtools}
 \usepackage{graphicx,epsfig,amsfonts,amssymb}
\usepackage{bm} 
\usepackage{times}
 \usepackage{lipsum}
 \usepackage{verbatim}
\usepackage{enumitem}
\usepackage{hyperref}

\hypersetup{
  linktocpage,
    colorlinks,
    citecolor=blue,
    filecolor=blue,
    linkcolor=blue,
    urlcolor=blue
}

\setcitestyle{compress}

\newcommand{\Tr}{\mathrm{Tr}\,}
 
\newcommand{\beg}{\begin{equation}}
\newcommand{\en}{\end{equation}}

\newcommand{\lam}{\lambda}
\newcommand{\eps}{\varepsilon}

\newcommand{\eref}[1]{Eq.~(\ref{#1})}
\newcommand{\re}[1]{(\ref{#1})}

\newcommand{\esref}[1]{Eqs.~(\ref{#1})}

\newcommand{\avg}[1]{\langle #1\rangle}
\newcommand{\ket}[1]{\lvert #1 \rangle}
\newcommand{\bra}[1]{\langle #1 \rvert}

\newcommand{\norm}[1]{\left\lVert#1\right\rVert}

\newcommand{\dn}{\downarrow}
\newcommand{\up}{\uparrow}
\newcommand{\dis}{\displaystyle}

\begin{document}

\setlength{\parskip}{0pt}
 
\title{Nonlocality as the source of purely quantum dynamics of BCS superconductors}

\author{Aidan Zabalo}
\affiliation{Department of Physics and Astronomy, Center for Materials Theory, Rutgers University, Piscataway, NJ 08854 USA}

\author{Ang-Kun Wu}
\affiliation{Department of Physics and Astronomy, Center for Materials Theory, Rutgers University, Piscataway, NJ 08854 USA}
  
\author{J. H. Pixley}
\affiliation{Department of Physics and Astronomy, Center for Materials Theory, Rutgers University, Piscataway, NJ 08854 USA}

\author{Emil A. Yuzbashyan}
\affiliation{Department of Physics and Astronomy, Center for Materials Theory, Rutgers University, Piscataway, NJ 08854 USA}


 \begin{abstract}
 
 We show that the classical (mean-field)  description of far from equilibrium superconductivity is \textit{exact} in the thermodynamic limit for local observables but breaks down for global quantities, such as the entanglement entropy or Loschmidt echo.
 We do this by solving for and comparing  \textit{exact quantum}  and \textit{exact classical}  long-time dynamics of a BCS superconductor with   interaction strength  inversely proportional to time and evaluating local observables explicitly. Mean field is exact
 for both normal and anomalous averages (superconducting  order) in the thermodynamic limit. However,   for anomalous expectation values, this limit does not commute with adiabatic and strong coupling limits and, as a consequence, their quantum fluctuations can be unusually  strong. The  long-time steady state of the system is a gapless superconductor  whose superfluid properties  are only accessible through energy resolved measurements. This state is nonthermal but conforms to an emergent generalized Gibbs ensemble. Our study clarifies the nature of symmetry-broken many-body states in and out of equilibrium and fills a crucial gap in the theory of time-dependent quantum integrability.
 
 \end{abstract}

 \maketitle

\makeatletter
\def\l@subsection#1#2{}
\def\l@subsubsection#1#2{}
\makeatother
 
 \tableofcontents

\newpage
 
\section{Introduction}

Superconductivity is one of the best-known  examples of quantum  phenomena on the macroscopic scale that is  conventionally understood in terms of a many-body wave function  with a well-defined relative phase.   But to what extent is quantum mechanics necessary to describe  superconductivity? After all, the celebrated Bardeen-Cooper-Schrieffer (BCS) theory~\cite{bcs} is a mean-field theory where the wave function of a superconductor is a product state  with no entanglement  between Cooper pairs in momentum space. 
Within  the mean-field framework,  Cooper pairs are equivalent to classical Anderson pseudospins (i.e., classical angular momentum variables) and the BCS model can be mapped to a classical spin Hamiltonian~\cite{pseudo,enolskii}. Physical properties of the superconductor can thus  be explained in terms of  classical spins, including the excitation spectrum and thermodynamics~\cite{pseudo,coleman,note1}, Josephson effect~\cite{angular}, topological properties of $p$-wave superconductors~\cite{read}, etc.   Moreover, this BCS mean field  is exact for the ground state and low energy excitations in bulk  superconductors~\cite{richardson,roman,baytin}. From this perspective, there are simply  no observable quantum or non-mean-field  effects in equilibrium superconductivity in the thermodynamic limit. 

In the past two decades, there has been considerable theoretical and experimental interest in  coherent far from equilibrium  superconductivity~\cite{barankov1,andreev,simons,kuznetsov1,osc,barankov3,dzero1,dzero2,axt,tomadin,nahum,victor,faribault1,faribault2,matsunaga1,faribault3,beck,matsunaga2,foster2,matsunaga3,manske,foster3,aoki,foster,density,kolya1}. A natural question to ask therefore is, do superconductors exhibit any purely quantum, beyond mean-field effects in their far from equilibrium dynamics? This is the question we address in this paper. Nearly all studies  of the BCS  dynamics employ the   mean-field approach, i.e., exclude quantum fluctuations
 from the beginning and investigate classical Hamiltonian spin dynamics.  However, there is a priori no guarantee that  mean field is  accurate   far from equilibrium because highly excited states contribute to the dynamics and  their effect can accumulate in time.  Numerical studies of   the quantum BCS      dynamics  indeed  suggest that there are deviations from mean field at long times~\cite{faribault1,faribault2,faribault3}. Such studies, however,   cannot conclusively determine the status of mean field in the thermodynamic limit as they  are limited to    small numbers of Cooper pairs  -- quantum dynamics  is  
  essentially
 impossible   to simulate on a classical computer for a macroscopic number of interacting particles.

In this paper, we determine  the  \textit{exact quantum} and \textit{exact classical} long-time dynamics of the BCS Hamiltonian 
with 
interaction strength
inversely proportional to time and compare them  in the thermodynamic limit.  We focus on superconductors   smaller than the coherence length  
that are effectively zero-dimensional  and  whose dynamics is therefore spatially uniform~\cite{turbulence}.
We prepare the system in the  ground state at $t=t_0\to0^+$ and evolve it to $t\to+\infty$, i.e., the interaction strength, $g(t)\propto\frac{\eta}{t}$,  decreases from infinity to zero.  We show that the   classical and quantum dynamics of \textit{local} observables    \textit{coincide exactly}.  Local observables  are sums of quantum averages of operators that contain a finite (in the thermodynamic limit) number of fermion creation and annihilation operators, such as single-particle level occupancies,  superconducting order parameter, and all other $n$-point normal and anomalous averages with finite $n$ and their sums~\cite{clarify_finite}.   The time-dependent mean field is \textit{exact} for such observables in the thermodynamic limit.
At the same time, for  non-local measures, such as the von Neumann entanglement entropy and Loschmidt echo~\cite{echo},  the mean field breaks down  both in and out of equilibrium.

For averages of local operators that commute with the total fermion number operator $\hat N_\mathrm{f}$, the mean field  is  
exact regardless of whether  the initial state is the true quantum ground state, which is  an eigenstate of $\hat N_\mathrm{f}$, or the BCS (mean-field) ground state, which is not.
The situation with observables that change the fermion number, such as the BCS order parameter and other anomalous averages, is more subtle and  sensitive to the way in which the thermodynamic limit is taken.  The BCS  wave function is a sum over states with all possible  $N_\mathrm{f}$,
which makes the anomalous  expectation values
nonzero.   For initial states of this type, we find that while quantum and classical dynamics of anomalous averages coincide 
when we take the thermodynamic limit first, 
if we instead take either the $t_0\to 0^+$  or $\eta\to+\infty$ (adiabatic) limit  first, the anomalous averages do not agree (i.e., these limits do not commute).
This is an inherently quantum mechanical effect (dephasing of sectors with different $N_\mathrm{f}$) that is noticeable in  the far from equilibrium dynamics    already for relatively large superconductors  for a suitable choice of the parameters as we will see.   

We believe our predictions can    be tested in several experimental setups. Ultracold  atoms or ions
interacting via an optical cavity   or   lattice  vibrational (phonon)  mode seem particularly  promising.  Several studies  explained  how to simulate far from equilibrium quantum   BCS dynamics similar to ours  in       these systems~\cite{cavityQED1,cavityQED2,cavityQED3,cavityQED4}.  In particular, it appears  simple to prepare the system
in the $t=0^+$ (infinite coupling) BCS  ground state.   Superconducting coupling    inversely proportional to time  is probably  achievable  as well but this requires further investigation.
 It is possible to  realize  the $\frac{1}{t}$ time dependence of the coupling     in an ultracold atomic Fermi gas near a Feshbach resonance by varying the external magnetic field linearly in time, however, in this scenario our model kicks in not at $t=0^+$ but at a later time as we describe  below. Also promising are various other quantum simulators and quantum computation devices.

 Anomalous averages are matrix elements of operators that contain unequal numbers of fermion creation and annihilation operators.
 Consider, for example, the equal time anomalous Green's function $\langle \hat c_\dn \hat c_\up\rangle$, where $\hat c_\dn$ and
 $\hat c_\up$ are two fermion annihilation operators.     The expectation value of  $\hat c_\dn \hat c_\up$ is zero in any state with definite fermion number  $N_\mathrm{f}$, such as the solution of the nonstationary Schr\"odinger equation that starts  in the exact    ground state at $t=0^+$.  In this case, we  interpret    $\langle \hat c_\dn \hat c_\up\rangle$ at time $t$ as the matrix element  $\langle\Psi_2(t)| \hat c_\dn \hat c_\up|\Psi_1(t)\rangle$  between two solutions $\Psi_1(t)$ and $\Psi_2(t)$, where $\Psi_1(t=0^+)$  is the  ground state    with $N_\mathrm{f}$  fermions and $\Psi_2(t=0^+)$ with  $N_\mathrm{f}-2$.   Throughout this paper we retain the standard  notion of the ``expectation value'', while  using  ``average''  in a more general sense as explained above. We show that   averages of local operators obtained from the exact solutions for the quantum and mean-field (classical) dynamics coincide in the thermodynamic limit.  The above adjustment of the initial condition is  redundant when the initial state of the quantum dynamics is the BCS ground state. In this case, $\langle \hat c_\dn \hat c_\up\rangle=\langle\Phi(t)| \hat c_\dn \hat c_\up|\Phi(t)\rangle$, where
 $\Phi(t=0^+)$ is the $t=0^+$   BCS ground state, i.e., averages are the same as expectation values.  It is these anomalous expectation values that dephase and disagree with mean field for the ``wrong'' order of limits. 
 
 Our conclusions about the domain of applicability of the mean-field treatment   have important implications for the nature of symmetry-broken many-body states. They  hold in and out of equilibrium and we expect them to apply much more generally then to superconductivity alone. While mean-field wave functions are often able to capture the order parameter  and other local observables, the  entanglement properties and many-particle structure are out of reach. The success of mean-field theories to date has thus secretly relied on the symmetry breaking order parameter being a ``classical'' object and not caring  about the nature of the entanglement of the state.

We will see that at long times our system enters a state where the    BCS order parameter and  superfluid density vanish due to dephasing, but energy-resolved anomalous correlation functions are nonzero for any finite $\eta$  -- in the adiabatic limit ($\eta\to+\infty$)  the system evolves  to the zero temperature Fermi gas. More generally, the asymptotic state is a non-Fermi-liquid and  best described as a gapless superconductor whose superfluid features can only be observed through energy resolved quantities such as the  spectral supercurrent density~\cite{supercurrent1,supercurrent2}. In addition,  we  will show that there is an emergent generalized Gibbs ensemble~\cite{gge,gge2} that reproduces exact time averaged values of local observables in the thermodynamic limit.

Solving for the far from equilibrium dynamics of a macroscopically large number of interacting quantum particles is normally an unrealistic task, both with numerical and analytical methods. Moreover, methods based on conventional integrability~\cite{mehta,nba,qa1,qa2}, such as nonequilibrium Bethe ansatz,    quench action, etc., do not work for nonautonomous  (time-dependent) Hamiltonians 
such as the one considered in this work. Fortunately, building on a previous result~\cite{me}, we were able to overcome this obstacle for a class of physically relevant nonautonomous    Hamiltonians. 
The first important observation is that the nonstationary Schr\"odinger equation for the BCS Hamiltonian with coupling inversely  proportional to time is integrable via the off-shell Bethe Ansatz~\cite{me}, a technique~\cite{babujian2}  of  solving      Knizhnik-Zamolodchikov (KZ) equations~\cite{KZ} that describe correlation functions   in the SU(2) Wess-Zumino-Witten model. 
However, the off-shell Bethe Ansatz produces an immensely complicated formal solution, much more complex than the regular Bethe Ansatz,  that does not provide any explicit information about the physical observables or any obvious way to evaluate them. 
The major  technical breakthrough of this work is in the development of  a systematic method to extract the  \textit{exact} and \textit{explicit} late-time wave function of
the quantum problem \textit{and} its semiclassical version from the formal solution. Our method is general and applies equally well to other   integrable
time-dependent Hamiltonians~\cite{kolya,kitaev,volodya,me,aniket}, e.g., to the problem of molecular production in an atomic Fermi gas swept through a Feshbach resonance.

The remaining content   is organized as follows. Section~\ref{sum_sec} is a brief summary of the entire paper. In Sec.~\ref{model_sec} we introduce the quantum and mean-field BCS models, equations of motion, and  initial conditions. In Sec.~\ref{formal_sec} we review
the integral representation of solutions of the nonstationary Schr\"odinger equation for the BCS Hamiltonian with coupling $g(t)\propto\frac{\eta}{t}$. In Sec.~\ref{late_time_sec} we obtain our  first key result  --  exact late time   wave function    for the quantum BCS dynamics launched from the exact ground state at $t=0^+$. Section~\ref{QvsC_sec} presents the exact late-time solution for the corresponding mean-field dynamics -- our second key result. We demonstrate in Sec.~\ref{thm_qm_sec} that exact quantum and mean-field averages of arbitrary local operators coincide in the thermodynamic limit -- the third key result. In Sec.~\ref{gge_sec}
we discuss  the physical properties of the  steady state  our system enters at long times and show that it  conforms to an emergent generalized Gibbs ensemble. We establish in Sec.~\ref{order_sec} that various limits commute for dynamics with definite fermion number.
 However,  they do not commute   for anomalous   expectation values when the initial state is not a fermion number eigenstate, such as the BCS (mean-field) ground state, as we  show in Sec.~\ref{deph_sec}.    In Sec.~\ref{approach_sec} we analyze numerically the approach to the steady state and find that  it  is accessible  in finite time in the  thermodynamic limit. We study the early-time dynamics and   the growth of the   entanglement entropy in Sec.~\ref{early_sec}. We conclude and outline possible directions for future research in Sec.~\ref{conclusion}.
 
\section{Summary of the paper}
\label{sum_sec}

This section is a condensed version of the present paper. We first summarize our key results and then separately list several notable complementary findings. We obtain  four key results in this paper.

\begin{enumerate}[label=(\arabic*), left=0pt, itemsep=0pt, topsep=0pt]

\item The exact long-time solution of the nonstationary Schr\"odinger equation for the BCS Hamiltonian,  see Eq.~\eqref{eq:bcs_model},  with superconducting coupling   $g(t)= \frac{1}{\nu t}$. The initial condition is the exact ground state at $t=0^+$.

\item Exact long-time solution of the mean-field equations of motion for the same time-dependence of the interaction.

\item We show that the far from equilibrium superconductivity   is semiclassical for local observables. We prove this for $g(t)= \frac{1}{\nu t}$  but expect it to be valid much more generally. The semiclassical  picture (mean field) breaks down     for global measures, such as the entanglement entropy and Loschmidt echo.

\item We provide two crucial ingredients for the emergent theory of time-dependent quantum integrability.  First, we show  that the off-shell Bethe Ansatz~\cite{babujian2} is the primary framework in which to study integrable nonautonomous Hamiltonians. We determine if a given time-dependent Hamiltonian is integrable by checking if it goes through  this ansatz~\cite{me,kitaev,volodya}. If yes, we derive an integral representation for  solutions of its nonstationary  Schr\"odinger equation, which is our main tool for answering   physics questions.
 Second, we delineate a method   based on the integral representation to evaluate the solution \textit{explicitly} in relevant limits, such as $t\to0$ and $t\to\pm\infty$. This also solves the many-body Landau-Zener problem for the  Hamiltonian in question by determining  transition probabilities between various asymptotic states.

\end{enumerate}

Let us also     overview  the first three results quantitively  including the main formulas.
The first  one is the exact $t\to+\infty$ asymptotic solution of the nonstationary Schr\"odinger equation for the BCS Hamiltonian~\re{eq:bcs_model} with superconducting coupling   $g(t)= \frac{1}{\nu t}$,
\beg
 |N_\mathrm{f}\rangle_\infty = C\! \sum_{\{\alpha\}} e^{i\Lambda_{\{\alpha\}} } \prod_\alpha\left[ e^{-2it\eps_\alpha} e^{-\frac{\pi\alpha}{\nu}} e^{-i\theta_\alpha} \right] |\{\alpha\}\rangle,
\label{psiinf_sum}
\en
where $C$ is a  normalization constant, $N_\mathrm{f}$ is the number of fermions, $\{\alpha\}$ is the set of doubly occupied single-fermion levels $\eps_\alpha$ (remaining levels are empty), the summation is over all such states with given $N_\mathrm{f}$, and
\beg
  \theta_\alpha=\frac{1}{\nu} \sum_{j\ne\alpha} \ln|\eps_j-\eps_\alpha|,\quad 
\Lambda_{\{\alpha\}}=\frac{1}{\nu} \sum_{\beta\ne\alpha} \ln|\eps_\beta-\eps_\alpha|.
\en
Summation over $j$ is over all levels except $\eps_\alpha$;  $\Lambda_{\{\alpha\}}$ is a double sum over all $\alpha$ and $\beta$ from the set $\{\alpha\}$ such that $\alpha\ne\beta$.  
The initial condition is the exact ground state at $t=0^+$. 

The second key result is an exact solution of the late-time mean-field dynamics [mean-field equations of motion~\re{ceom}] in the thermodynamic limit. The initial state is the BCS (mean-field)   ground state at $t=0^+$ for the same $g(t)$. The mean-field wave function at $t\to+\infty$ is
\beg
\Psi_\mathrm{mf}=\prod_{k=1}^N \left( u_k  +     v_k \hat{c}^{\dagger}_{k\uparrow}\hat{c}^{\dagger}_{k\downarrow}\right) |0\rangle.
\label{psimf_sum}
\en
Here  $\hat{c}^{\dagger}_{k\sigma} (\hat{c}_{k\sigma})$ are the fermionic creation (annihilation) operators  
 for spin projection $\sigma$  and single-particle   level $\eps_{k}$,  $N$ is the number of $\eps_{k}$,  
\beg 
u_k=\frac{ e^{\frac{\zeta_k-i\varphi_k}{2}} }{\sqrt{2\cosh \zeta_k}},\quad v_k=\frac{e^{-2i\eps_k t} e^{\frac{-\zeta_k+i\varphi_k}{2}}}{\sqrt{2\cosh \zeta_k}},
\label{bogol_sum}
\en 
are  the   Bogoliubov amplitudes, 
\beg
\varphi_k=-\frac{1}{\nu}\sum_{j\ne k} \tanh\zeta_j  \ln|\eps_j-\eps_k|,\quad
 \zeta_k= \frac{  \pi (k-\mu) }{\nu},
 \en
 and 
 \begin{equation}
  \mu  
  = \frac{N+1}{2}
  + \frac{N}{2\pi\eta}\ln\left\{
    \frac{\sinh\left[   \frac{ \pi \eta N_\mathrm{f} }{2N}  \right]}
    {\sinh\left[ \pi\eta  -  \frac{ \pi \eta N_\mathrm{f} }{2N}  \right]}\right\}.
  \label{eq:mu_sum}
\end{equation}
 is the chemical potential.
    
 The third key result is that the mean field is exact for local observables in the thermodynamic limit.  
 Consider a product of
 $n$ operators
 \beg
 \hat O=\hat o_{k_1}\dots \hat o_{k_n},
 \en
where $k_1,\dots,k_n$ are any $n$ distinct single-particle labels and each $\hat o_{k}$  is either of the following three operators:  fermion pair creation ($\hat{c}^{\dagger}_{k\uparrow}\hat{c}^{\dagger}_{k\downarrow}$),  annihilation ($\hat{c}_{k\dn}\hat{c}_{k\up}$), or level occupancy  ($\hat n_k=\hat{c}^{\dagger}_{k\up}\hat{c}_{k\up} + \hat{c}^{\dagger}_{k\dn}\hat{c}_{k\dn}$). We say that $\hat O$ is local if $\frac{n}{N}\to0$   in the thermodynamic limit -- the limit $N\to\infty$  keeping the fermion number  density fixed~\cite{local}. 

Suppose $\hat O$ changes the fermion number by $2l$, e.g., $\hat{c}^{\dagger}_{k\uparrow}\hat{c}^{\dagger}_{k\downarrow}$ changes it by $+2$. We claim that the average of $\hat O$ in the exact asymptotic state~\re{psiinf_sum} coincides with its expectation value in
the mean-field wave function~\re{psimf_sum} in the thermodynamic limit, i.e.,
\beg
\langle N_\mathrm{f}+2l | \hat O | N_\mathrm{f}\rangle_\infty= \langle \hat O\rangle_\mathrm{mf}= \langle \hat o_{k_1}\rangle_\mathrm{mf}\dots \langle \hat o_{k_n}\rangle_\mathrm{mf},
\en
where $\langle\dots\rangle_\mathrm{mf}\equiv \langle \Psi_\mathrm{mf} | \dots | \Psi_\mathrm{mf}\rangle$. The  expectation value of a product in   $\Psi_\mathrm{mf}$   is a product of the expectation values, since  it is a product state.      $\langle \hat o_{k}\rangle_\mathrm{mf}$, in turn, are straightforward to evaluate:
$$
\langle \hat{c}^{\dagger}_{k\uparrow}\hat{c}^{\dagger}_{k\downarrow}\rangle_\mathrm{mf}=u_k v_k^*,\\\, \langle\hat{c}_{k\dn}\hat{c}_{k\up}\rangle_\mathrm{mf}=u_k^* v_k, \\\, \langle \hat n_k\rangle_\mathrm{mf}=2|v_k|^2.
$$
Therefore, not only do we show that the time-dependent BCS mean field is exact in the thermodynamic limit, but also evaluate  quantum averages of arbitrary local operators in this limit.

\subsection{Complementary results}

In addition to the above key results, we obtain a number of other interesting results.

\begin{enumerate}[label=(\alph*), left=0pt, itemsep=0pt, topsep=0pt]

\item The steady state of the exact time evolution of the BCS Hamiltonian with coupling $g(t)= \frac{1}{\nu t}$ is a gapless superconductor similar to Phase I in interaction quenched superconductors~\cite{foster}.  Indicators of fermionic superfluidity integrated over  the single-particle energy, such as the BCS order parameter, energy gap for pair-breaking excitations and superfluid density vanish in this state. Nevertheless,   it is a superfluid state,
which  is seen in  energy resolved measures, e.g.,   the  spectral supercurrent density.

\item This steady state is nonthermal, but is described by an emergent generalized Gibbs ensemble (GGE)  in the thermodynamic limit
with   level occupation numbers $\hat n_k$ emerging as the integrals of motion at $t\to+\infty$. This is a nontrivial  property of the steady state as it means that expectation values of local operators   can be expressed in terms of only $N$ GGE parameters as opposed to $2^N$ for a generic state.

\item We find through  numerical analysis that a suitably defined distance to the  steady state tends to zero as $\frac{R}{t^3}$, where $R$
is finite in the thermodynamic limit. Therefore, even in this limit the system is able to approach the steady state arbitrarily closely in finite time.  

\item Consider the time evolution with the nonautonomous quantum BCS Hamiltonian launched from an initial state that is not an 
an eigenstate of the total fermion number  operator $\hat N_\mathrm{f}$ at $t=t_0$ and a local operator $\hat O$ that does not commute with $\hat N_\mathrm{f}$. We find that the parameter that controls the ratio of the exact and mean-field expectation values of $\hat O$   at $t\to+\infty$ is  
\beg
\mathtt{Q}=\frac{ \eta^2 \ln^2 \frac{t_*}{t_0} }{ 2N}
\label{Q_sum}
\en
as opposed to $\frac{1}{N}$, which controls other quantum fluctuations (finite size corrections) in and out of equilibrium. 
Here $\eta=\frac{N}{\nu}$ is the dimensionless  coupling constant that remains finite in the thermodynamic limit, $t_*\sim \frac{1}{W}$, and $W$ is the bandwidth of $\eps_k$. \eref{Q_sum} shows that the thermodynamic limit $N\to\infty$ does not commute with the $t_0\to 0^+$  
and  adiabatic ($\eta\to+\infty$) limits. At the same time, these limits mutually commute for local operators that conserve  $N_\mathrm{f}$ and   initial states with definite $N_\mathrm{f}$.
  
\item  We  determine  the short-time   dynamics of the bipartite von Neumann entanglement entropy   in the thermodynamic limit,
\begin{equation}
  \begin{gathered}
    \mathtt{S}_{\mathrm{ent}}
    = \sqrt{1 + \frac{\tau^{2}}{4}}
    \coth^{-1}\left[\sqrt{1 + \frac{\tau^{2}}{4}}\right] + \ln  \frac{\tau}{4},
  \end{gathered}
  \label{eq:coh_state_ent_sum}
\end{equation}
where $\tau=\eta\ln \frac{t}{t_0}$. 
This result is for the quantum evolution launched from the BCS product state at $t=t_0$. The entropy grows monotonically from $ \mathtt{S}_{\mathrm{ent}}=0$ at $t=t_0$. It remains finite in the thermodynamic limit emphasizing once more the failure of mean field   for global quantities such as  $\mathtt{S}_{\mathrm{ent}}$ (within mean-field approach $ \mathtt{S}_{\mathrm{ent}}=0$ at all times). Interestingly, the entire growth of $\mathtt{S}_{\mathrm{ent}}$  is due to the interaction part of the BCS Hamiltonian. For finite $N$, the monotonous growth stops at $\tau\sim \sqrt{N}$.  After this  the entropy   shows recurrences    with a maximum value $\mathtt{S}_{\mathrm{ent}}\sim \frac{1}{2}\ln N$.

\end{enumerate}

\section{Quantum and classical BCS Models} 
\label{model_sec}

We study two related models in this paper. One is the quantum BCS Hamiltonian with  interaction strength  inversely proportional to time and the other is its classical (mean-field) counterpart. We start with the quantum model, introduce Anderson pseudospin-$\frac{1}{2}$ operators, and review how the classical  Hamiltonian arises in the $\hbar\to0$ limit and,  independently, in the mean-field approach.   

The quantum BCS model  describes  pairing interactions between fermions moving in a given single-particle potential~\cite{dirty},  
 \begin{equation}
   \hat{H} =
   \sum_{j,\sigma} 2\eps_j \hat{c}^{\dagger}_{j\sigma}\hat{c}_{j\sigma}
   -g(t)\sum_{j,k} \hat{c}^{\dagger}_{j\uparrow}\hat{c}^{\dagger}_{j\downarrow}\hat{c}_{k\downarrow}\hat{c}_{k\uparrow},
   \label{eq:bcs_model}
 \end{equation}
where  $\hat{c}^{\dagger}_{j\sigma} (\hat{c}_{j\sigma})$ creates (annihilates) a fermion
with spin projection $\sigma$  on  the single-particle level $\eps_{j}$. The   superconducting coupling
$g=g(t)$  has  dimensions of energy and is usually a constant but  will  depend on time in the present paper.  The pairing is between  the states $\ket{j\!\up}$ and   $\ket{j\!\dn}=T_R\ket{j\!\up}$ of the same energy $\eps_j$, where $T_R$ is the time reversal operation.  When the single-particle potential is zero,
the momentum $\bm p$ is a good  quantum number and therefore   $\ket{j\!\up}\to \ket{\bm p\!\up}$ and   $\ket{j\!\dn}\to \ket{-\bm p\!\dn}$. With these replacements the more general \eref{eq:bcs_model} becomes the original BCS Hamiltonian~\cite{bcs}.

We consider a nonautonomous (driven) BCS model where the coupling is inversely proportional to time,
\beg
g(t)=\frac{1}{\nu t}\equiv \frac{\eta}{N t}.
\label{g}
\en
Here $\hbar=1$ making both $\nu$ and $\eta=\frac{N}{\nu}$ dimensionless. The ``rate" $\nu$ must be proportional to the number $N$ of single-particle levels $\eps_j$, so that the kinetic and interaction terms in \eref{eq:bcs_model}  both scale as $N$ in the thermodynamic limit. 

The time dependence~\re{g} can be realized in ultracold atomic Fermi gases at least for sufficiently small values of $\frac{\eta}{t}$. Most Feshbach resonances experimentally realized to date  are broad. In the broad resonance limit and at sufficiently weak coupling,
\eref{eq:bcs_model} is a good description of the gas~\cite{victor1}. The coupling constant $g$ is inversely proportional to a linear function of  the detuning from the resonance, which, in turn, is  linear in the external magnetic field. Varying the magnetic field linearly with time, we make  $g\propto \frac{\eta}{t}$. The weak coupling condition means that $\frac{\eta}{t}$ has to be small, i.e.,
  we have to start our dynamics at a sufficiently large $t_0$.   Since the only
  energy scale not related to the interaction is the Fermi energy $\eps_F$, the more precise condition is $\eps_F t_0\gtrsim \eta$, see Ref.~\onlinecite{victor1} for   the relationship between $g$ and the magnetic field and criteria of applicability
of the BCS model~\re{eq:bcs_model}.  In Introduction, we also mentioned other experimental platforms where our setup can potentially be realized.

Consider a quantum spin Hamiltonian
\begin{equation}
  \hat{H}(t) =
  \sum_{j=1}^N 2\eps_j \hat{s}_j^z
  -g(t)\sum_{j,k=1}^N\hat{s}_j^+\hat{s}_k^-.
  \label{eq:H}
\end{equation}
When the magnitude of spins is $s=\frac{1}{2}$ this is the BCS Hamiltonian~\re{eq:bcs_model} recast in terms of Anderson pseudospins~\cite{pseudo}
\begin{subequations}
  \begin{align}
 & \hat s_{j}^{z} = \frac{1}{2}\left(\hat{c}^{\dagger}_{j\uparrow} \hat c_{j\uparrow} + \hat{c}^{\dagger}_{j\downarrow} \hat c_{j\downarrow}- 1\right),\label{zfermions}\\
  &\hat s_{j}^{+} = \hat{c}^{\dagger}_{j\uparrow}\hat{c}^{\dagger}_{j\downarrow},\quad \hat s_{j}^{-} = \hat{c}_{j\dn}\hat{c}_{j\up}.
  \end{align}
  \label{z+}
\end{subequations}
Pseudospin  operators satisfy the usual  SU(2) commutation relations.
  On the subspace of unoccupied and doubly occupied (unblocked)   levels $\eps_j$, the magnitude of spins
   $s=\frac{1}{2}$. Singly occupied (blocked) levels decouple and do not participate in the dynamics and we exclude  them
from \eref{eq:bcs_model}.  Sometimes, it is helpful to study the model~\re{eq:H}  for general $s$. In such cases, we will often refer to
 it as the ``generalized BCS Hamiltonian".

We  obtain the classical counterpart of the quantum  Hamiltonian~\re{eq:H} by replacing  quantum spins $\hat{\bm s}_j$ with
classical angular momentum variables (classical spins) $\bm S_j$  of length $S$
\begin{equation}
  H(t) =
  \sum_{j=1}^N 2\eps_j S_j^z
  -g(t)\sum_{j,k=1}^N S_j^+S_k^-,
  \label{eq:class_H}
\end{equation}
where $S_j^\pm=S_j^x\pm i S_j^y$. The variables $\bm S_j$ are equipped with the standard angular momentum Poisson brackets 
$\left\{S^{a}_{j}, S^{b}_{k}\right\} = \delta_{j,k}\epsilon^{abc}S_{j}^{c}$. By the quantum-to-classical correspondence principle, the classical BCS Hamiltonian~\re{eq:class_H} is the $\hbar\to0$ and $s\to\infty$ limit  with $S=\hbar s =\mbox{fixed}$ of the quantum Hamiltonian~\re{eq:H}. In this approach, the length $S$ of the classical spins is arbitrary.
 
\subsection{Mean-field equations of motion}

There is an alternative route leading from  the quantum~\re{eq:H} to the classical~\re{eq:class_H} Hamiltonian that fixes the length $S$ of $\bm S_j$ -- the mean-field approximation. Consider the Heisenberg equations of motion for $\hat{\bm s}_j$
\begin{equation}
  \frac{ d\hat{ \bm s }_j }{dt}= i[\hat{H}(t),\hat{\bm{s}}_j]=2(\eps_j \bm z -\hat{\bm\Delta})\times \hat{\bm{s}}_j,
\label{heom}
\end{equation}
where  $\hat{\bm\Delta}=\hat\Delta_x  {\bm x} + \hat\Delta_y  {\bm y}$,
\beg
\hat\Delta_x =g \sum_{k=1}^N \hat s_k^x, \quad \hat\Delta_y= g  \sum_{k=1}^N \hat s_k^y,
\en 
and $ \bm x$, $\bm y$, and $\bm z$ are unit vectors along the coordinate axes. Since $\hat{\bm\Delta}$ is a sum of a large number of spin-$\frac{1}{2}$ operators, it is natural to expect  it to behave classically~\cite{pseudo}, $\hat{\bm\Delta}\approx \langle \hat{\bm\Delta}\rangle$,
in the thermodynamic limit. The replacement of $\hat{\bm\Delta}$ with  $\langle \hat{\bm\Delta}\rangle$ in \eref{heom} is
the   \textit{mean-field approximation}. Note that this is the only approximation involved in deriving the classical Hamiltonian.

Making this replacement and then taking the quantum average with respect to the time-dependent state of the system, we obtain equations of motion for $\langle \hat{\bm s}_j\rangle$ identical to Hamilton's equations of motion  with the Hamiltonian~\re{eq:class_H}
when we set  $\langle \hat{\bm s}_j\rangle=\bm S_j $,
\begin{equation}
  \frac{ d \bm S_j }{dt}= \left\{ \bm S_j, H(t) \right\} =2(\eps_j \bm z -\bm\Delta )\times \bm S_j,
\label{ceom}
\end{equation}
where $\bm\Delta=\langle \hat{\bm\Delta}\rangle=\Delta_x  \bm x + \Delta_y  \bm y$,
\beg
\Delta_x =g \sum_{k=1}^N  S_k^x,\quad \Delta_y = g  \sum_{k=1}^N  S_k^y,
\en 
and the usual  BCS order parameter reads
\beg
\Delta =g\sum_{k=1}^N S_k^-=g\sum_{k=1}^N \langle \hat s_k^-\rangle=\Delta_x - i\Delta_y.
\label{usualdelta}
\en
The length $S$ of $ \langle \hat{\bm s}_j\rangle =\bm S_j$  is conserved by the mean-field time evolution. 

\subsection{BCS and projected BCS wave functions}

Suppose we start the mean-field time evolution   in a BCS-like product state  
\beg
\Psi_\mathrm{BCS}=\prod_k (u_k+v_k \hat s_k^+) |0\rangle= \prod_k\left(u_k|\!\dn\rangle+v_k|\!\up\rangle\right).
\label{psibcs}
\en
 Then, the wave function will remain  a product state of this form at all times and  
 \beg
 S=|{\bm S}_{j}|=\frac{1}{2}.
 \label{length89}
 \en
  In \eref{psibcs} the vacuum $|0\rangle=|\!\dn\dn\dn\dots\rangle$ is the state with all spin-$\frac{1}{2}$ down (all levels $\eps_k$ empty),   $|\!\!\up\rangle$ and $|\!\!\dn\rangle$ are the up and down states of spin $\hat {\bm s}_k$, and $(u_k, v_k)$ is a pair of  complex numbers (Bogoliubov amplitudes).
    While the mean-field approximation generally appears very reasonable for large $N$, its validity is questionable, e.g., when  $\langle \hat{\bm\Delta}(t)\rangle$ vanishes as  in the normal state and for certain interaction quenches~\cite{dzero1,T}. In such cases, quantum fluctuations of $ \hat{\bm\Delta}(t) $ can be important.
    
We will also need the projection 
\beg
\Psi_\mathrm{PBCS}=P_{N_\up}\Psi_\mathrm{BCS}=P_{N_\up}\prod_k \left(u_k|\!\dn\rangle+v_k|\!\up\rangle\right)
\label{projBCS}
\en 
of the BCS wave function onto a fixed fermion number $N_\mathrm{f}=2N_\up$ subspace, where $N_\up$ is the   number of up spins [see \eref{zfermions}]. It is convenient to write $\Psi_\mathrm{PBCS}$ as
\beg
\Psi_\mathrm{PBCS}=\frac{1}{2\pi}\int_0^{2\pi}\!\!\!\!\! d\phi e^{i\phi N_\up} \prod_k  \left(u_k|\!\dn\rangle+e^{-i\phi} v_k|\!\up\rangle\right).
\label{pbcs}
\en
 Using $\Psi_\mathrm{PBCS}$ instead of $\Psi_\mathrm{BCS}$ produces corrections of order $N_\up^{-1}$ for large $N_\up$ to the low energy equilibrium properties~\cite{pseudo}.

We discussed above two ways to obtain the classical BCS Hamiltonian~\re{eq:class_H}. One is to 
send  the magnitude of the quantum spins $s\to\infty$ and the other is the mean-field approach. The end Hamiltonian and equations of motion are the same   due to Ehrenfest's theorem and the nature of  mean-field approximation which replaces $\langle \hat A_1 \hat A_2\rangle\to \langle \hat A_1 \rangle \langle \hat A_2\rangle$. The difference is that
in the large spin limit the lengths of the classical spin vectors are arbitrary, while in the mean-field approach they are determined by the initial quantum wave function. Another distinguishing feature of the mean-field approach is its connection to an approximate (mean-field) solution of the Schr\"odinger equation. Indeed, assuming a product initial state and given the solution  ${\bm S}_{j}(t)=\langle \hat {\bm s}_{j}(t)\rangle$ of classical equations of motion, we can reconstruct the many-body product wave function   at time $t$ because for   spin-$\frac{1}{2}$ the average $\langle \hat {\bm s}_{j}(t)\rangle$   determines its wave function up to an overall phase. In what follows, we set $S=\frac{1}{2}$ and  identify classical and mean-field dynamics, i.e., treating the classical variables as quantum averages of the corresponding operators we 
associate a product BCS wave function with the classical spin distribution. 

\subsection{Initial conditions}
\label{ic_sec}

Both  quantum and classical BCS Hamiltonians conserve the $z$-component of their total spins
\beg
\hat{\bm \jmath}=\sum_{k=1}^N \hat{\bm s}_k,\quad  \bm J=\sum_{k=1}^N \bm S_k,
\en
\eref{zfermions} implies that the total fermion number operator $\hat N_\mathrm{f}=2\hat N_\up$, where $\hat N_\up$ counts the number of up pseudospins. 
  In terms of  $\hat N_\mathrm{f}$ and $\hat N_\up$,
the $z$-components of total quantum and classical spins read
 \begin{subequations}
\begin{align}
&  \hat\jmath_z=\frac{\hat N_\mathrm{f}-N}{2}=\hat N_\up-\frac{N}{2},\\ 
& J_z=  \frac{\langle \hat N_\mathrm{f}\rangle  -N}{2}=\langle \hat N_\up\rangle -\frac{N}{2}= \langle \hat\jmath_z\rangle\label{classJzNup}.
\end{align}
 \end{subequations}
 
 We initiate the quantum evolution with fixed  fermion number (fixed $\jmath_z$)  in the  ground state of the Hamiltonian~\re{eq:bcs_model}, or equivalently Hamiltonian~\re{eq:H} for $s=\frac{1}{2}$, at $t=0^+$, which  up to a diverging multiplicative constant takes the form 
 \beg
   \hat H_\mathrm{int}\propto - \hat \jmath_+ \hat\jmath_-=- \jmath(\jmath+1) +\hat\jmath_z^2-\hat\jmath_z,
   \en
    where $\jmath(\jmath+1)$ is the eigenvalue of $\bm\jmath^2$. The  ground state of $\hat H_\mathrm{int}$ with $\jmath_z=N_\up-\frac{N}{2}$ is a symmetric combination of all states with $N_\up$ up and $N-N_\up$ down spins
 \beg
 \Psi_0(N_\up)={\binom{N}{N_\up} }^{-\frac{1}{2}}\sum_{ \{ \alpha\} } | \{ \alpha \} \rangle\propto \hat \jmath_+^{N_\up}|0\rangle,
 \label{psi0}
 \en
 where $ | \{ \alpha \} \rangle$ is a state with spins at positions $\{ \alpha \}=\{\alpha_1,\alpha_2,\dots,\alpha_{N_\up}\}$ up and the remaining spins down. The summation is over all such states, i.e., over all sets  $\{\alpha \}$.   The ground state maximizes the magnitude $\jmath$ of the total spin,  $\jmath=\frac{N}{2}$.     Note that $\Psi_0(N_\up)$ is a projected BCS state of the form
 \beg
  \Psi_0(N_\up)\propto P_{N_\up}\prod_k \left(|\!\dn\rangle+ |\!\up\rangle\right).
  \label{projpsi0}
  \en
  
The classical Hamiltonian~\re{eq:class_H} at $t=0^+$ is  
\beg
H_\mathrm{int}\propto -J_+J_-=-\bm J^2 +J_z^2.
\en
In the minimum energy spin configuration, all spins are aligned in the same direction and $|\bm J|=\frac{N}{2}$.  Up to a nonessential rotation around the $z$-axis, this spin configuration is
\beg
S_j^z=\frac{J_z}{N},\quad S_j^x=\frac{J_\perp}{N}, \quad S_j^y=0,
\label{classini}
\en
where $J_z^2+J_\perp^2=\frac{N^2}{4}$.   \eref{classini} is our initial condition for the classical dynamics.  

Consider, in particular, the classical ground state~\re{classini} for   $J_z=0$. In this state, all spins are along the $x$-axis, $\langle \hat {\bm s}_j\rangle=\bm S_j=\frac{ \bm x }{2}$. The corresponding
BCS wave function is
\beg
\Psi_\mathrm{BCS}(t=0^+)=|\rightarrow \rightarrow \rightarrow \dots\rangle=\frac{1}{2^{\frac{N}{2}} } \prod_k \left(|\!\dn\rangle+|\!\up\rangle\right),
\label{alongx}
\en
where $\rightarrow$ indicates  spin-$\frac{1}{2}$ pointing along the positive $x$-axis.
This is the ground state predicted by the BCS theory at infinite coupling for $\langle \hat N_\mathrm{f} \rangle=N$ (number of fermion pairs is half the number of available single-particle states). We note that this value of $\langle \hat N_\mathrm{f} \rangle$ is most relevant and most frequently   studied for $s$-wave superconductors, where the pairing interaction is between fermions in a narrow window around the Fermi level. Since the density of states is approximately constant and the window is centered at the Fermi energy, the number of fermion pairs is half the number of levels involved in  superconductivity.
The BCS state~\re{alongx} corresponds to $u_j=v_j=1$ in \eref{psibcs}. These are indeed the values of the Bogoliubov amplitudes in the BCS ground state for infinite coupling [$g(t)=+\infty$ for $t=0^+$].  It is not an eigenstate of  the quantum Hamiltonian and does not possess   a definite
number of fermions. However, the average fermion number is equal to $N$ as in the exact ground state with $N$ fermions and, moreover, it reproduces   the exact ground state energy to the leading order in $\frac{1}{N}$. We use $\Psi_\mathrm{BCS}$ as another choice of the initial condition at $t=t_0$, which is especially important for observables that do not conserve $N_\mathrm{f}$.

Throughout this paper we support the analytic calculations   against exact numerical simulations of the classical and quantum models.
The   classical dynamics is obtained by directly solving \eref{ceom} with the numerical  ODE solver within MATLAB. 
Similarly, the quantum dynamics is obtained by direct simulation of the nonstationary  Schr\"odinger equation 
 for the Hamiltonian~\re{eq:H} with $s=\frac{1}{2}$ (i.e., the quantum BCS Hamiltonian) and $\hbar=1$.   Working in the eigenbasis of  $\hat s_j^z$ and
identifying $\uparrow$ with $1$ and $\downarrow$ with $0$,
we represent each basis vector as a binary number of digital size $N$, which we then convert to an integer index~\cite{ed}. 
Employing this basis and a PDE solver,  we compute the time-dependent components of $\Psi(t)$ and  evaluate various expectation values and the  entanglement entropy. We use the same initial conditions~\re{psi0} and \re{classini} for quantum and classical dynamics in  numerical  simulations and   analytical calculations, except in simulations we   set the initial $t$ to a very small nonzero value $t_0$ in the Hamiltonian  and carefully handle the limit $t_0\to 0^+$.

\section{Formal solution for quantum dynamics}
\label{formal_sec}

 In this section, we review the ``formal'' exact  solution~\cite{me} of the nonstationary Schr\"odinger equation for the generalized BCS Hamiltonian~\re{eq:H} with $g(t)$ given by \eref{g} and spins of arbitrary magnitude $s$. We dub  this solution ``formal'' as it  is  extremely   complicated, inexplicit, and  superficially  appears useless for obtaining concrete physical information.  This superficial impression turns out to be incorrect, and, with some additional work, we will derive from this solution   explicit
  answers  for the late-time wave function and observables for the quantum BCS model ($s=\frac{1}{2}$) later in this paper. Furthermore, in Appendix~\ref{classApp} we  derive the late-time   classical (mean-field)  BCS dynamics with this $g(t)$ by taking the $s\to\infty$ limit of the formal solution.
 
 Amazingly, there are three different kinds of integrability of the BCS model:  quantum, classical, and time-dependent.  The first one is the regular Bethe Ansatz integrability that implicitly provides the exact many-body eigenstates and energies of the quantum BCS Hamiltonian at fixed value of the interaction constant $g$~\cite{rich1,rich2,gaudin1,gaudin,sklyanin,dukelsky}. Classical integrability, also known as  Liouville-Arnold integrability, guarantees an exact solution of the Hamilton's equations of motion for the classical BCS model~\cite{kuznetsov1,enolskii}, also at a fixed (time-independent)  $g$. Most important for us here is the third kind --  integrability of the nonstationary Schr\"odinger equation for the nonautonomous BCS Hamiltonian with  $g=g(t)=\frac{1}{\nu t}$. We associated  this type of integrability with the off-shell Bethe Ansatz in the Introduction.  Here it is worthwhile to emphasize that the name ``off-shell Bethe Ansatz'' is somewhat  misleading because, unlike the usual Bethe Ansatz, this is not as of now a general technique   applicable to many different models, but a sequence of steps that work   only for the BCS and closely related models that originate from the Gaudin algebra~\cite{me,dukelsky,babujian2}. It is not unusual when both  quantum and classical versions of a model are integrable or even superintegrable, such as the  harmonic oscillator or the Coulomb potential. However, it is much more rare when in addition there is an integrable nonautonomous version of the same model.

The general solution $\Psi(t)$~\cite{me} of the nonstationary Schr\"odinger equation for  the nonautonomous generalized BCS Hamiltonian~\re{eq:H} with spins of magnitude $s$ and $z$-projection of the total spin $\jmath_z=N_\up-\frac{N}{2}$   (when $s=\frac{1}{2}$, this value of $\jmath_z$ corresponds to $N_\up$ up spins and $2N_\up$ fermions) is an $N_\up$--fold contour integral over variables $\lam_1,\dots,\lam_{N_\up}$,  
 \beg
 \Psi(t) = \oint_\gamma d\bm\lambda \exp\left[{-\frac{i {\cal S}(\bm\lam,\bm\eps, t)}{\nu}}\right]   \Xi(\bm \lambda,\bm\eps), 
\label{psi}
\en
where $\bm \eps=(\eps_1,\dots,\eps_N)$, $\bm\lambda=(\lambda_1,\dots,\lambda_{N_\up})$, $d\bm\lam= d\lam_1\dots d\lam_{N_\up}$, 
and
\beg
  \Xi(\bm \lambda,\bm\eps)=\prod_{\alpha=1}^{N_\up} \hat L^+(\lambda_\alpha)|0\rangle,\quad \hat L^+(\lambda)=\sum_{j=1}^N \frac{\hat s_j^+}{\lambda - \eps_j}.
\label{phi}
\en
  The quantity ${\cal S}(\bm\lam,\bm\eps, t)$ is known as the Yang-Yang action,
\beg
\begin{split}
{\cal S}(\bm\lam,\bm\eps, t)= 2\nu t&\sum_\alpha \lam_\alpha +
2s\sum_j\sum_\alpha \ln(\eps_j-\lam_\alpha)\\
&- \sum_\alpha\sum_{\beta\ne\alpha}\ln(\lam_\beta-\lam_\alpha),
\end{split}
\label{S}
\en
 where we dropped the terms that contribute only to the time-independent overall (global) phase of   $\Psi(t)$. The choice of the contour $\gamma$ in \eref{psi} must be such that the integrand is single-valued and $\Psi(t)$ satisfies the initial condition.
 
 \section{Exact late-time quantum BCS dynamics}
 \label{late_time_sec}
 
 Here we use the formal solution from the previous section to evaluate the late-time wave function $\Psi_\infty(N_\up)$ and observables $\langle \hat s_k^+ \hat s_j^-\rangle$ and $\langle \hat s_j^z\rangle$  for the quantum BCS dynamics with the time-dependent Hamiltonian~\re{eq:H} with spins of magnitude $s=\frac{1}{2}$ and $\jmath_z=N_\up-\frac{N}{2}$  [or equivalently  the time-dependent BCS Hamiltonian~\re{eq:bcs_model} with $N_\mathrm{f}=2N_\up$ fermions]. We check our analytical answers  against direct numerical simulations. 
 In Sec.~\ref{thm_qm_sec}, we will obtain the late-time asymptotic behavior of general $n$-point quantum averages in  the thermodynamic limit.

 At large $t$ the  integrand in \eref{psi} is highly oscillatory. The integral therefore localizes to the vicinity of the stationary points of the
 Yang-Yang action. The stationary point equations $\frac{\partial \mathcal{S}}{\partial \lam_\alpha}=0$ read
 \beg
2\nu t+\sum_j \frac{1}{\lam_\alpha-\eps_j}=\sum_{\beta\ne\alpha} \frac{2}{\lam_\alpha-\lam_\beta},\quad \alpha=1,\dots,N_\up.
\label{richi}
\en
These are the well-known Richardson equations that determine the exact spectrum of the BCS Hamiltonian~\cite{rich1,rich2,gaudin1,gaudin,sklyanin,dukelsky}. In our context, they provide the instantaneous spectrum at time $t$. In the instantaneous ground state at $t=0^+$ all $\lam_\alpha$ diverge as $(\nu t)^{-1}$,
see Ref.~\onlinecite{sasha}. This implies that we must choose  integration contours $\gamma$ in \eref{psi} so that  the contour for  each $\lam_\alpha$  can be deformed to infinity without encountering essential singularities, i.e., $\gamma$ must enclose
all $\eps_j$.
 
 When $t \to +\infty$, each $\lam_\alpha$ approaches one of the $\eps_j$ to keep the left hand side of \eref{richi} finite. This means that the instantaneous spectrum approaches that of the noninteracting Fermi gas.
  Let $\lam_\alpha\to\eps_\alpha$. 
  The set of $N_\up$ integers $\{\alpha\}$ specifies which spins are flipped (up).  \eref{richi} implies  that for large $t$
\beg
\lam_\alpha=\eps_\alpha+\frac{1}{2\nu t}.
\label{roots}
\en
 It now follows from \eref{phi}    that at the stationary point  for $t \to +\infty$ 
 \beg
  \Xi(\bm \lambda,\bm\eps) \to |\{ \alpha \}\rangle,
  \label{phi1}
  \en
up to an overall constant. Here $|\{ \alpha \}\rangle$ is the state obtained from the vacuum by flipping $N_\up$ spins at positions $\{ \alpha \}$, i.e., the same state as in \eref{psi0}.
For example, for $N=4$ and $\{ \alpha \}=\{2,4\}$,  we have $| \{2,4\} \rangle=|\dn\up\dn\up\rangle$.

Now let us evaluate the Yang-Yang action on the stationary points. Substituting  \eref{roots} into \eref{S} and neglecting terms of order
$t^{-1}$, we find
$$
{\cal S}_{\{ \alpha\}}=    \sum_\alpha \sum_{j\ne\alpha}\ln (\eps_j-\eps_\alpha)-
2\sum_{\beta>\alpha}\ln|\eps_\beta-\eps_\alpha|+2\nu t\sum_\alpha\eps_\alpha,
$$
where we also dropped a constant that is the same for all $\{ \alpha \}$ and therefore only contributes to the global phase of the wave function, which we do not seek to determine.
  Greek indices $\alpha$ and $\beta$ here and below are from the set $\{ \alpha \}$ and $j$ takes all values from $1$ to $N$.
We rewrite the  first term on the right hand side as
$$
 \sum_\alpha \sum_{j\ne\alpha}\ln (\eps_j-\eps_\alpha)= \sum_\alpha\Bigl[ -i\pi\alpha+ \sum_{j\ne\alpha}\ln |\eps_j-\eps_\alpha| \Bigr].
 $$
Here we used $\ln(-1)=\ln e^{-i\pi}=-i\pi$. This choice of the branch of the logarithm is dictated by the physical requirement that
in the adiabatic limit $\nu\to0^+$ the system stays in the ground state at $t\to +\infty$.
  The  $-i\pi \alpha$ in the above equation arises from counting the number of $\eps_j$ smaller than $\eps_\alpha$. Each such term contributes $\ln(-1)$.
There are $\alpha-1$ terms and replacing $\alpha-1\to\alpha$   here only changes the norm of the wave function.
Therefore,
\beg
\begin{split}
{\cal S}_{\{ \alpha \}}= &2\nu t\sum_\alpha \eps_\alpha - i\pi \sum_\alpha\alpha\\
 + &
 \sum_\alpha  \sum_{j\ne\alpha}\ln |\eps_j-\eps_\alpha|
   -
2 \sum_{\beta>\alpha}\ln|\eps_\beta-\eps_\alpha|.
\end{split}
\label{Seq}
\en
A  compact and useful way to write this expression is
\beg
{\cal S}_{\{ \alpha \}}= 2\nu t\sum_\alpha \eps_\alpha - i\pi \sum_\alpha\alpha-2\sum_{k>j} \hat s_j^z  \hat s_k^z  \ln |\eps_j-\eps_k|.
\label{optS}
\en
 This ${\cal S}_{\{ \alpha \}}$ is equivalent to \eref{Seq}  when applied to the state $|\{ \alpha \}\rangle$, up to a constant that is independent of $\{\alpha\}$.  

The asymptotic wave function   is a sum over all stationary points
\beg
 \Psi_\infty(N_\up)\equiv\Psi(t\to+\infty)=\!\sum_{\{ \alpha\}} e^{-\frac{ i{\cal S}_{\{ \alpha \}} }{\nu} }| {\{ \alpha \} }\rangle.
 \label{psi_sum}
 \en
Using \eref{Seq}, we obtain
 up to an overall complex constant  (normalization and the global phase of the wave function)
\beg
\Psi_\infty(N_\up) =  \sum_{\{\alpha\}} e^{i\Lambda_{\{\alpha\}} } \! \prod_\alpha\left[ e^{-2it\eps_\alpha} e^{-\frac{\pi\alpha}{\nu}} e^{-i\theta_\alpha}  \right] |\{\alpha\}\rangle,
\label{psiinf}
\en
where
\beg
  \theta_\alpha=\frac{1}{\nu} \sum_{j\ne\alpha} \ln|\eps_j-\eps_\alpha|,\quad 
\Lambda_{\{\alpha\}}=\frac{1}{\nu} \sum_{\beta\ne\alpha} \ln|\eps_\beta-\eps_\alpha|.
\en
The Hessian arising from integrating over the vicinity of stationary points goes into this constant as well. Note that $\Lambda_{\{\alpha\}}$ is a double sum over all $\alpha$ and $\beta$ from the set $\{\alpha\}$ such that $\alpha\ne\beta$. This phase is  one of the two sources of quantumness   in the late-time dynamics, the other source being 
the difference between the BCS and projected BCS wave functions, \eref{psibcs} and \eref{pbcs}, respectively. Without $\Lambda_{\{\alpha\}}$, the late-time wave function $\Psi_\infty$ is of the form of a projected BCS state.   

\begin{figure}[t!]
\begin{center}
\includegraphics[width = 0.48\textwidth]{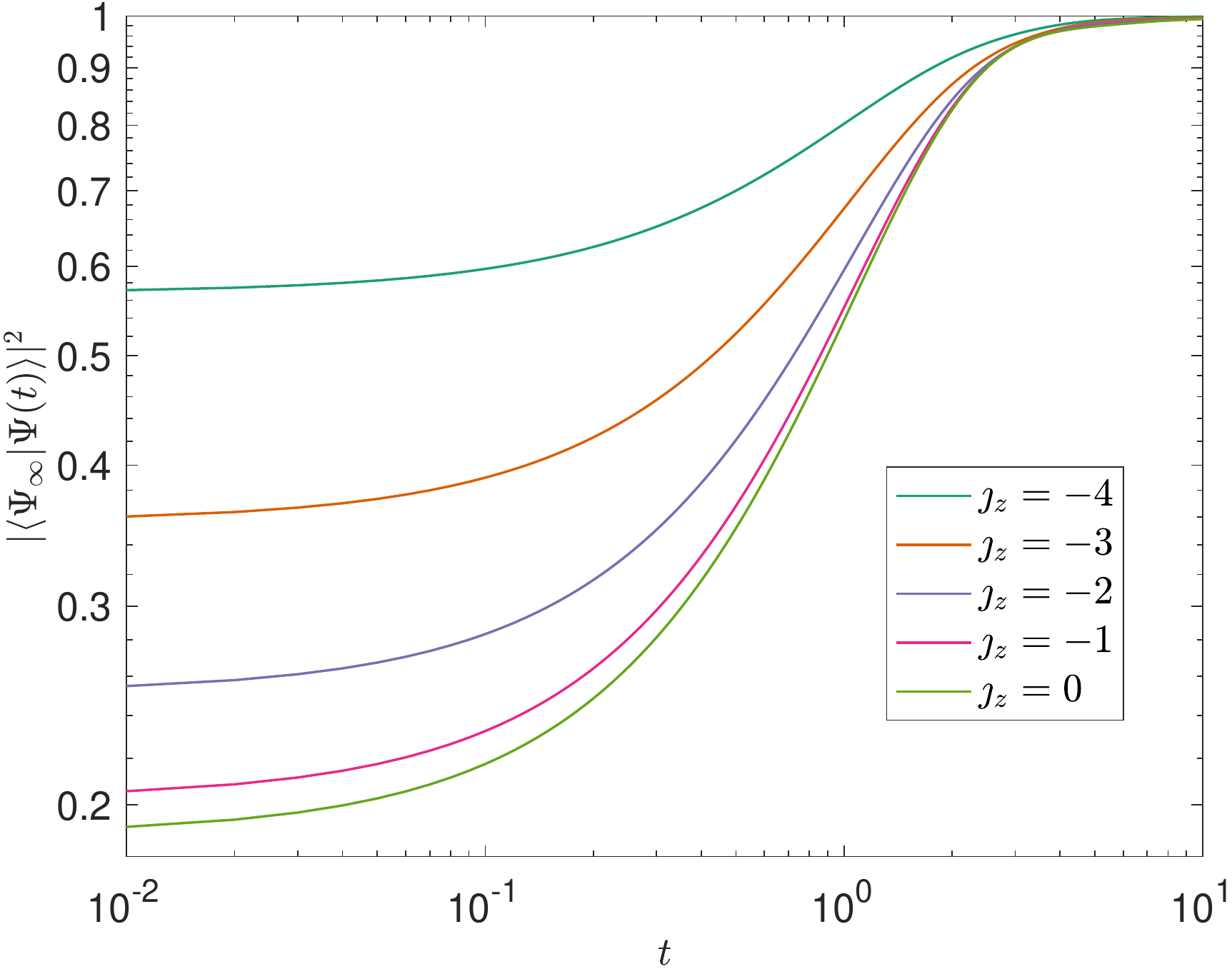}
\caption{The absolute square of the overlap $\langle\Psi_\infty|\Psi(t)\rangle$ between the asymptotically exact analytical answer
$\Psi_\infty$   and direct numerical solution $\Psi(t)$ of the nonstationary Schr\"odinger equation for the BCS 
Hamiltonian 
with time-dependent coupling constant $g(t)=\frac{\eta}{Nt}$. The number of energy levels $\eps_j=\frac{j}{N}$  is $N=10$, $\eta=1$, and $\jmath_z$ is the $z$-component of the total spin ($2\jmath_z+N$ is the total number of fermions). Here and in all remaining figures we start the evolution from the ground state at $t=0^+$, but set the initial value of $t$ in the Hamiltonian to $t_0$. By default $t_0=10^{-5}$  in all figures.  We see that $|\langle\Psi_\infty|\Psi(t)\rangle|\to1$ at large $t$ confirming the exact answer~\re{psiinf}. }
\label{overlap_fig}
\end{center}
\end{figure}

We double check our result numerically by evaluating the absolute square of the overlap, $|\langle\Psi_\infty|\Psi(t)\rangle|^2$,
where we compute $\Psi_\infty$ using \eref{psiinf}  and $\Psi(t)$ via a direct numerical simulation of the nonstationary Schr\"odinger equation. If \eref{psiinf} is valid, we must have $|\langle\Psi_\infty|\Psi(t)\rangle|^2\to 1$   as $t\to+\infty$, which is what we indeed observe in Fig.~\ref{overlap_fig}. See also Figs.~\ref{fig:szquantum} and \ref{fig:spsm} for further confirmation of \eref{psiinf}.

\subsection{Observables}

We first use the late-time wave function~\re{psiinf} to evaluate several  basic observables for finite $N$ before  turning our attention to the thermodynamic limit of general $n$-point equal time correlation functions   in Sec.~\ref{thm_qm_sec}. We also compare these asymptotically exact finite $N$ results with direct numerical simulations  and  mean-field answers.

\begin{figure}[t!]
  \centering
  \includegraphics[width=0.48\textwidth]{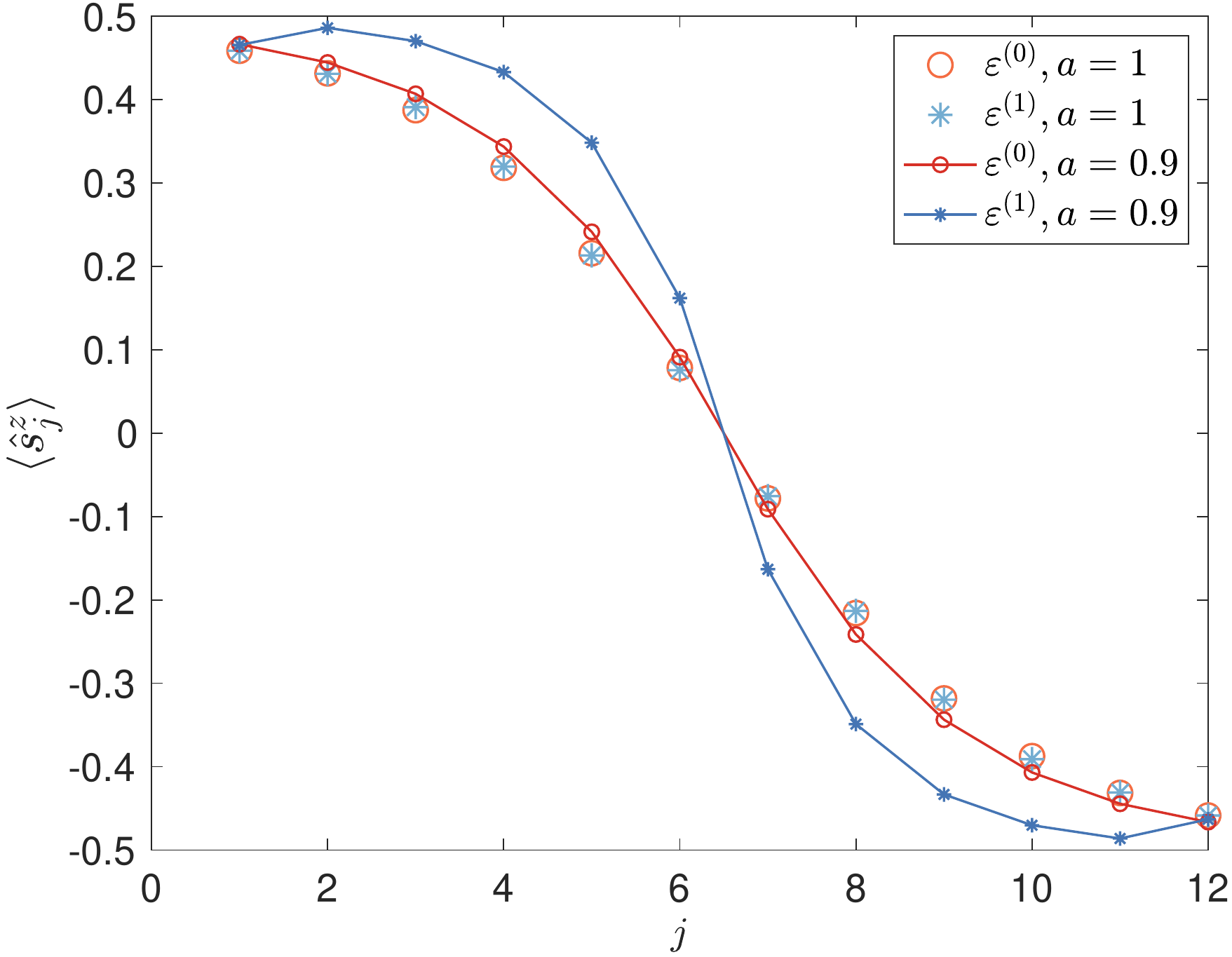}
  \caption{\label{fig:integ_break} The effect of breaking the time-dependent integrability. Here we compare  $\langle \hat s_j^z\rangle$ (equivalent to fermion  occupation numbers) at $t\to+\infty$ for two single-particle level  distributions and
  different   $g(t)$. The system evolves with the quantum BCS Hamiltonian with  coupling $g(t)\propto\frac{\eta}{t^a}$ starting from the ground state at $t=0^+$.  The number of   levels and fermions   is $N=12$, $\eta=1$,   and the two level distributions are: $\eps^{(0)}_j=\frac{j}{N}$ and $\eps^{(1)}_j=0.49+0.002j$ for all $j\ne 1$  and $\eps^{(1)}_1=0.1$. The case $a=1$ is integrable and we see that $\langle \hat s_j^z\rangle$ does not depend on the level distribution (unlike for $a=0.9$), which is characteristic of time-dependent integrability. 
  }
\end{figure}

Easiest to write is the probability distribution $P(\{s^z\})$ of finding the configuration $\ket{\{s^z\}} =
\ket{s_1^z s_2^z \ldots}$. This distribution does not depend on the phases $\theta_\alpha$ and $\Lambda_{\{\alpha\}}$ and therefore
is independent of $\eps_j$ and insensitive to the entanglement due to $\Lambda_{\{\alpha\}}$. In fact, $P(\{s^z\})$ has already been found in Ref.~\onlinecite{kolya1} via a different approach~\cite{kolya}, namely, by exploiting commuting multi-time Hamiltonian flows. In time-dependent integrability, such commuting flows play  a role similar to  integrals of motion for autonomous quantum integrable systems~\cite{kolya,me,aniket}.
We observe from \eref{psiinf} that the ratio of the probability of the spin  at $\eps_\alpha$ being up to the probability of it being down  is $e^{\frac{-2\pi\alpha}{\nu}}$. Equivalently,
we can say that the probabilities of $s_\alpha^z=\pm\frac{1}{2}$ are proportional to  $e^{\frac{-2\pi \alpha s_\alpha^z}{\nu}}$ and therefore
 \begin{equation}
 \label{eq:quantum_dist}
  P(\{s^z\}) = Z^{-1}
  e^{-\frac{2\pi}{\nu}\sum_k k s_k^z}
  \delta\Bigl[
    {\textstyle \sum_k\! s_k^z},  \jmath_z\Bigr],
\end{equation}
where $\jmath_z$ is the $z$-component of the total spin as before, $\delta[a,b]\equiv \delta_{ab}$ is the Kronecker delta, and $Z^{-1}$ is the normalization constant. The independence of $P(\{s^z\})$ from the distribution of the single-particle energies $\eps_j$ is a distinguishing  characteristic of time-dependent integrability~\cite{kolya}. We confirm this in Fig.~\ref{fig:integ_break} where we plot the late-time $\langle \hat s_j^z \rangle$
as a function of $j$ for $g(t)\propto t^{-a}$ with $a=0.9$ and $a=1$. Notice that $\langle \hat s_j^z \rangle$ does not change with the distribution of  $\eps_j$ for $a=1$ and does  for $a=0.9$.

\begin{figure}[t!]
  \centering
  \includegraphics[width=0.48\textwidth]{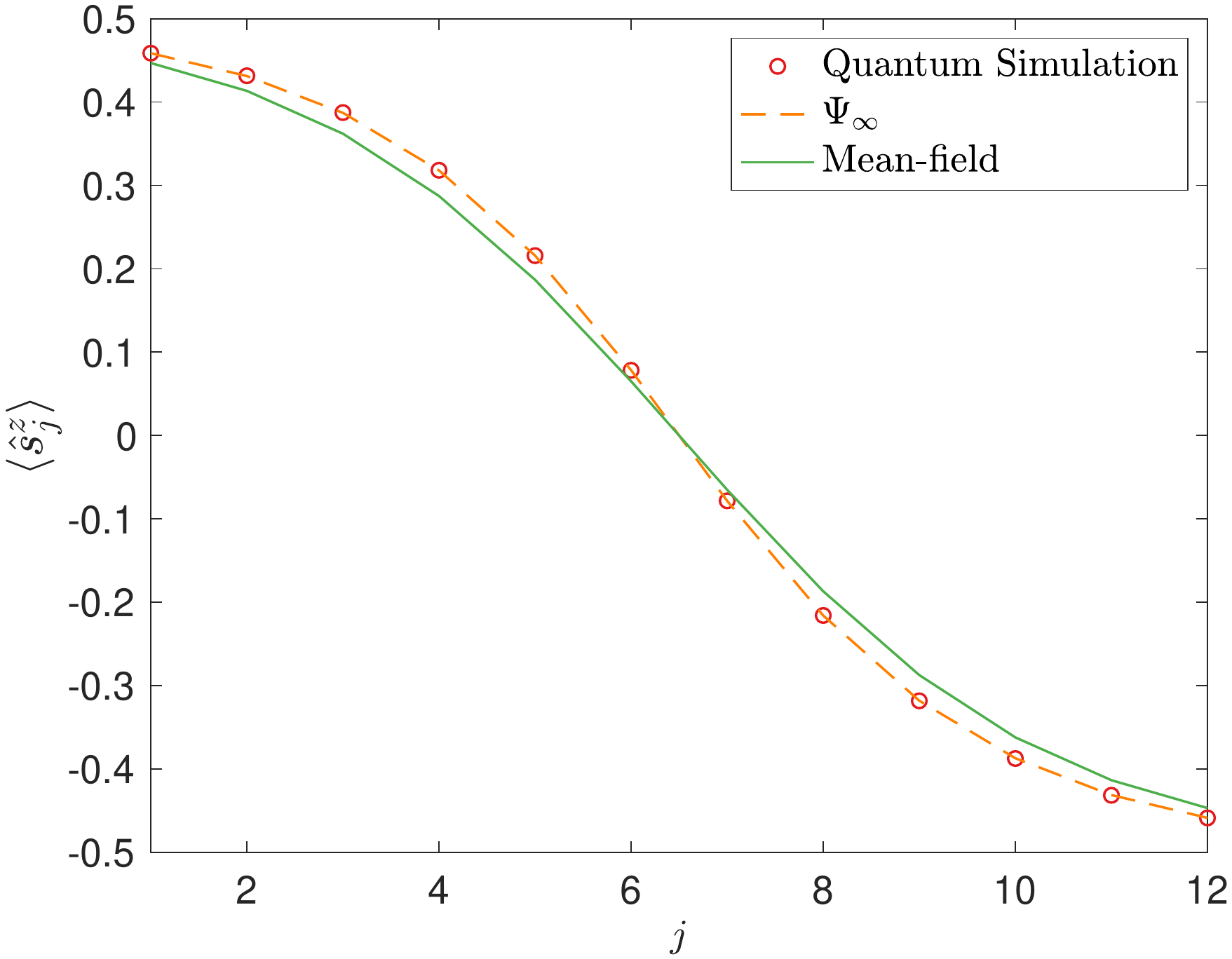}
  \caption{\label{fig:szquantum} Comparison of three answers for the late-time   distribution of $\langle \hat s_j^z\rangle$  for   the time-dependent quantum BCS Hamiltonian for $N=12$   energy levels $\varepsilon_j$ and  6  fermion pairs.  Other parameters are as in Fig.~\ref{overlap_fig}.
 One answer is from a direct numerical  simulation of the quantum dynamics. The other is the average $\langle \hat s_j^z\rangle_{\Psi_\infty}$ evaluated using the exact analytical   late-time wave function $\Psi_\infty$ in \eref{psiinf}. These two answers are indistinguishable. The third is the exact analytical answer for the mean-field   dynamics  in the thermodynamic limit.  
     }
\end{figure}

\begin{figure}
 \includegraphics[width=0.48\textwidth]{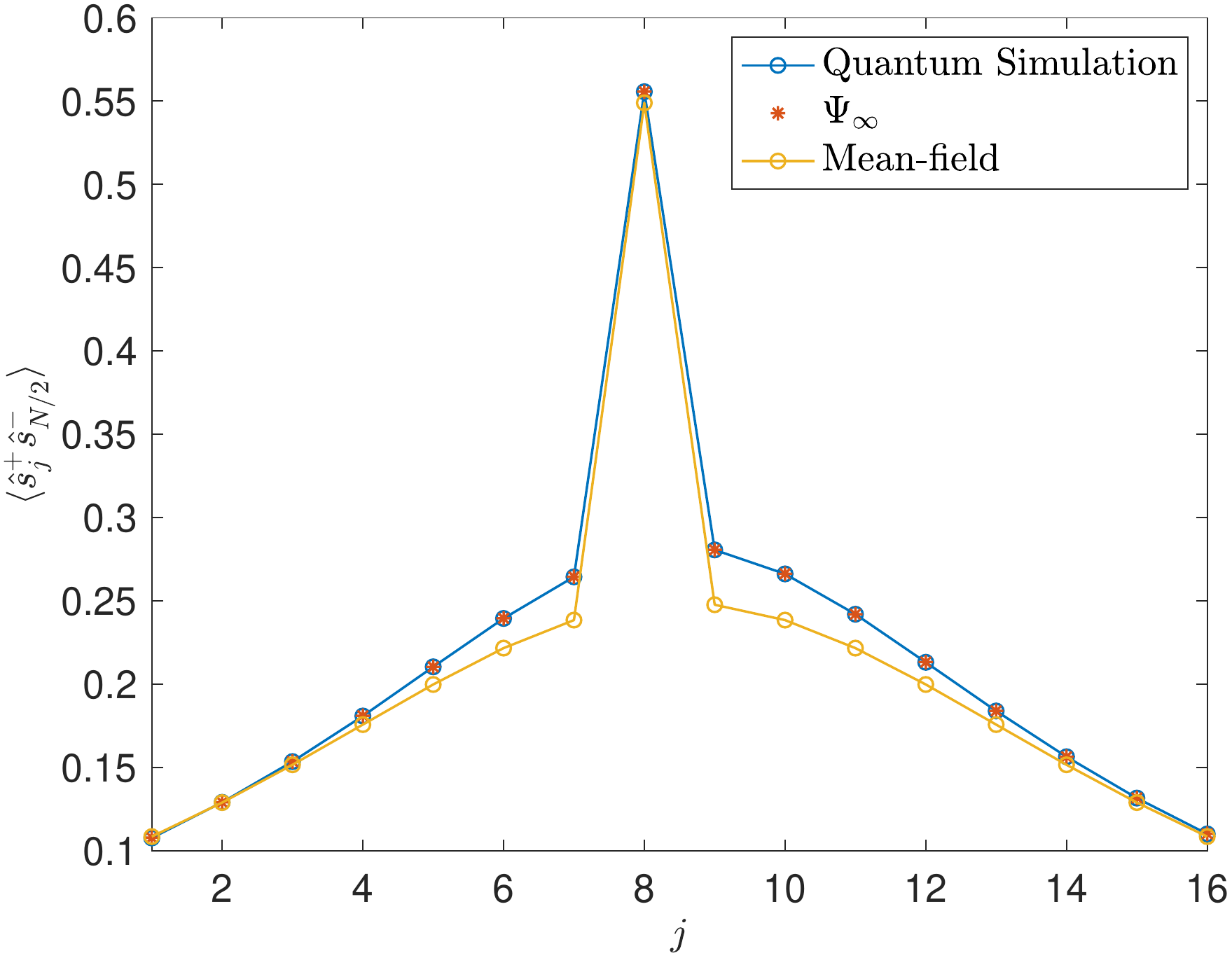}
 \caption{\label{fig:spsm} Same as Fig.~\ref{fig:szquantum} but for the   average $\avg{\hat s^{+}_{j} \hat s^{-}_{N/2}}$
 for $N=16$ levels and  eight fermion pairs.  To evaluate the  same-level value $\avg{\hat s^{+}_{N/2} \hat s^{-}_{N/2}}$, we   use
 the spin-$1/2$ identity $\avg{\hat s^{+}_{N/2} \hat s^{-}_{N/2}}=1/2 +\avg{\hat s_{N/2}^z}$ as explained below \eref{npoint1}. }
\end{figure}

The expectation value of the $z$-component of a spin in the state $\Psi_\infty$   is
\begin{equation}
    \avg{\hat{s}_j^{z}}_{\Psi_\infty}
    = C_\infty
    \sum_{\{\alpha\}  }
    \left[
        I_{\{\alpha\}}(j) - \frac{1}{2}
    \right]
    \prod_{\alpha}
        e^{-\frac{2\pi\alpha}{\nu}},
    \label{eq:sz_lam}
\end{equation}
where the indicator function $I_{\{\alpha\}}(j)$  is 1 if $j$ belongs to the set $\{\alpha\}$ and zero otherwise, and $C_\infty$ is the inverse norm of the late-time wave function squared,
\beg
 \frac{1}{C_\infty} =  \sum_{\left\{\alpha\right\}}\prod_{\alpha}e^{{-\frac{2\pi\alpha}{\nu}}}.
\en
Similarly, we evaluate the correlation function
  \begin{gather}
  \avg{\hat{s}_{k}^{+}\hat{s}_{j}^{-}}_{\Psi_\infty} =
 C_\infty
  e^{-2it(\epsilon_{j}-\epsilon_{k})}
  e^{-\frac{\pi(j + k)}{\nu}}
  \prod_{q\neq j,k}
  \biggr\lvert
  \frac{\epsilon_{q} - \epsilon_{j}}
  {\epsilon_{q} - \epsilon_{k}}
  \biggr\rvert^{-\frac{i}{\nu}}\nonumber\\
    \times\sum_{\substack{\left\{\beta\right\} \\
        \beta \neq j,k}}
  \prod_{\beta}
  e^{-\frac{2\pi\beta}{\nu}}
    \biggr\lvert
    \frac{\epsilon_{\beta} - \epsilon_{j}}
    {\epsilon_{\beta} - \epsilon_{k}}
    \biggr\rvert^\frac{2i}{\nu}.\label{splsmin}
  \end{gather}
Here, the set $\{\beta\}$ corresponds to all configurations with $N_\up - 1$ up spins. Importantly, the correlation function $  \avg{\hat{s}_{k}^{+}\hat{s}_{j}^{-}}$ depends on $\eps_j$, unlike $\langle \hat s_j^z\rangle$ and $P(\{s^z\})$.  In Figs.~\ref{fig:szquantum} and
\ref{fig:spsm} we compare \esref{eq:sz_lam} and \re{splsmin} with direct numerical simulations of the nonstationary Schr\"odinger equation and   the corresponding late-time dynamical variables in the BCS mean-field (classical) dynamics that we obtain in  the next section.

\section{Exact classical BCS dynamics}
\label{QvsC_sec}

We saw above that  in the BCS mean-field approximation  the averages $\bm S_k=\langle \hat{\bm s}_k\rangle$ evolve according to Hamilton's equations of motion for the classical counterpart of the BCS Hamiltonian, 
\begin{equation}
  H(t) =
  \sum_{j=1}^N 2\eps_j S_j^z
  - \frac{\eta}{Nt} \sum_{j,k=1}^N S_j^+S_k^-,
  \label{eq:class_H1}
\end{equation}
  with  standard angular momentum Poisson brackets for components of $\bm S_j$. Launched from a BCS product state, the mean-field time evolution   keeps the system in a product state at all times, the length of vectors $\bm S_j$ is $S=\frac{1}{2}$, and knowing $\langle \hat{\bm s}_j\rangle$ at time $t$, we also know the corresponding  BCS product wave function   up to a global phase.  
    
Here we present the \textit{exact solution} for the long time dynamics of the classical Hamiltonian~\re{eq:class_H1}.   We derive this from the formal solution of Sec.~\ref{formal_sec} by taking the classical limit, where $\hbar\to0$ and the magnitude of quantum spins $s\to\infty$ in the generalized BCS Hamiltonian~\re{eq:H}   so that $\hbar s=S=\frac{1}{2}$.  Before the classical limit, we take the long time limit where the multivariable  contour integral~\re{psi} localizes to its stationary points. Our treatment is similar to that in Sec.~\ref{late_time_sec} but now solutions of the stationary point equations are highly degenerate and as a result the calculations are more complicated.  

\begin{figure}[t!]
\begin{center}
\includegraphics[width = 0.49\textwidth]{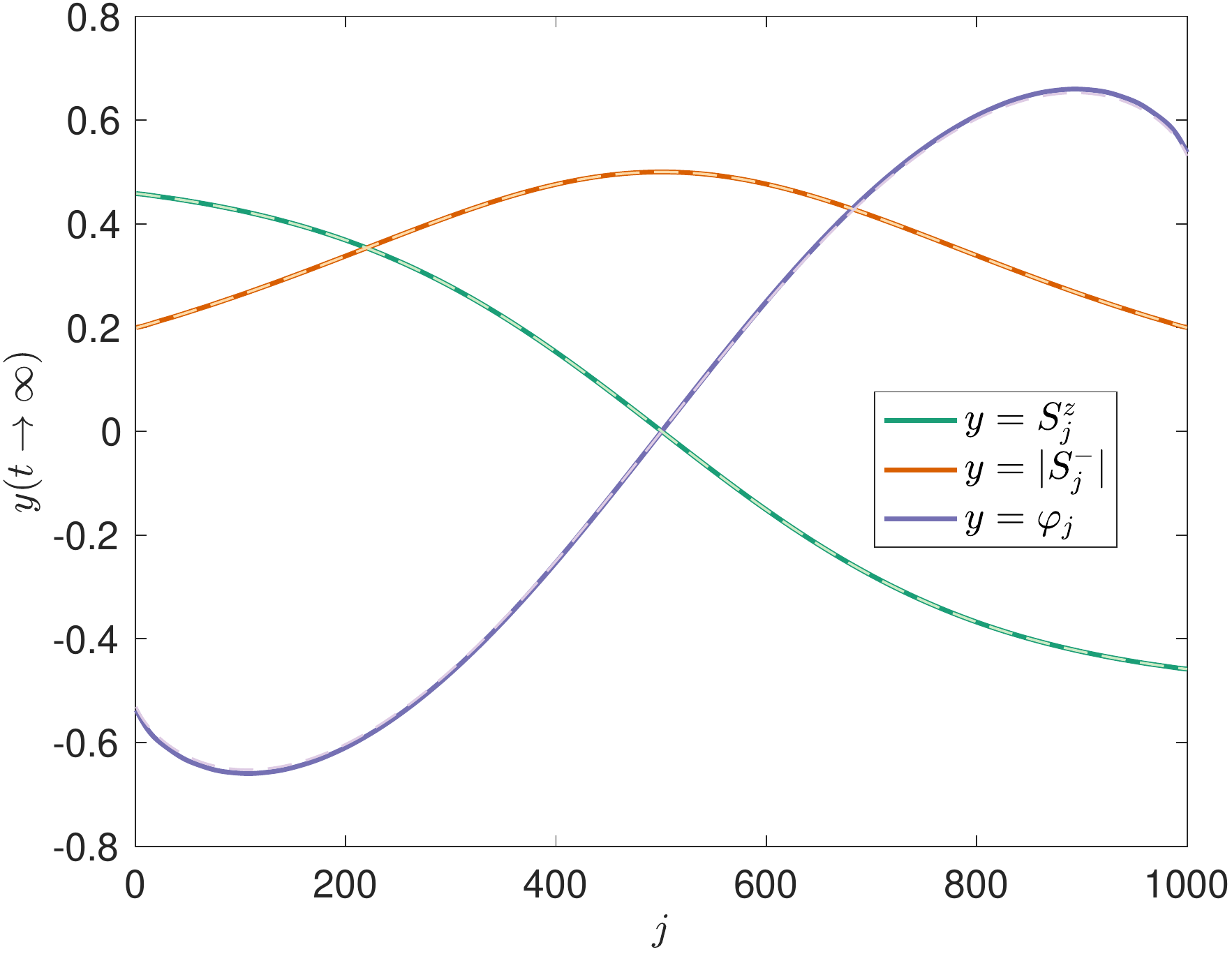}
\caption{ Classical (mean-field) dynamics of the BCS Hamiltonian with  coupling $g(t)\propto \frac{\eta}{t}$ asymptotes to a state  $S_j^z=\text{const}$ and $S_j^-=|S_j^-| e^{-2i\eps_j t} e^{i\varphi_j}$ 
at late times. In the text, we  derived exact analytic expressions 
for  $S_j^z$, $|S_j^-|$, and $\varphi_j$ in the limit $N\to\infty$ [dashed curves, see \eref{full1}]. Here we compare them  with direct numerical simulation of the mean-field equations of motion  [solid curves] for  $N=10^3$ classical spins $\bm S_j$ and    $z$-component of the total spin $J_z=0$. Maximum relative errors for $S_j^z$, $|S_j^-|$, and $\varphi_j$ are
  $0.1\%, 0.4\%$, and $2\%$, respectively. Other  parameters are as in Fig.~\ref{overlap_fig}.
}
\label{Fig1}
\end{center}
\end{figure}

\begin{figure}[htb!]
\begin{center}
\includegraphics[width = 0.49\textwidth]{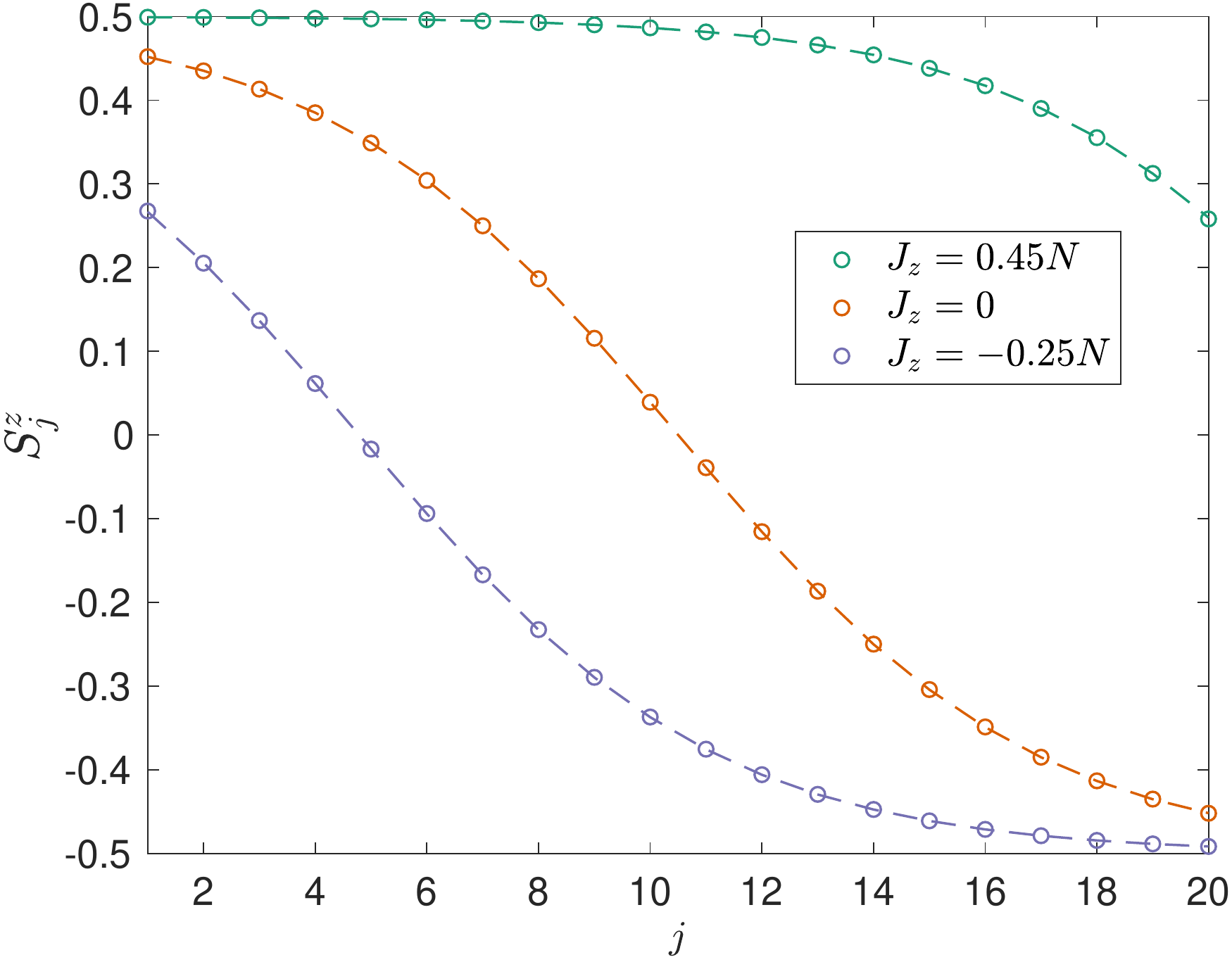}
\caption{Comparison of our analytic answer (dashed curves) for $S_j^z=\langle \hat s_j^z\rangle$  in the late-time asymptotic state   of the mean-field BCS dynamics   with   coupling $g(t)\propto \frac{\eta}{t}$ and direct numerical simulations (circles) of the equations of motion for  $N=20$ classical spins and three different values of $J_z=\sum_k S_k^z$. The significance of $S_j^z$ is in its relation to the average fermion occupation number for level $\eps_j$, $\langle \hat n_j\rangle = 2S_j^z+1$. Other parameters are the same as in Fig.~\ref{overlap_fig}. Note how close  our $N\to\infty$ answer is  to the results of simulations with only 20 spins. This is consistent with our numerical observation in Sec.~\ref{approach_sec} that the corrections to this limit \textit{within} mean-field scale as $N^{-\frac{3}{2}}$ rather than $N^{-1}$. 
}
\label{Fig2mus1}
\end{center}
\end{figure}

We relegate the details of the derivation   to Appendix~\ref{classApp}  and just state the answer here: the $t\to+\infty$  asymptote of classical spins $\bm S_j$ for $N\to\infty$  is 
\begin{subequations}
\begin{align}
&S_j^-=\frac{  e^{-2i\eps_j t +i\varphi_j} }{2\cosh \zeta_j },\quad
 S_j^z = -\frac{1}{2} \tanh\zeta_j,\label{11}\\
 &\varphi_j=-\frac{\eta}{N}\sum_{k\ne j} \tanh\zeta_k  \ln|\eps_k-\eps_j|,\label{12}\\
& \zeta_j= \frac{  \pi\eta (j-\mu) }{N}\label{zetaj12},
\end{align}
\label{full1}%
\end{subequations}%
where $\mu$ is a Lagrange multiplier (chemical potential) given by \eref{eq:mu} below. We check the analytical results~\re{full1} against direct numerical simulation of Hamilton's (mean-field) equations of motion~\re{ceom} for
$N=10^3$ and $N=20$ classical spins in Figs.~\ref{Fig1} and \ref{Fig2mus1} and find excellent  agreement.

The chemical potential $\mu$ is set by the condition that the conserved $z$-component of the total spin $\bm J$ be equal to its initial value,
\beg
J_z=\sum_{k=1}^N S_k^z=-\frac{1}{2}\sum_{k=1}^N \tanh\left[ \frac{\pi\eta(k-\mu)}{N} \right].
\label{mu12}
\en
 In the thermodynamic limit  $N\to\infty$, the sum turns into  an integral. Integrating and solving for $\mu$, we find
 \begin{equation}
  \mu  
  = \frac{N+1}{2}
  + \frac{N}{2\pi\eta}\ln\left\{
    \frac{\sinh\left[ \pi \eta\left(\frac{1}{2}+  \frac{ J_z }{N}\right) \right]}
    {\sinh\left[ \pi\eta \left(\frac{1}{2}-  \frac{ J_z }{N}\right)  \right]}\right\}.
  \label{eq:mu}
\end{equation}
We kept the subleading correction ($N+1$ instead of simply $N$ in the first term on the r.h.s.) because it reproduces $\mu=\frac{N+1}{2}$ for $J_z=0$ which is exact for any even $N$ and  significantly improves the agreement with finite $N$ numerics.

Within the mean-field treatment, each quantum spin $\hat {\bm s}_j$ evolves individually  in an effective magnetic field $2(\eps_j \bm z -\bm\Delta )$. The wave function of the system is thus of the BCS product form 
\beg
\Psi_\mathrm{mf}=\prod_{k=1}^N \left( u_k |\!\dn\rangle+     v_k |\!\up\rangle\right).
\label{psimf}
\en
 at all times provided it was of this form at $t=0$. Normalization requires $|u_k|^2+|v_k|^2=1$ and  \esref{psimf} and \re{full1} together with $\bm S_k=\langle \hat{\bm s}_k\rangle$
imply 
\begin{subequations}
\begin{align}
\langle \hat s_k^+\rangle_\mathrm{mf}=u_k v_k^*=\frac{e^{2i\eps_k t} e^{-i\varphi_k} }{2\cosh\zeta_k}, \\
 \langle \hat s_k^-\rangle_\mathrm{mf}=u_k^* v_k=\frac{e^{-2i\eps_k t} e^{i\varphi_k} }{2\cosh\zeta_k},\label{s-n-1n1}\\
\langle \hat s_k^z\rangle_\mathrm{mf}=\frac{ |v_k|^2-|u_k|^2}{2}=-\frac{1}{2}\tanh \zeta_k.\label{szzzz}
  \end{align}
\label{indv_spins}
\end{subequations}
The subscript ``mf'' indicates the expectation values in the late-time mean-field wave function~\re{psimf}.
Using these equations, we reconstruct the   Bogoliubov amplitudes
\beg 
u_k=\frac{ e^{\frac{\zeta_k-i\varphi_k}{2}} }{\sqrt{2\cosh \zeta_k}},\quad v_k=\frac{e^{-2i\eps_k t} e^{\frac{-\zeta_k+i\varphi_k}{2}}}{\sqrt{2\cosh \zeta_k}},
\label{bogol}
\en 
up to a common phase which only affects the global phase of $\Psi_\mathrm{mf}$. Equations~\re{psimf} and \re{bogol} provide the exact   
late-time mean-field wave function   for the time evolution with the BCS Hamiltonian  with interaction strength inversely proportional to time starting from the mean-field ground state at $t=0^+$ in the thermodynamic limit. 

Due to the product form of the   mean-field wave function, it is simple to determine the expectation value  of an arbitrary product of spin
operators
\beg
 \biggl\langle \prod_{m=1}^n  \hat s_{j_m}^{r_m} \biggr\rangle_\mathrm{mf} = 
\prod_{k=1}^n \bigl\langle \hat s_{j_m}^{r_m} \bigr\rangle_\mathrm{mf},
\label{npoint1}
\en
where the upper indices $r_m$ take values $+, -$, or $z$ and individual spin averages $\bigl\langle \hat s_{j_m}^{r_m} \bigr\rangle_\mathrm{mf}$ are given by \eref{indv_spins}.    It is understood that there is \textit{only one spin operator per energy level} $\varepsilon_j$ in \eref{npoint1}. In other words, any operator nonlinear in the components of $\bm s_j$
must be reduced to a linear one before comparing the averages. This can always be done for spin-$\frac{1}{2}$, e.g., $\hat s_j^+ \hat s_j^-=\frac{1}{2}+\hat s_j^z$, $(\hat s_j^z)^2=\frac{1}{4}$, etc.  Otherwise, \eref{npoint1} may not hold because, for example,  $\langle(\hat s_j^z)^2\rangle_\mathrm{mf}=\frac{1}{4}\ne (\langle\hat s_j^z\rangle_\mathrm{mf})^2$.

\section{Thermodynamic limit of quantum dynamics}
\label{thm_qm_sec}

In this section, we show how a BCS product state emerges in the thermodynamic limit   from the entangled finite $N$   wave function $\Psi_\infty$ -- the late-time asymptotic solution~\re{psiinf} of the nonstationary Schr\"odinger equation for the quantum BCS Hamiltonian~\re{eq:bcs_model}  with interaction strength inversely proportional to time.   The precise statement is that any   local equal time correlation function
of  fermionic or spin operators evaluated in the exact asymptotic state $\Psi_\infty$  is identical to that in the product state $\Psi_\mathrm{mf}$ of the mean-field (classical) dynamics in this limit.   Recall that we say a quantity is local if the  number $n$ of  single-particle levels $\eps_k$ (``points'') it involves is such that
$\frac{n}{N}\to0$ when $N\to\infty$~\cite{local}.
All correlators of this type will be straightforward to  evaluate once we establish this  correspondence between quantum and classical dynamics.  

However,   $\Psi_\infty \ne \Psi_\mathrm{mf}$   even in the thermodynamic limit. This manifests itself in non-local quantities involving an infinite number of points in the  thermodynamic limit, such as, e.g., 
$\langle\hat s_1^-\dots \hat s_{N_\up}^- s_{ N_\up+1}^+\dots \hat s_{2N_\up}^+\rangle$, where $N_\up$ is the number of up spins, or the von Neumann entanglement entropy. 
The values of  quantities of this type are generally different for $\Psi_\infty$ and $\Psi_\mathrm{mf}$. This   is not specific to the nonautonomous setup as these  quantities similarly disagree between the exact and BCS ground states for the time-independent BCS Hamiltonian. An even more interesting example of the breakdown of the classical picture for global observables   is the Loschmidt echo~\cite{echo}. Within mean-field approach we obtain the classical Loschmidt echo, i.e., the echo of the classical spin Hamiltonian~\re{eq:class_H}, which   is qualitatively  different from the true quantum echo, see Ref.~\onlinecite{echo} for further details.

The crucial step in deriving the thermodynamic limit of the late-time quantum dynamics is to notice by inspecting \esref{optS} and \re{psi_sum} that we can write $\Psi_\infty$ in the form of a generalized projected BCS state [cf. \eref{projBCS}]
\beg
\Psi_\infty= P_{N_\up}\prod_{k=1}^N \left(\hat U_k |\!\dn\rangle+   \hat V_k |\!\up\rangle\right),
\label{gen_pbcs}
\en
where $P_{N_\up}$ is the projector onto the subspace with $N_\up$ up spins and
\begin{subequations} 
\begin{eqnarray}
\hat U_k= e^{-\frac{i\hat \varphi_k}{2}}, \quad \hat V_k = e^{\frac{i\hat \varphi_k}{2}-2i\eps_kt-\frac{\pi k}{\nu}},\\
 \hat \varphi_k =\frac{2}{\nu} \sum_{j\ne k} \hat s_j^z \ln|\eps_j-\eps_k|.
\label{xk}
\end{eqnarray}
\end{subequations} 
It is understood that when the product~\re{gen_pbcs} is expanded, all ket vectors are placed to the right of the operators $e^{\mp \frac{i\hat \varphi_k}{2}}$.
The projector ensures that we end up with the summation over the same basis states $|\{\alpha\}\rangle$ with $N_\up$ up spins as in
\eref{psi_sum}. The terms $-2i\eps_kt-\frac{\pi k}{\nu}$ then add up to the first two sums in \eref{optS} multiplied by $-\frac{i}{\nu}$.
Similarly, $\hat \varphi_k$ correspond to the last sum in \eref{optS}.  States $|\!\dn\rangle$ and $|\!\up\rangle$ come with $e^{\mp i\frac{\hat \varphi_k}{2}}$ in \eref{gen_pbcs} because 
for them $\hat s_k^z \to \mp\frac{1}{2}$ in \eref{optS}.

\subsection{Local operators}

First, we study   local operators in the thermodynamic limit.  As usual in the theory of superconductivity, we understand  the \textit{thermodynamic limit} as  $N\to\infty$ so that the single-particle   levels $\varepsilon_j$ fill a finite energy interval with a  piecewise continuous density of states  and the number of fermions per level $\frac{N_\mathrm{f}}{N}=\frac{2N_\up}{N}$ stays finite. The latter condition is equivalent to a finite  
density of fermions.

  We start with $\langle  \hat{s}_{k}^{+} \hat{s}_{j}^{-}\rangle_{\Psi_\infty}$ for $j\ne k$ and then generalize to arbitrary  products.
 It is helpful to rewrite \eref{gen_pbcs} as an integral [cf. \eref{pbcs}]
\beg
\Psi_\infty=\frac{1}{2\pi}\int_0^{2\pi}\!\!\!\!\! d\phi e^{i\phi N_\up} \prod_k  \left(\hat U_k|\!\dn\rangle+e^{-i\phi} \hat V_k|\!\up\rangle\right).
\label{pbcs1}
\en
Consider    $\langle\Psi_\infty|\Psi_\infty\rangle$. This is a double integral over $\phi$ and $\phi'$.   The integrand depends only on $\xi=\phi-\phi'$, so one integration simply gives $2\pi$. Taking this overlap converts $\hat s_j^z\to \langle \hat s_j^z \rangle_{\Psi_\infty}$ because $e^{c \hat s_z}$ is linear in $\hat s_z$ for spin-$\frac{1}{2}$ and $\hat s_j^z$ mutually commute. We find
\beg
\langle\Psi_\infty|\Psi_\infty\rangle=\frac{1}{2\pi}\int_{-2\pi}^{2\pi} d\xi e^{G(\xi)}, 
\label{int_norm}
\en
where
\beg
G(\xi)=i\xi N_\up+\sum_k \ln\left( | U_k|^2+ e^{-i\xi} | V_k|^2 \right),
\label{Fxi}
\en
and
\begin{subequations} 
\begin{eqnarray}
 U_k= e^{-\frac{i \varphi_k}{2}}, \quad  V_k = e^{\frac{i \varphi_k}{2}-2i\eps_kt-\frac{\pi k}{\nu}},\\
\varphi_k =\frac{2}{\nu} \sum_{j\ne k}\langle \hat s_j^z\rangle_{\Psi_\infty} \ln|\eps_j-\eps_k|.
\end{eqnarray}
\label{xkk}
\end{subequations} 

Similarly, we can evaluate various matrix elements. Take, for example, $\langle\Psi_\infty|\hat{s}_{k}^{+}\hat{s}_{j}^{-}|{\Psi_\infty}\rangle$. 
Here it is important to realize that operators $\hat{s}_{k}^{+}$ and $\hat{s}_{j}^{-}$ commute with $\hat U_l$ and $\hat V_l$ up to terms
of order $\frac{1}{N}$. Keeping this in mind, we 
go
through the same steps as for $\langle\Psi_\infty|\Psi_\infty\rangle$ and obtain
\beg
\begin{split}
 \langle\Psi_\infty|&\hat{s}_{k}^{+}\hat{s}_{j}^{-}|{\Psi_\infty}\rangle=\\
=&\frac{1}{2\pi}\!\!\!\int\limits_{-2\pi}^{2\pi} \frac{U_k V_k^* U_j^* V_j e^{-i\xi} e^{G(\xi)} d\xi }{\left( | U_k|^2+ e^{-i\xi} | V_k|^2 \right)\left( | U_j|^2+ e^{-i\xi} | V_j|^2 \right) }.
\end{split}
\label{corr_int}
\en
The additional factors in this equation as compared to \eref{int_norm}   result from the action of $\hat{s}_{j}^{-}$ on the state $U_j|\!\dn\rangle+e^{-i\phi}  V_j|\!\up\rangle$ and the analogous  action of $\hat{s}_{k}^{+}$. \eref{corr_int} is only valid when $j\ne k$ and only up to terms of order $\frac{1}{N}$.  

Integrals of the form~\re{int_norm} and \re{corr_int} have been   analyzed extensively in  studies of the equilibrium projected BCS wave function and  its equivalence to the regular BCS product state in the thermodynamic limit~\cite{nucl1,nucl2}.   It is known
that the saddle point method becomes exact in  the thermodynamic limit  because $G(\xi)$ is of order $N$. For the same reason, the saddle point $\xi_0$ is the same for both integrals~\re{int_norm} and \re{corr_int}. Evaluating the integrals  with this method, we find
that   $\langle  \hat{s}_{k}^{+} \hat{s}_{j}^{-}\rangle_{\Psi_\infty}$ is identical to the average of the same operator $\hat{s}_{k}^{+} \hat{s}_{j}^{-}$
 in a product state
\beg
 \Psi_\mathrm{thd}=C_\mathrm{n}\prod_{k=1}^N \left( U_k |\!\dn\rangle+    e^{\frac{\pi \mu}{\nu}} V_k |\!\up\rangle\right),
 \label{unnormprod}
\en
where $\mu=-\frac{i\nu\xi_0}{\pi}$ and $C_\mathrm{n}$ is a normalization constant.  
The subscript ``thd'' stands for ``thermodynamic'' indicating that this wave function is exact for evaluating certain correlation functions in the thermodynamic limit. Evaluating $C_\mathrm{n}$ 
and recalling that $\nu=\frac{\eta}{N}$ [see \eref{g}],
we  see that $ \Psi_\mathrm{thd}$ is identical to the late-time mean-field wave function~\re{psimf},  
\beg
 \Psi_\mathrm{thd}=\Psi_\mathrm{mf}=\prod_{k=1}^N \left( u_k |\!\dn\rangle+     v_k |\!\up\rangle\right),
\label{psimf1}
\en
where $u_k$ and $v_k$ are given by \eref{bogol}.

There is nothing special about $\hat{s}_{k}^{+} \hat{s}_{j}^{-}$. The same logic applies to   general products   of $\hat s^+, \hat s^-,$ and $\hat s^z$,
\beg
 \hat O= \prod_{m=1}^n  \hat s_{j_m}^{r_m},
\label{npoint}
\en
 where $r_m =+, -$, or $z$ and as before there is no more than one spin operator for each $j_m$.  The number of   operators $n$ must be such that $\frac{n}{N}\to0$ when $N\to\infty$, i.e. $\hat O$ must be local. Otherwise, terms  of the order $N^{-1}$ of the type we neglected in deriving \eref{unnormprod} can add up to a contribution of order one.   
 Nonzero matrix elements of $\hat O$ between states $\Psi_\infty(N_\up+\Delta N_\up)$ and $\Psi_\infty(N_\up)$   coincide with its expectation value   in the product state~\re{psimf} in the thermodynamic limit ($\Delta N_\up$ is the number of $\hat s^+$ minus  number of $\hat s^-$ in $\hat O$, i.e., the amount by which it increases the number of up spins). 
Therefore, using \eref{npoint1} we have to the leading order in $\frac{n}{N}$,   
\beg
\Bigl\langle  \widetilde{N}_\up \Bigl| \prod_{m=1}^n  \hat s_{j_m}^{r_m}\Bigr| N_\up \Bigr\rangle_\infty = 
\prod_{k=1}^n \bigl\langle \hat s_{j_m}^{r_m} \bigr\rangle_\mathrm{mf}.
\label{exp123}
\en
  Here $|N_\up\rangle_\infty$ is the normalized version of $\Psi_\infty(N_\up)$, 
\beg
|N_\up\rangle_\infty=\frac{ \Psi_\infty(N_\up)}{{\norm{\Psi_\infty(N_\up)}}},
\label{normvers}
\en
and $ \widetilde{N}_\up =  N_\up+\Delta N_\up$. Note also that   the matrix elements of arbitrary products  of fermionic creation $\hat c^\dagger_{j\sigma}$ and annihilation $\hat c_{k\sigma'}$  operators    are either zero  or reduce to matrix elements  of the form~\re{exp123}.  

The quantity $\mu$ in \eref{unnormprod} is determined by the equation
$N_\up-\frac{N}{2}=\jmath_z= \sum_{k=1}^N\langle \hat s_k^z\rangle_\mathrm{mf}$, 
  which is a consequence of the conservation of the $z$-projection of the total spin $\hat{\bm\jmath}$. Simultaneously it is the equation
for the stationary point $\xi_0$ of $G(\xi)$ defined in \eref{Fxi} as it should be because we   defined $\mu$ in this section as $\mu=-\frac{i\nu\xi_0}{\pi}$.
Since $J_z=\langle \hat \jmath_z\rangle=\jmath_z$ and $S_k^z=\langle \hat s_k^z\rangle_\mathrm{mf}$, this equation is equivalent to \eref{mu12} and $\mu$ in \eref{unnormprod} is therefore  the same as the chemical potential~\re{eq:mu} of the mean-field dynamics. 

 Equations~(\ref{exp123},  \ref{indv_spins}, \ref{12}), and (\ref{zetaj12})  determine explicitly the exact thermodynamic   limit of any  matrix element  on the left hand side of \eref{exp123}. In particular,
\begin{subequations}
\begin{align}
&\langle \hat s_k^+ \hat s_j^-\rangle_{\Psi_\infty}=\frac{e^{2i(\eps_k-\eps_j)t}  e^{i(\varphi_j-\varphi_k)} }{2\cosh\zeta_k \cosh\zeta_j},\quad
j\ne k,\label{skplsjmin1}\\
&\langle \hat s_k^+ \hat s_k^-\rangle_{\Psi_\infty}=\frac{1}{2}-\frac{1}{2}\tanh \zeta_k,\\
&\langle \hat s_k^z \rangle_{\Psi_\infty}=-\frac{1}{2}\tanh \zeta_k.
\label{test1}\\
&\langle N_\up -1| \hat s_k^-| N_\up\rangle_\infty = \frac{e^{-2i\eps_k t} e^{i\varphi_k} }{2\cosh\zeta_k}.
\label{s-n-1n}
\end{align}
\label{skplsjmin}
\end{subequations}
Note that $\langle\dots\rangle_{\Psi_\infty}\equiv \langle N_\up|\dots | N_\up\rangle_\infty$.
As a check on our results, we also derived the thermodynamic limit of $\langle \hat s_k^z\rangle_{\Psi_\infty}$ and $\langle \hat s_k^+ \hat s_j^-\rangle_{\Psi_\infty}$  directly from \esref{eq:sz_lam} and~\re{splsmin} by writing them as integrals and  using the saddle point method, which is exact in this limit. The answers are precisely  \esref{test1} and \re{skplsjmin1}.
Instead of the BCS-like product $\Psi_\mathrm{mf}$ we can equally well employ the projected version of this state
\beg
\Psi_\mathrm{pmf}=P_{N_\up} \Psi_\mathrm{mf}= P_{N_\up} \prod_{k=1}^N \left( u_k |\!\dn\rangle+     v_k |\!\up\rangle\right).
\label{pmf}
\en
We see this in the same way as we showed the equivalence of $\Psi_\mathrm{mf}$ and $\Psi_\infty$ only without the complication of
$\hat U_k$ and $\hat V_k$ being operators, see also Refs.~\onlinecite{nucl1,nucl2}.

We conclude that the thermodynamic limits  of  averages of local operators $\hat O$ in the  late-time asymptotic state of the exact quantum BCS dynamics and in the late-time asymptotic  state of mean-field (classical) BCS dynamics \textit{coincide exactly }. Let us emphasize once more that when $\hat O$ does not conserve the total  fermion number,  we define its  average  in the   asymptotic solution of quantum dynamics with definite fermion number   as the  nonzero matrix element between   solutions  with   different fermion numbers.  Its expectation value in  the state  $\Psi_\infty$    is zero and not useful for comparison to mean field.  When $\hat O$ commutes with $N_\mathrm{f}$,  its average and  expectation  value in any state are the same.  

Note that   local correlators are the ones most readily  accessible in experiment.  In this sense,  BCS mean field is exact far from equilibrium for the evolution launched from the exact quantum ground state with a definitive number of fermions. This is true despite the fact that   the BCS order parameter vanishes 
at late times, see the discussion below \eref{length89}. However, the status of mean-field changes when: (1) the initial state
of the quantum BCS evolution is not a particle number eigenstate \textit{and} $\hat O$ does not commute with the total fermion number operator or (2) for  non-local quantiles, as we will see shortly.

\subsection{Entanglement entropy and other non-local quantities} 
\label{ent_sec}

Even though  matrix elements of   local operators in the exact late-time state $\Psi_\infty$ of quantum time evolution from the exact ground state   and in
the BCS product state $\Psi_\mathrm{mf}$  of the mean-field (classical)  evolution as well as in  the projected BCS state $\Psi_\mathrm{pmf}$ are identical in the thermodynamic limit,  $\Psi_\infty\ne \Psi_\mathrm{mf}$ and $\Psi_\infty\ne \Psi_\mathrm{pmf}$.   We see this by comparing coefficients at   basis states in \esref{pmf} and \re{psiinf}. The von Neumann entanglement entropy $\mathtt{S}_\mathrm{ent}$ is zero for $\Psi_\mathrm{mf}$ and of order $\ln N$ for
both $\Psi_\infty$ and $\Psi_\mathrm{pmf}$, as we will see below. 
 Moreover, we observe numerically that  $\mathtt{S}_\mathrm{ent}(\Psi_\mathrm{pmf})\approx \mathtt{S}_\mathrm{ent}(\Psi_\infty)$.

It is not difficult to write an operator whose quantum average is different in  $\Psi_\infty$   and in  $\Psi_\mathrm{mf}$  or $\Psi_\mathrm{pmf}$. The number of spins involved in such an operator is necessarily proportional to $N$ in the thermodynamic limit.
  Consider, for example,
operators $|\bm e_2\rangle\langle\bm e_1|$ that convert a basis state $|\bm e_1\rangle$ into a basis state $|\bm e_2\rangle$. One of these operators is
\beg
\hat s_1^-\dots \hat s_{N_\up}^- s_{ N_\up+1}^+\dots \hat s_{2N_\up}^+.
\en
Evaluating its expectation value  in the state $\Psi_\infty$  using \eref{psi_sum} or \eref{psiinf} and in the state $\Psi_\mathrm{mf}$ (or equivalently in $\Psi_\mathrm{pmf}$) using \esref{psimf} or \re{bogol}, we see that they generally do not agree even in the thermodynamic limit $N\to \infty$ keeping $\frac{N_\mathrm{f}}{N}=
\frac{2N_\up}{N}$ constant. 

\begin{figure}[t!]
\begin{center}
\includegraphics[width = 0.49\textwidth]{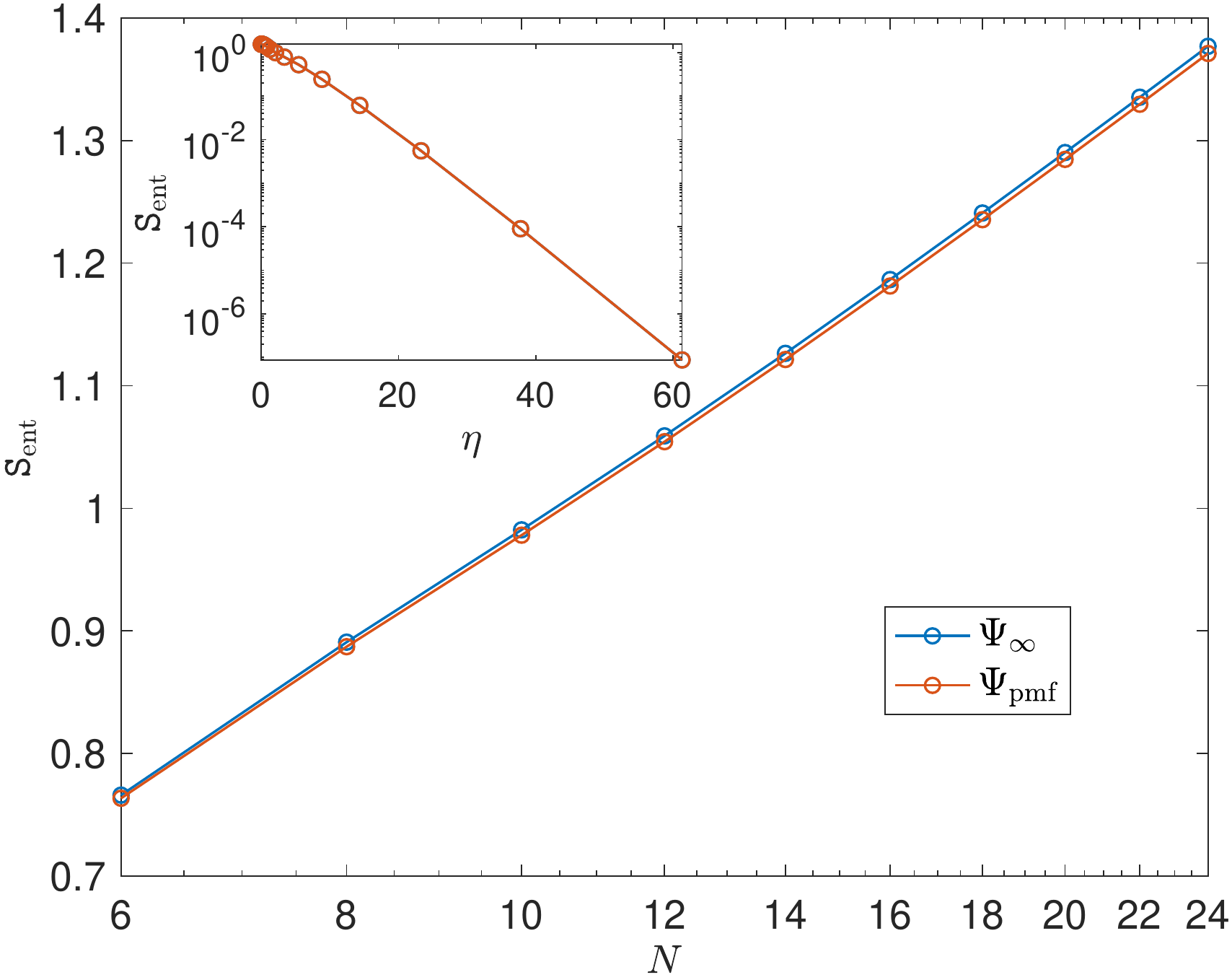}
\caption{  Entanglement  entropies $\mathtt{S}_\mathrm{ent}$ for the   states $\Psi_\infty$ and   $\Psi_\mathrm{pmf}$  vs. $\ln N$, where $N$ is the number of the energy levels $\eps_k$,  $\Psi_\infty$ is the exact solution of the nonstationary Schr\"odinger equation at $t\to+\infty$ for the quantum BCS Hamiltonian with coupling strength $g(t)\propto \frac{\eta}{t}$ starting from the exact ground state at $t=0^+$ and  $\Psi_\mathrm{pmf}$ is the exact solution of the mean-field version of the same problem projected onto a fixed particle number subspace.  Here the number of fermion pairs (number of up pseudospins) is $N_\up=\frac{N}{2}$, $\eta=1$, and $\eps_k=\frac{k}{N}$. Inset: $\mathtt{S}_\mathrm{ent}$ for the same two states as functions of $\eta$; the two curves are indistinguishable on this scale.   $\mathtt{S}_\mathrm{ent}(\Psi_\mathrm{pmf})$  and  $\mathtt{S}_\mathrm{ent}(\Psi_\infty)$ agree well already for small $N$ and are approximately linear in $\ln N$ with slopes 0.438 and 0.441, respectively. Note also that $\mathtt{S}_\mathrm{ent}$ rapidly decreases with $\eta$ consistent with $\mathtt{S}_\mathrm{ent}\to0$ in the adiabatic limit $\eta\to+\infty$. 
}
\label{Fig2mus}
\end{center}
\end{figure}

A popular example of a non-local quantity is the bipartite von Neumann entanglement entropy
\beg
\mathtt{S}_\mathrm{ent} = - \Tr[ \rho_A \ln \rho_A],
\en
where $\rho_A=\Tr_{\!\!\bar A}\, \rho$ is the reduced density matrix of the subsystem $A$: the trace of the system  density matrix over the complement of $A$. Suppose $N$ is even and consider the  most interesting case $N_\up=\frac{N}{2}$. Our choice of $A$ is the spins $\hat{\bm s}_j$  corresponding to the bottom half of the energies $\eps_j$. We find  that the entanglement  entropy for the nonautonomous quantum  BCS dynamics   plateaus  at late times and for large $N$  at
\beg
\mathtt{S}_\mathrm{ent}= c(\eta)\ln N,
\label{bb}
\en
 where $c(\eta)$ is a function of $\eta$ of order one.  This formula holds for both   initial conditions we analyzed: the exact and the BCS ground states at $t=0^+$.  We consider  the latter initial condition in Sec.~\ref{renyi_sec}. The  asymptotic state of the quantum dynamics launched from the exact ground state at $t=0^+$ is $\Psi_\infty$. We   plot  $\mathtt{S}_\mathrm{ent}$ versus $\ln N$ for $\Psi_\infty$ for a range of $N$ and $\eta=1$ in Fig.~\ref{Fig2mus}. A linear fit to this plot gives $c(\eta=1)=0.441$. Remarkably, the entanglement entropy of the projected mean-field state $\Psi_\mathrm{pmf}$ closely matches that of the exact asymptotic state $\Psi_\infty$.
 
 We do not prove \eref{bb} for $\Psi_\infty$ and $\Psi_\mathrm{pmf}$ in general but restrict ourselves to the diabatic, $\eta\to0$, limit.
 We see from Eqs.~(\ref{bogol}, \ref{pmf}), and \re{psiinf} that in this limit
 $\Psi_\mathrm{pmf}=\Psi_\infty=\Psi_0$, where $\Psi_0$ is the exact $t=0^+$ ground state given by \eref{projpsi0}.
 To determine the entanglement entropy for $\Psi_0$, we employ the Schmidt decomposition $\Psi_0=\sum_{i=1}^m w_i | p_i\rangle_A \otimes | q_i\rangle_{\bar A}$, where $| p_i\rangle_A$ and $| q_i\rangle_{\bar A}$ are orthonormal vectors in $A$ and $\bar A$.  The entanglement entropy
can be expressed in terms of the coefficients $w_i$ as 
\beg
\mathtt{S}_\mathrm{ent} = -\sum_{i=1}^m |w_i|^2 \ln |w_i|^2.
\label{bbS}
\en
We have
\beg
\Psi_0(N_\up)=\sum_{N_\up^A=  N_\mathrm{m} }^{ N_\up-N_\mathrm{m} } \frac{ \mbinom{\frac{N}{2}}{N_\up^A} }{ \sqrt{\vphantom{\prod}\smash[b]{\mbinom{N}{N_\up} } } }\left|  N_\up^A \right\rangle_A \otimes \left|   N_\up-N_\up^A \right\rangle_{\bar A},
\label{arb}
\en
where $N_\mathrm{m}=N_\up -\min( \frac{N}{2}, N_\up)$ and $|  N_\up^A \rangle_A$ is the state of the subsystem $A$ with a definite number $N_\up^A$ of up spins and symmetric with respect to an arbitrary permutation of spins. In other words, the subsystem $A$ has the maximum total spin $\frac{N}{4}$ and a definite $z$-projection $N_\up^A-\frac{N}{4}$ of the total spin.  Similarly,   $|  N_\up^{\bar A} \rangle$ is the state of the subsystem $\bar A$ with the maximum total spin and $N_\up^{\bar A}$ up spins. In our  case $N_\up=\frac{N}{2}$, but  we wrote \eref{arb} for arbitrary $N_\up$ for later use.

 Reading off $w_i$ from this equation  and substituting them into \eref{bbS}, we see
 that the values of $N_\up^A$ close to $\frac{N_\up}{2}=\frac{N}{4}$ dominate the summation. Now using the following precise asymptotic expression~\cite{asymptotia} valid for large $N_1$ and $\left|N_2-\frac{N_1}{2}\right|=o(N_1^\frac{2}{3})$:
\beg
2^{-N_1} \binom{N_1}{N_2}=\sqrt{ \frac{2}{\pi N_1} } e^{-2N_1 x^2}, \quad x=\frac{N_2}{N_1}-\frac{1}{2},
\label{gauss}
\en
and converting the summation in \eref{bbS} into an integration, we obtain the leading large $N$ asymptotic behavior of $\mathtt{S}_\mathrm{ent}$,
\beg
\mathtt{S}_\mathrm{ent}=\frac{1}{2} \ln N.
\label{sentinf}
\en
Therefore, $\lim_{\eta\to0} c(\eta)=\frac{1}{2}$. In the opposite (adiabatic) limit $\eta\to+\infty$,
the late-time asymptotic state is the ground state of a noninteracting Fermi gas with no entanglement, i.e., $\lim_{\eta\to+\infty} c(\eta)=0$. Generally, we expect $c(\eta)$ in \eref{bbS} to decrease monotonically  from $\frac{1}{2}$ to 0 as $\eta$ increases from 0 to $+\infty$, see also the inset
in Fig.~\ref{Fig2mus}. 

Entanglement entropy that scales as $\ln N$ (see also the end of Sec.~\ref{renyi_sec}) is  a purely quantum phenomenon that survives the thermodynamic limit. Indeed, the asymptotic state of the mean-field (classical) dynamics is the product state 
 $\Psi_\mathrm{mf}$ with zero entanglement, $\mathtt{S}_\mathrm{ent} =0$. However,  the projected mean-field state $\Psi_\mathrm{pmf}$ appears to  match $\mathtt{S}_\mathrm{ent}$ of the exact $t\to+\infty$ asymptotic state $\Psi_\infty$. Note also
 that $N\propto V$ and $N\propto N_\mathrm{f}$, where $V$ is the system volume and $N_\mathrm{f}$ is the fermion number. Therefore, we can equally well  say that  $\mathtt{S}_\mathrm{ent}$ scales as $\ln V$ or $\ln N_\mathrm{f}$ in the thermodynamic limit.
 
 We see again that the mean-field approximation breaks down for non-local  measures, even though it is exact  for local observables in the thermodynamic limit even for far from equilibrium dynamics that involves highly excited states. Note that the many-body wave function  of the system is itself non-local  and therefore cannot be reproduced by mean field precisely. It is important to emphasize that this obvious breakdown  of the BCS mean-field theory for global quantities is not in any way specific to the nonautonomous BCS Hamiltonian.  For example, we similarly expect the entanglement entropy
 of the exact ground state of the quantum BCS Hamiltonian~\re{eq:bcs_model}  to be proportional to $\ln N$ at $N_\up=\frac{N}{2}$ etc.

\section{Steady State Properties}
\label{gge_sec}

Before    comparing quantum and mean-field dynamics for initial states that are not particle number eigenstates, let us
discuss several physical properties of the steady  state.    Two main results of this section are: (1) the late-time steady state is a gapless superconductor whose superconducting properties can only be revealed through energy resolved measurements and (2)   it conforms to the generalized Gibbs ensemble. The first result holds for a general protocol of turning off the superconducting coupling $g(t)$. The second one is similarly general as long as the mean field remains exact in the thermodynamic limit (we established this above for dynamics with $g(t)\propto \frac{\eta}{t}$ starting in the $t=0^+$ ground state, but it is reasonable to assume the scope of validity of mean field is much broader).

Suppose the interaction  vanishes at $t\to+\infty$, such as our $g(t)\propto \frac{\eta}{t}$.
 Then, at long times  the system evolves with the noninteracting part of the Hamiltonian,
\beg
\hat H(t\to+\infty)=\hat H_0  =
  \sum_{j=1}^N 2\eps_j \hat s_j^z.
\en
Regardless of the  history,  expectation values of spin components at late times  are of the form
\begin{subequations}
\begin{align}
\langle \hat s_k^z\rangle = \ell_k,\quad \langle \hat s_k^-\rangle = b_k e^{-2i \eps_k t},\label{stst1} \\
  \langle \hat s_k^+ s_j^-\rangle =B_{kj} e^{2i (\eps_k -\eps_j) t}.\label{stst2} 
  \end{align}
  \label{ststboth} 
\end{subequations}
For $\Psi_\mathrm{mf}$, we have $B_{kj}=b_k^* b_j$, $b_k=e^{i\varphi_k}\sqrt{ \vphantom{  \sum  }  \smash[b]{ \frac{1}{4}  -\ell_k^2 } }$,
where $\ell_k$ and $\varphi_k$ are given by \esref{szzzz} and \re{12}, respectively. In the thermodynamic limit, $\ell_k$ and $B_{kj}$ for
$\Psi_\infty$ are the same as for $\Psi_\mathrm{mf}$ but   $b_k=0$ , since $\Psi_\infty$  is an eigenstate of the total fermion number operator. However, the \textit{average} of $\hat s_k^-$ in the state $\Psi_\infty$  [the l.h.s. of \eref{s-n-1n}] is the same as its expectation
value in $\Psi_\mathrm{mf}$.

The conventionally defined superconducting order parameter  is zero  in the steady state because it is proportional to the coupling $g=g(t)$, which vanishes as $t\to+\infty$. Consider instead
\beg
\Delta_1=\frac{1}{N}\sum_{k=1}^N \langle \hat s_k^- \rangle,\quad \Delta_2= \frac{1}{N} \sqrt{ \vphantom{\sum}\smash[b]{\sum\nolimits_{k\ne j} }\langle \hat s_k^+ s_j^-\rangle}.
\label{2deltas}
\en
The first of these quantities is the usual BCS order parameter~\re{usualdelta} divided by $gN$. The second is useful for
 the description of off-diagonal long-range order  in
states with definite particle number~\cite{faribault2,delft} as for them $\Delta_1=0$. For a BCS-like product state such as  $\Psi_\mathrm{mf}$, $\Delta_2=\Delta_1$ in the thermodynamic limit. Even though we stripped $\Delta_1$ and $\Delta_2$ of the coupling, they still decay to zero at large times in the continuous  limit  due to dephasing. Indeed, in this limit sums in \eref{2deltas} become integrals that tend to zero as $\frac{1}{t}$ 
when $t\to+\infty$ by the Riemann-Lebesgue   lemma. 

We can  learn more about the properties of  asymptotic states $\Psi_\infty$ and $\Psi_\mathrm{mf}$  from the mean-field dynamics of 
BCS superconductors   quenched via a sudden change of the coupling $g_i\to g_f$. For sufficiently small but generally nonzero $g_f$~\cite{dzero1}, these systems too go into a steady state of the   form~\re{stst1}  at long times, which is known as phase I in this context and is one of the three asymptotic states (nonequilibrium phases) that the superconductor can end up in depending on $g_i$ and $g_f$~\cite{foster}.  Energy averaged indicators  of fermionic superfluidity, such as the superconducting order parameter, energy gap for pair-breaking excitations~\cite{dzero1}, and superfluid density~\cite{density}, vanish in this state due to dephasing (anomalous averages at different energies in \eref{stst1} oscillate with different frequencies). These conclusions rely  on the general form of the steady state~\re{stst1} only and are therefore valid in our case as well. 

 Nevertheless,    exact asymptotic states $\Psi_\infty$ and $\Psi_\mathrm{mf}$ we derived above for  quantum and mean-field dynamics  for $g(t)\propto \frac{\eta}{t}$ do exhibit superconducting correlations, e.g.,
the equal time anomalous Green's functions $\langle \hat s_k^-\rangle$ for $\Psi_\mathrm{mf}$ and  
$ \langle \hat s_k^+ s_j^-\rangle$ for $\Psi_\infty$ are nonzero. 
However, to reveal these correlations, we need energy resolved measures,   such as the  spectral supercurrent density~\cite{supercurrent1,supercurrent2}. Any complete discussion of prospects of  experimental observation and characterization of these asymptotic states is beyond the scope of this paper. At this point, it suffices  to say that physically our system  is a gapless fermionic superfluid with vanishing energy averaged superfluid characteristics at long times.

Time averaged expectation values of observables in asymptotic states   $\Psi_\infty$ and $\Psi_\mathrm{mf}$ and, in particular, the distribution $P(\{s^z\})$ of $z$-components of spins  are nonthermal, and  there is no   reason to expect isolated systems with infinite range interactions such as ours to thermalize~\cite{dont1,dont2,dont3}. 
Instead, these states are   described by the generalized Gibbs ensemble~\cite{gge,gge2}, as we now show. 

Since at  $t\to+\infty$ the system evolves with $\hat H_0$, its wave function is of the form~\cite{note75}
\beg
\Psi(t)=\sum_{ \{s^z\}} C_{ \{s^z\}} e^{-i E\left(\{s^z\}\right) t} | \{s^z\}\rangle,
\en
where $\{s^z\}=\{s_1^z,\dots,s_N^z\}$ is a set of eigenvalues of $\hat s_k^z$, the sum is over all such sets, and $| \{s^z\}\rangle$ are simultaneous eigenstates of all $\hat s_k^z$. The eigenergies $E\left(\{s^z\}\right)$ of $\hat H_0$ are generally nondegenerate and as
a result the time averaged expectation values of observables are given by the diagonal ensemble~\cite{diagonal},  
\beg
\langle \Psi(t)| \hat O |\Psi(t) \rangle= \sum_{ \{s^z\}} \left|C_{ \{s^z\}}\right|^2  \left\langle \{s^z\}| \hat O | \{s^z\}\right\rangle.
\label{diagonalE}
\en
In the present case,
\beg
 \left|C_{ \{s^z\}}\right|^2=P(\{s^z\}),
\en
i.e., the diagonal ensemble $ \left|C_{ \{s^z\}}\right|^2$ is the same as the distribution $P(\{s^z\})$ of $s_k^z$ -- the probability of finding the system in the state $ | \{s^z\}\rangle$. 
Therefore, to demonstrate  that GGE describes our asymptotic 
states,  it is enough to prove that it is equivalent to  $P(\{s^z\})$. 

No constants of motion are known for the Hamiltonian~\re{eq:H} with $g(t)\propto \frac{\eta}{t}$ at finite $t$. However, a set of local integrals  of motion  obviously emerges at $t\to+\infty$, namely, $\hat s_k^z$ that commute with $\hat H_0$ and among themselves. GGE by definition is the density matrix $\hat \rho_\mathrm{GGE}=e^{-\sum_k \vartheta_k \hat s_k^z}$. In the $ | \{s^z\}\rangle$ basis $\hat \rho_\mathrm{GGE}$ is diagonal with diagonal matrix elements, 
\beg
\rho_\mathrm{GGE}(\{s^z\})=C_\rho e^{-\sum_k \vartheta_k  s_k^z},
\label{gge}
\en
 i.e.,
$\Tr( \hat \rho_\mathrm{GGE} \hat O)$ is given by \eref{diagonalE} with $\rho_\mathrm{GGE}(\{s^z\})$ in place of $ \left|C_{ \{s^z\}}\right|^2$.
Crucially, we need only $N$ parameters $\vartheta_k$ to specify the GGE, while  for the diagonal ensemble we need to 
specify every $ \left|C_{ \{s^z\}}\right|^2$, which is $2^N$ parameters.  Furthermore, the general belief that GGE should be a valid   description of the  long-time dynamics in the thermodynamic limit extends only to  Hamiltonian systems with local  interactions. We see that whether or not GGE reproduces the diagonal ensemble  is a nontrivial question.

We already know the distribution $P(\{s^z\})$ for $\Psi_\infty$, see \eref{eq:quantum_dist}. The distribution for $\Psi_\mathrm{mf}$ is the same but without the Kronecker delta because $\Psi_\mathrm{mf}$ is a product state, see \eref{psimf1}. Further, it is not difficult to show that  fixing
the average $z$-projection of the total spin $\langle \hat \jmath_z\rangle$ instead of its eigenvalue $\jmath_z$ introduces corrections of order $\frac{1}{N}$ to the expectation values of local operators in the thermodynamic limit. It follows that in this limit  we have for both $\Psi_\infty$ and $\Psi_\mathrm{mf}$
\begin{equation}
 \label{eq:quantum_dist123}
  P(\{s^z\}) = C_P
  e^{-\frac{2\pi}{\nu}\sum_k (k-\mu) s_k^z},
\end{equation}
where $\mu$ is the chemical potential  that determines $\langle \hat \jmath_z\rangle$ as we already discussed in the previous two sections. Comparing \esref{eq:quantum_dist123}
and \re{gge}, we see that  in the thermodynamic limit the GGE with
\beg
\vartheta_k=\frac{2\pi}{\nu} (k-\mu)=2\pi \eta \frac{k-\mu}{N}.
\label{invTs}
\en
is an exact description of the steady states of quantum and classical dynamics of the BCS model with interaction inversely proportional to time.   Usually, we determine the parameters $\vartheta_k$ in $\rho_\mathrm{GGE}$  from the expectation values of the integrals of motion
in the initial state~\cite{gge,diagonal}. This is impossible in our case as $\hat s_k^z$ are conserved only at $t\to+\infty$.

Let us also comment on the relevance of the thermal distribution $\hat \rho_T=e^{-\beta_T( \hat H_0 - \tilde\mu \hat N_\mathrm{f} )}=
e^{- \sum_k 2\beta_T(\eps_k -\tilde\mu) \hat s_k^z}$ for our steady states. We observe with the help of \eref{invTs} that the GGE is identical to the thermal distribution for certain $\beta_T$ and $\tilde\mu$ when and only when the single-fermion levels $\eps_k$ are equidistant.   Such an exceptional point always exists   in the multi-dimensional parameter space of a general integrable system~\cite{gme} and    
  should be  regarded as a degenerate  instance of GGE   rather than a case of  thermalization.
  
   We expect the GGE description  to be valid  for nonautonomous BCS Hamiltonians more generally, including for nonintegrable  time dependence and  a broad class of initial conditions.  Note that GGE is valid whenever the mean field is, since the mean-field wave function is a product state. Then, $ P(\{s^z\})$ is a product of individual spin distributions and  the distribution for an unentangled spin-$\frac{1}{2}$ can always be written as $e^{-\vartheta \hat s^z}$. Nevertheless, it is interesting to investigate the relationship between the emergent GGE and time-dependent integrability as well as the scope of the validity of GGE for nonautonomous BCS dynamics more thoroughly.

\section{Order of limits}
\label{order_sec}

Let us discuss various limits of the quantum BCS dynamics with time-dependent coupling $g(t)=\frac{\eta}{Nt}$. 
Above we worked out the late-time   asymptotic behavior followed by the large   $N$ behavior at fixed fermion density. 
 We launched the time
evolution from the exact $t=0^+$ ground state with definite fermion number $N_\mathrm{f}$. Since the interaction diverges at $t=0$, we interpreted this   as starting in the ground state at $t=t_0$ and then taking the limit $t_0\to0^+$. In this section, we show that the three limits: $t\to+\infty$, thermodynamic, and $t_0\to0^+$  mutually commute as long as our time evolving state
is a particle (fermion) number eigenstate. 
   Of interest is also the adiabatic limit   $\eta\to+\infty$ and we show that it commutes with the thermodynamic and $t_0\to0^+$ limits. In stark contrast, we will see in Sec.~\ref{deph_sec} that several of these commutativity properties do not hold  for observables that do not conserve the total fermion number, such as the Cooper pair annihilation operator $\hat s_k^-$,  when the time-dependent wave function is not a particle number eigenstate.

The time enters the Hamiltonian~\re{eq:H} in the combination $\frac{t}{\eta}$.
The limit $\eta\to 0^+$ (taken after  $t_0\to 0^+$) is the diabatic (quantum quench) limit. In this limit, the Hamiltonian changes instantaneously from 
\begin{equation}
 \hat H_\mathrm{int}= - \frac{\eta}{Nt}\sum_{j,k=1}^N\hat{s}_j^+\hat{s}_k^-.
  \label{h0hint}
\end{equation}
 with infinite coupling to the noninteracting Hamiltonian $\hat H_0$.
The opposite limit $\eta\to+\infty$ is the adiabatic limit where the Hamiltonian changes infinitely slowly. First, let us  take this limit before the thermodynamic one.  The ground state of the BCS Hamiltonian is nondegenerate, therefore   the system stays in it at all times in the adiabatic limit for any finite $N$ by the adiabatic theorem. The late-time wave function $\Psi_\infty$ we derived above confirms this. Recall that the fermion number is $N_\mathrm{f}=2N_\up$, i.e., twice the number of up pseudospins. The Hamiltonian at $t\to+\infty$ is $\hat H_0$: the Hamiltonian of noninteracting fermions. In its   ground state,  the first $N_\up$ spins are up and the rest are down ($\eps_j$ are arranged in ascending order).    \eref{eq:quantum_dist} shows that $\lim_{\eta\to+\infty}\Psi_\infty$ is indeed    the noninteracting ground state for any $N$, including $N\to\infty$, because the  probability of any other spin configuration relative to the ground state vanishes.

 Now let us take the thermodynamic limit before the adiabatic one. The quantum average of $\hat s_k^z$ in the thermodynamic limit  is given by \eref{test1} and it is not difficult to see  that taking the adiabatic limit next we end up  in the same noninteracting ground state again. Therefore, the thermodynamic and adiabatic limits commute. This makes sense physically as our instantaneous energy spectrum is that of the BCS superconductor and there is  a finite (in the thermodynamic limit) gap between the ground state and the first excited state at any finite $t$.

The dependence of the wave function on $t_0$ (at any $t$)  follows from elementary quantum mechanics.     At early times the system evolves adiabatically  with $\hat H_\mathrm{int}$    because there is a diverging gap between the ground state and the first excited state~\cite{verify22}. In the adiabatic evolution, the wave function merely accumulates an overall phase. As a result, the entire dependence on $t_0$ comes from early times and is confined to the global phase.  This can be seen from the exact solution~\re{psi} as well. A small change in $t_0$ is a small change in the initial condition. This translates into a small deformation of the contour $\gamma$,  which has no affect on the saddle-point calculation in Sec.~\ref{late_time_sec}. As a result, the late-time wave function $\Psi_\infty$  in \eref{psiinf}  is valid for any sufficiently small $t_0$. Moreover,  $\Psi_\infty$  being defined up to a global phase only does not depend on $t_0$ at all.

 Solving the nonstationary Schr\"odinger equation for $\hat H_\mathrm{int}$ at small $t$ (see also Sec.~\ref{early_sec}), we determine the $t_0$-dependence of the solution $\Psi(t)$ of the nonstationary  Schr\"odinger equation for the  BCS Hamiltonian~\re{eq:bcs_model} [equivalently \eref{eq:H} for $s=\frac{1}{2}$] at any $t\gg t_0$,
 \beg
 \Psi(t) = e^{-i  E_0(N_\up) \tau_*} F(t) ,
 \label{psi_sum1}
 \en
 where   $F(t)$ is independent of $t_0$, $E_0(N_\up)$ is the rescaled ground state energy at $t=0^+$,
 \beg
 E_0(N_\up) =\frac{ N_\up^2- N_\up N -N_\up}{N},\quad \tau_*=\eta\ln \frac{t_*}{t_0},
 \label{e0}
 \en
 and $t_*$ is a  function of $\eta$, $N$, and $N_\up$ only (see below).  
 
 In Sec.~\ref{late_time_sec} we worked out the exact late-time asymptotic solution $\Psi_\infty(N_\up)$ for the quantum BCS time evolution   up to a  global phase.  \eref{psi_sum1} provides this phase. In other words,
   \beg
 \widetilde{\Psi}_\infty(N_\up) =   e^{-i  E_0(N_\up) \tau_*} \Psi_\infty(N_\up)
 \label{one2other}
 \en
  is the late-time asymptotic solution  including the full overall phase.    $\Psi_\infty(N_\up)$ captures the time dependence of the global phase at large $t$ as it solves  the nonstationary Schr\"odinger equation in this limit~\cite{factor12}. Therefore, $t_*$ is independent of $t$.  We do not attempt to determine $t_*$ exactly but provide an order of magnitude estimate in the thermodynamic limit.  Physically, $t_*$ is the time until which $\hat H_0$ is negligible and the system evolves adiabatically with $\hat H_\mathrm{int}$. It separates the strong coupling (early-time) regime where the dimensionless BCS coupling $\frac{g(t)}{\delta_1}\gg 1$ from the weak coupling (late-time) regime where
  $\frac{g(t)}{\delta_1}\ll 1$. Here 
  $\delta_1\approx \frac{W}{N}$ is the mean spacing between single-particle energy levels $\eps_j$ and $W$ is the bandwidth. The dimensionless coupling is equal to 1 at $t=\frac{\eta}{W}$. Therefore, we expect $t_*\sim \frac{\eta}{W}$. In Sec.~\ref{early_sec} we estimate $t_*$ for equally spaced   $\eps_j$ more accurately  as 
  \beg
  t_*\approx\frac{0.1 \eta}{W}.
  \en
  Note that $t\sim t_*$ is also the time when  the global phase stops accumulating.   The precise form of $t_*$ is unimportant for our purposes. The only  assumption we will be making is that the variation in $t_*$  due to changing $N_\up$ by a finite integer is negligible in the thermodynamic limit.

 Consider the expectation value  in the state $\widetilde{\Psi}_\infty(N_\up)$  of a  product $\hat O$   of spin operators that commutes with the total fermion number operator.   First, we see  from \eref{psi_sum1} that $\langle \hat O\rangle_{\widetilde{\Psi}_\infty}$ is independent of $t_0$. Therefore, the limit $t_0\to 0^+$ commutes with all the other limits. Further,  regardless of  the order in which we calculate the large $t$ and large $N$ asymptotic  behaviors of $\langle \hat O\rangle_{\widetilde{\Psi}_\infty}$, the answer is an $N$-dependent number of order one, which has a definite $N\to\infty$ value,  times a set of exponents $e^{\pm 2 i \eps_k t}$. It is clear from the derivation of Sec.~\ref{late_time_sec} that we obtain the same  value regardless of whether we take the thermodynamic limit before or after the stationary point calculation. We see that  $t\to+\infty$ and thermodynamic limits also commute. To summarize the results of this section,   $t\to+\infty$ (late-time),  $N\to\infty$  (thermodynamic),  $\eta\to+\infty$ (adiabatic),
 and $t_0\to0^+$ limits mutually commute when we launch the evolution from a state with a definite total fermion number $N_\mathrm{f}$, with the exception of the late-time and adiabatic limits which of course  do not commute.

\section{Quantum evolution from BCS ground state}
\label{deph_sec}

Above we examined the time evolution with the quantum BCS Hamiltonian~\re{eq:bcs_model} with coupling $g(t)=\frac{\eta}{N t}$ starting from the exact
$t=0^+$  ground state, which is an eigenstate of the total fermion number operator $\hat N_\mathrm{f}$. We saw  that averages of local operators coincide with those in  the time-dependent mean-field   state supplied by the classical BCS dynamics. This is equally true for operators that conserve the fermion number and those that do not, such as the pair annihilation operator $\hat s_k^-$.   In the latter case, the average is defined as the matrix element between two solutions of the nonstationary Schr\"odinger equation with different $N_\mathrm{f}(=2N_\up)$, see \eref{exp123}.

Now let us investigate the evolution   from the $t=0^+$ (infinite superconducting coupling) mean-field BCS ground state, which is a superposition of states with all possible $N_\mathrm{f}$.  Even though it is distinct from the exact   ground state, the thermodynamic limits   of various observables are the same~\cite{nucl1,nucl2}.  However, the status of the BCS mean field changes in the course of evolution.   Phases of components of the many-body wave function corresponding to different $N_\mathrm{f}$  evolve at different   rates.   As a result, the entanglement entropy grows and   expectation values of operators that  do not commute with the total  fermion number, e.g., the equal time anomalous Green's function $\langle \hat s_k^-\rangle$, dephase at late times as their nonzero matrix elements are between sectors with different $N_\mathrm{f}$.  We will see that the agreement of such expectation values with their mean-field counterparts is more fragile than that  of particle-number-conserving observables and depends on $\eta$ and $t_0$ in addition to $N$.

For simplicity, we focus on the most interesting case when the average $z$-component of the total spin $\langle \hat \jmath_z\rangle=0$. This corresponds to half of the spins being up on average, $\langle \hat N_\up\rangle=\frac{N}{2}$, and average fermion number $\langle \hat N_\mathrm{f}\rangle=N$.
  The BCS ground state in this case is [see \eref{alongx}]
\beg
\Psi_\mathrm{BCS}=|\rightarrow \rightarrow \rightarrow \dots\rangle=2^{-\frac{N}{2}}\prod_k\left(|\!\dn\rangle+|\!\up\rangle\right).
\label{alongx1}
\en
As discussed in Sec.~\ref{ic_sec}, in mean-field approach this state corresponds to the lowest energy classical spin configuration for $J_z=0$ where all classical spin vectors are along the $x$-axis.

To obtain the quantum evolution launched from $\Psi_\mathrm{BCS}$, we decompose this state into exact $t=0^+$   ground states~\re{psi0} with varying number of up spins (fermions),
\beg
\Psi_\mathrm{BCS}=2^{-\frac{N}{2}}\sum_{N_\up=0}^N {\binom{N}{N_\up}}^{\frac{1}{2}}\Psi_0(N_\up).
\label{bcsdecomp}
\en
In Sec.~\ref{late_time_sec}, we derived the $t\to+\infty$ asymptotic solution $\Psi_\infty(N_\up)$ [\eref{psiinf}] of the nonstationary Schr\"odinger equation with the initial condition $\Psi(t=0) =\Psi_0(N_\up)$ up to an overall phase. In the previous section, we  obtained the  dependence of the overall phase on   $t_0$, see \eref{one2other}. 
 By linearity of the Schr\"odinger equation, the asymptotic solution  for the quantum evolution with the time-dependent BCS Hamiltonian
starting from $\Psi_\mathrm{BCS}$ is 
\beg
\Phi_\mathrm{\infty}=2^{-\frac{N}{2}}\sum_{N_\up=0}^N {\binom{N}{N_\up}}^{\frac{1}{2}} e^{-i  E_0(N_\up) \tau_* }|N_\up\rangle_\infty,
\label{phiinf1}
\en
where $|N_\up\rangle_\infty$ is  the normalized version of $\Psi_\infty(N_\up)$ defined in  \eref{normvers}.  

Consider an arbitrary product $\hat O_\mathrm{con}$ of $n$ spin operators  that conserves the number $2N_\up$ of fermions  or, equivalently, the number $N_\up$ of up spins, i.e., commutes with the $z$-projection of the total spin $\hat \jmath_z$. As before, we assume that $\frac{n}{N}\to0$  when $N\to\infty$. Matrix elements of $\hat O_\mathrm{con}$  between states with different $N_\up$ are zero and therefore its quantum average   in the evolved BCS state is
\beg
\langle \hat O_\mathrm{con} \rangle_{\Phi_\infty}= 2^{-N}\sum_{N_\up=0}^N \binom{N}{N_\up} \langle \hat O_\mathrm{con} \rangle_{\Psi_\infty(N_\up)}.
\label{BCSav}
\en
This summation localizes at $N_\up\approx \frac{N}{2}$ for large $N$.  We see this with the help of \eref{gauss} with $N_1=N$, $N_2=N_\up$, and $x=\frac{N_\up}{N}-\frac{1}{2}$.   In the thermodynamic limit, the average $\langle \hat O_\mathrm{con} \rangle_{\Psi_\infty(N_\up)}$ is of the form~\re{exp123}. It depends on $x$ through the chemical potential $\mu$ and 
 is generally a smooth function of $x$   of order one.  Equation~\re{BCSav} becomes
$$
\langle \hat O_\mathrm{con} \rangle_{\Phi_\infty}= \sqrt{ \frac{2N}{\pi} } \int_{-\frac{1}{2}}^{\frac{1}{2}} dx e^{-2N x^2} \left[\langle \hat O_\mathrm{con} \rangle_{\Psi_\infty(N_\up)}\right],
$$
where $N_\up= xN+\frac{N}{2}$.
In the limit $N\to\infty$, the weight function tends to the Dirac delta function, $\sqrt{ \frac{2N}{\pi} } e^{-2N x^2}\to \delta(x),$ and we have
\beg
\langle \hat O_\mathrm{con} \rangle_{\Phi_\infty}= \left. \langle \hat O_\mathrm{con} \rangle_{\Psi_\infty}\right|_{N_\up= \frac{N}{2}\dis. }
\label{abc1}
\en
Therefore,  expectation values of observables  conserving the  total fermion number     for the  evolution from the BCS ground state with  average  fermion  number $\langle \hat N_\mathrm{f}\rangle$ and    from the exact  ground state with   definite  $N_\mathrm{f}$ coincide when $\frac{ N_\mathrm{f} }{N}\to\frac{ \langle \hat N_\mathrm{f}\rangle }{N}$ as
$N\to\infty$.  

The behavior of observables $\hat O_\mathrm{nc}$ that do not commute with $\hat N_\mathrm{f}$   is   different. Consider, for example, $\hat s_k^-= \hat{c}_{k\uparrow}\hat{c}_{k\downarrow}$. Note that the expectation value of $\hat s_k^-$ in the state $\Phi_\infty$ is the equal time anomalous Green's function at $t\to+\infty$. Going through the same steps as for $\hat O_\mathrm{con}$, we find
\beg
\langle \hat s_k^- \rangle_{\Phi_\infty}=  \sqrt{ \frac{2N}{\pi} } \int_{-\frac{1}{2}}^{\frac{1}{2}} dx e^{2i \tau_* x-2N x^2}  \langle \hat s_k^-\rangle_\mathrm{mf},
\label{abc2}
\en
 where we used \esref{s-n-1n} and \re{s-n-1n1}. Applying the steepest descent method or simply completing the square in the exponent,
 we obtain
 \beg
 \langle \hat s_k^- \rangle_{\Phi_\infty}=\left. e^{-\frac{\tau_*^2}{2N} } \langle \hat s_k^-\rangle_\mathrm{mf} \right|_{N_\up= \frac{N}{2}\dis. }
 \label{007}
 \en
 Recall that $\tau_*=\eta\ln\frac{t_*}{t_0}$ and  $\langle \hat s_k^-\rangle_\mathrm{mf}$ is the expectation of $\hat s_k^-$ in the late-time asymptotic wave function~\re{psimf} for the mean-field  time evolution.   We see immediately that the thermodynamic limit $N\to\infty$ does not commute with  the $t_0\to0^+$  limit. Indeed, in the former limit $\langle \hat s_k^- \rangle_{\Phi_\infty}=\langle \hat s_k^-\rangle_\mathrm{mf}$, while in the latter limit $\langle \hat s_k^- \rangle_{\Phi_\infty}=0\ne \langle \hat s_k^-\rangle_\mathrm{mf}$. 
 
 Similarly, the prefactor $e^{-\frac{\tau_*^2}{2N} }$ in \eref{007} vanishes if we take 
 the $\eta\to+\infty$ (adiabatic) limit before the thermodynamic one and is equal to 1 if we take these limits in the reverse order. However,
 in the adiabatic limit $\langle \hat s_k^-\rangle_\mathrm{mf}=0$, i.e., both $ \langle \hat s_k^- \rangle_{\Phi_\infty}$ and $\langle \hat s_k^-\rangle_\mathrm{mf}$ vanish.   Nevertheless,    these two limits   do not commute for anomalous  averages as we will see more clearly in Sec.~\ref{early_sec} where we obtain an expression very similar to \eref{007} but for the early-time quantum dynamics of the BCS Hamiltonian.
 
 The non-commutation of the thermodynamic with adiabatic and $t_0\to0^+$ limits is a purely quantum effect because in  classical  (mean-field) dynamics with the same initial condition [\esref{classini} and \re{alongx1}] we by definition obtain $\langle \hat s_k^- \rangle=\langle \hat s_k^-\rangle_\mathrm{mf}$ at $t\to+\infty$, where $\langle \hat s_k^-\rangle_\mathrm{mf}$ is given by \eref{s-n-1n1}  regardless of  the order in which we take these   limits. The effect comes from the global phase of the wave function in \eref{one2other}: amplitudes of states with different $N_\up$ (different fermion numbers $N_\mathrm{f}$) are periodic in $\tau_*$ with a frequency $E_0(N_\up)$ that disperses with respect to $N_\up$ resulting in dephasing for observables that do not commute with $\hat N_\mathrm{f}$. 
 
 This is
 analogous to free particle wave packet spreading.    Indeed, using  \eref{gauss} in \eref{phiinf1}, we see that  $\Phi_\infty$ is of the form of the time-dependent wave function of a free particle initially prepared in a Gaussian wave packet  (in momentum representation). The variable $x=\frac{N_\up}{N}-\frac{1}{2}$  plays the role of particle's momentum   and $\tau_*$   the role of time.      The transverse part of the total spin  $\hat\jmath_\pm=\sum_k \hat s_k^\pm$    is roughly analogous to particle's position  and the uncertainty in it similarly grows. Indeed, it is straightforward to show that $\langle \hat \jmath_+\hat \jmath_-\rangle - \langle \hat \jmath_+\rangle \langle \hat \jmath_-\rangle$ is zero at $t=0$, because all spins are along the $x$-axis, and is equal to $N$ in the state $\Phi_\infty$ for large $N$ due to dephasing.  
  
Notice that the drastic difference in the late-time values of the anomalous average  $\langle\hat s_k^-\rangle= \langle\hat{c}_{k\uparrow}\hat{c}_{k\downarrow}\rangle$ for quantum and mean-field time evolution is also a  dynamical effect but is unrelated to the time dependence of the BCS coupling constant. It only requires an initial state that is  a superposition of a large number of the eigenstates of the Hamiltonian. In the present case, the magnitude of this
quantum dynamical effect (dephasing) is controlled by the parameter 
\beg
\mathtt{Q}=\frac{ \eta^2 \ln^2 \frac{t_*}{t_0} }{ 2N}
\en
 in contrast to the  parameter
$\frac{1}{N}$ that controls other quantum fluctuations (finite size corrections) of local observables~\cite{local}  in far from equilibrium dynamics as we saw above. Similarly, $\frac{1}{N}$ is  the parameter that  ensures the smallness of quantum fluctuations   in equilibrium~\cite{richardson,roman,baytin}.
Dephasing  is the dominant quantum effect for anomalous averages when $\eta^2 \ln^2 \frac{t_*}{t_0}\gg1$. For example, for $\eta=20$, $\frac{t_*}{t_0}=10^3$, and $N=10^4$ usual finite size corrections are of order $0.01\%$ and are negligible compared to dephasing, which is no longer a small correction as $\mathtt{Q}\approx 1$.

\section{Approach to the steady state}
\label{approach_sec}

So far we focused on the  $t\to+\infty$ asymptotic state. It is also important to understand how quickly the system reaches this state. Here we analyze this issue numerically for the  classical   time evolution, i.e., for the dynamics generated by the mean-field BCS Hamiltonian~\re{eq:class_H}
with interaction strength $g(t)\propto \frac{\eta}{t}$.

\begin{figure}
\begin{center}
 \includegraphics[width = 0.49\textwidth]{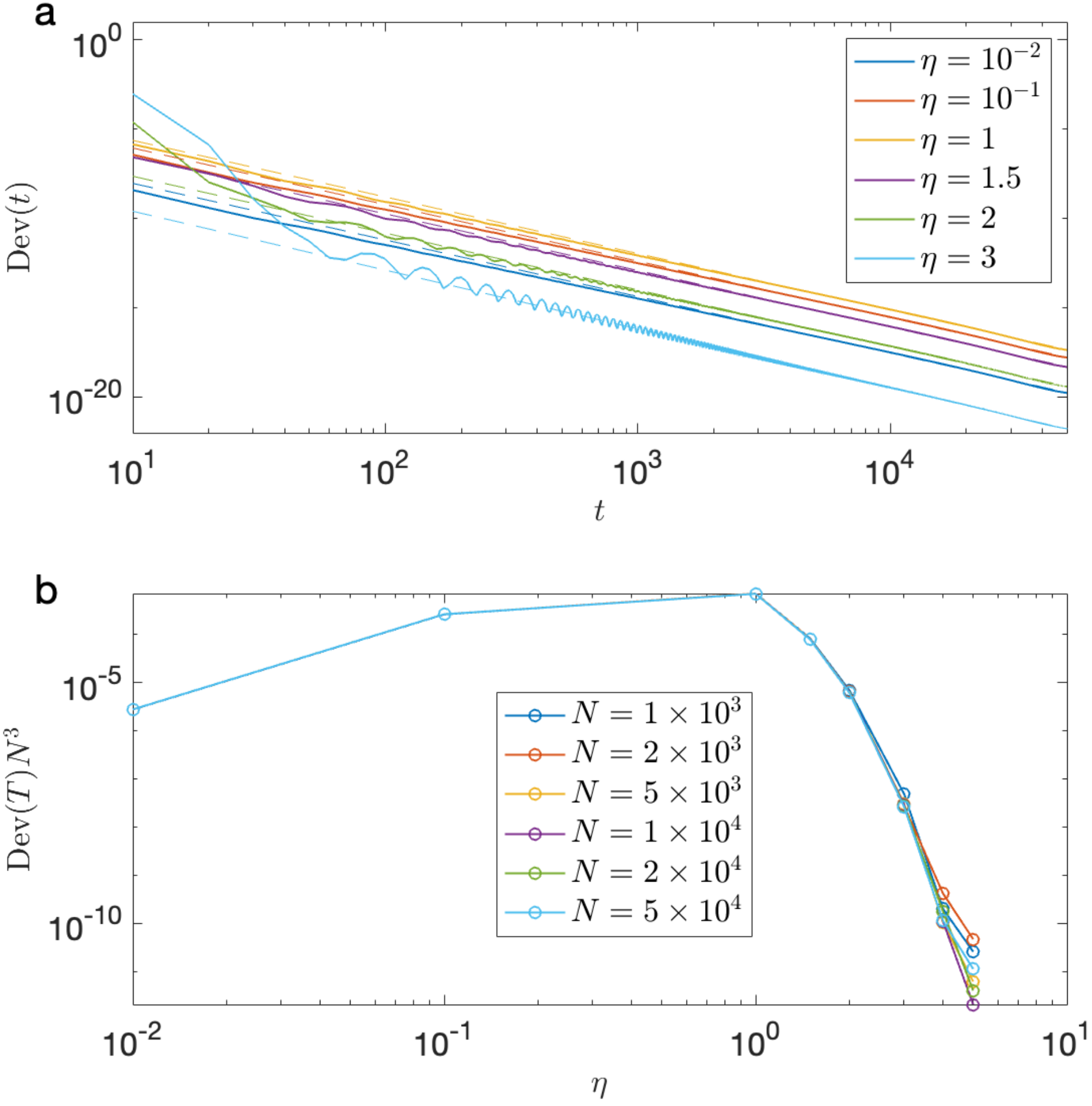}
\caption{Deviation~\re{dev}   from the asymptotic state of the mean-field   dynamics of the BCS Hamiltonian with coupling
$g(t)=\frac{\eta}{Nt}$ starting at $t=t_0=10^{-17}$ from the $t=0^+$ ground state.  $N$ is the number of energy levels $\eps_k=\frac{k}{N}$ and the number of fermions. Top panel: the deviation as a function of time for various values of $\eta$,  $N=5\times 10^4$, and tolerance $2.22\times 10^{-14}$.  The time dependence at  $1\ll t\lesssim N$ fits a power law  with exponent $-3.2$,
which is somewhat overestimated due to faster decay at earlier times. At  larger $t$ the deviation  saturates. Bottom panel: deviation scaled by $N^3$ at $t=T=1.57N$ as a function of   $\eta$ for various $N$. The  dependence   on $N$ is $\text{Dev}\propto N^{-3.0\pm 0.1}.$ We conclude from these data that the system can get arbitrarily close to the asymptotic state in a finite time in the $N\to\infty$ limit. }
\label{Fig2}
\end{center}
\end{figure}

Consider the average squared deviation of the $z$-components   of classical spins from their $t\to+\infty$ asymptote in the thermodynamic limit $S_k^z(\infty)$ given by \eref{11}
\begin{equation}
 \text{Dev}(t,\eta, N)=\frac{1}{N}\sum_{k=1}^N \left[S_k^{z}(t)-S_k^{z}(\infty)\right]^2.
 \label{dev}
\end{equation}
In Fig.~\ref{Fig2}, we plot this deviation as a function of $t$ for a range of  $\eta$ at a large fixed $N$ (upper panel) and as a function of $\eta$ for
a range of $N$ at fixed large $t=T=1.57N$  (lower panel).  Units of time in our simulation are set by our choice of single-particle energies $\eps_k=\frac{k}{N}$ with $k=1,\dots,N$. Therefore, the bandwidth $W\approx 1$ and the units of time are approximately $W^{-1}$. 

Analysis of  results of simulations  presented in  Fig.~\ref{Fig2} shows that at  times $1\ll t\lesssim N$ the deviation decays as
\beg
\text{Dev}\approx \frac{ R(\eta)}{ t^{3.0}},
\label{power_law}
\en
where $R(\eta)$ is a positive function of $\eta$ that is independent of $N$ for large $N$ as is evident from the lower panel of Fig.~\ref{Fig2}.  At earlier times the decay of the deviation is even faster.   For this reason, inclusion of earlier times in the analysis of $\text{Dev}(t)$ shown in the top panel of Fig.~\ref{Fig2} produces higher powers of $t$, namely, $\text{Dev}\propto t^{-3.2}$. At $t\approx N$ the deviation saturates to a constant   proportional to $N^{-3.0\pm0.1}$.  This result provides a more reliable way to determine
the power law dependence of the deviation. Substituting $t=N$ into $\text{Dev}\propto t^{-y}$, we find $y=3.0$ as in 
\eref{power_law}.

Most importantly, we see that the system can get arbitrarily close to the  late-time asymptotic state in  finite time in the thermodynamic limit, i.e.,    the properties of $\Psi_\infty$ in this limit that we established above are  accessible. Note that because the interaction vanishes at $t\to+\infty$, the system goes into an asymptotic state with $S_k^z=\text{const}$ for any $N$. Moreover,   the thermodynamic and $t\to+\infty$ limits commute~\cite{showed}. This implies independently of the  numerical evidence that the system reaches an arbitrarily small vicinity of the asymptotic state in finite time. Separately, we observe   that the deviation vanishes in the diabatic (noninteracting), $\eta\to0$, and adiabatic, $\eta\to+\infty$ limits as expected. 

We mentioned above that the deviation saturates at $t\propto N$ at which point $\text{Dev}\propto N^{-3.0}$. This along with \eref{dev} implies that corrections to our $N\to\infty$ analytic results in Sec.~\ref{QvsC_sec}  for $S_k^z$ and other spin components are
of order $N^{-\frac{3}{2}}$. It is interesting to understand this scaling   with $N$ as naively we would expect   $N^{-1}$ scaling. These corrections to the $N\to\infty$ limit  \textit{within} mean field are not to be confused with the corrections \textit{to} mean field  due to quantum fluctuations. The latter corrections to $S_k^z$ are indeed of order $N^{-1}$, see
Sec.~\ref{thm_qm_sec}.

\section{Early-time quantum dynamics}
\label{early_sec}

 Now let us investigate quantum effects at early times where the interaction
part of the Hamiltonian  dominates the dynamics being proportional to  $\frac{1}{t}$   and the kinetic term $\hat H_0=\sum_j 2\epsilon_j
\hat{s}_j^z$ is negligible. Of special interest   is the evolution starting from the BCS ground state at $t=0^+$ and observables that do not commute with the total fermion number operator, e.g., $\hat s_k^-$. We already saw in Sec.~\ref{deph_sec}, that  such observables dephase with time.   The  magnitude of this quantum effect is controlled by a parameter  distinct from the  one that controls quantum fluctuations (corrections to mean field) in equilibrium.  In this section, we   illustrate this  in a much simpler setting of the early-time dynamics. Then, we derive the von Neumann entanglement entropy $\mathtt{S}_\mathrm{ent}$ at early times and show that it monotonically increases with time and is $N$-independent in the thermodynamic limit.  On the other hand, at finite but large $N$  we argue that 
$\mathtt{S}_\mathrm{ent}$    saturates  at $\mathtt{S}_\mathrm{ent}=c(\eta)\ln N$, 
where $c(\eta)$ is a function of $\eta$ of order one. Dephasing and the growth of entanglement are  two sides of the same coin. With time the phases of components of the many-body wave function  with different particle numbers randomize, so that they no longer combine into a BCS product state and eventually saturate the entanglement entropy.

\begin{figure}[t!]
\includegraphics[width=0.48\textwidth]{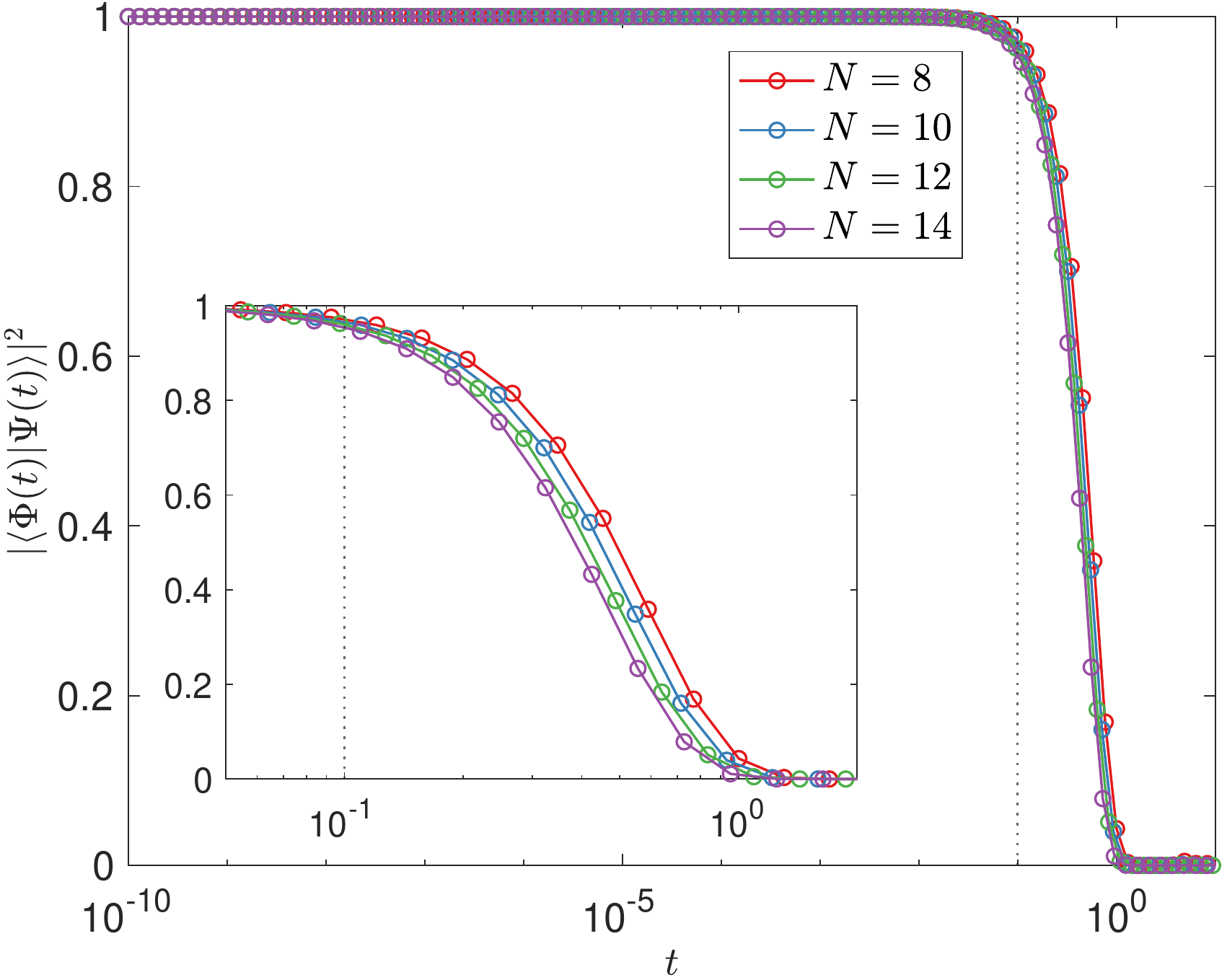}
\caption{\label{fig:earlytimeOL} The absolute square of the overlap $\langle \Phi(t)| \Psi(t)\rangle$ between the early-time wave function
$\Phi(t)$ [\eref{bcsdecomp1}] obtained  by neglecting the non-interacting part of the BCS Hamiltonian and the result of the direct simulation of the  nonstationary Schr\"odinger equation for the full   BCS Hamiltonian with coupling $g(t)\propto \frac{\eta}{t}$, $N$ energy levels $\eps_k=\frac{k}{N}$, and $\eta=1$. The initial condition is $\Phi(t_0)=\Psi(t_0)=\Psi_\mathrm{BCS}$, where $\Psi_\mathrm{BCS}$ is the infinite
coupling mean-field ground state
   and $t_0=10^{-10}$.  The doted  vertical line marks $t=0.1$ below which the overlap is close to 1. This plot supports our estimate that
   the early-time approximation is valid for $t< t_*\approx \frac{0.1\eta}{W}$, where $W$ is the bandwidth of $\eps_k$.
 }
\end{figure}

The Hamiltonian at early times approximately is
\begin{equation}
  \hat{H}(t\to 0)\approx \hat H_\mathrm{int} =
  -\frac{\eta}{N t} \hat\jmath_+\hat \jmath_-,
  \label{eq:H_et}
\end{equation}
where $\hat{\bm\jmath} =\sum_{k=1}^N \hat{\bm s}_k$ is the total spin, see the forth paragraph in Sec.~\ref{order_sec} for a brief discussion of the validity of this approximation. We rewrite the Schr\"odinger equation for $\hat H_\mathrm{int}$
as
\beg
i \frac{\partial \Psi}{\partial\tau} = -\frac{\hat \jmath_+ \hat \jmath_-}{N} \Psi,\quad \tau=\eta\ln \frac{t}{t_0}.
\label{evwrttau}
\en
Eigenvalues of $\hat \jmath_+ \hat \jmath_-$ are $\jmath(\jmath+1)-\jmath_z^2+\jmath_z$. The ground state of $\hat H_\mathrm{int}$ with $\jmath_z=N_\up-\frac{N}{2}$ and $\jmath=\frac{N}{2}$ is $\Psi_0(N_\up)$ in \eref{psi0}. The corresponding eigenvalue of $-\frac{\hat \jmath_+ \hat \jmath_-}{N}$  is $E_0(N_\up)$ in \eref{e0}. The infinite coupling BCS ground state~\re{alongx1} also has $\jmath=\frac{N}{2}$, but is not an
eigenstate of $\hat \jmath_z$ (and fermion number operator). It follows from \eref{bcsdecomp} that the solution of the nonstationary Schr\"ondiner equation at early times that starts in the   BCS ground state at $t=t_0$ is
\beg
\Phi(t)=2^{-\frac{N}{2}}\sum_{N_\up=0}^N {\binom{N}{N_\up}}^{\frac{1}{2}}e^{-i E_0(N_\up)\tau} \Psi_0(N_\up).
\label{bcsdecomp1}
\en
We compare  this early-time wave function, which we derived by neglecting the kinetic term $\hat H_0$, with the direct numerical solution of the nonstationary Schr\"odinger equation for the full quantum  Hamiltonian $\hat H=\hat H_0+\hat H_\mathrm{int}$,  same initial condition, for several $N$ in Fig.~\ref{fig:earlytimeOL}.
Repeating the procedure that took us from \eref{phiinf1} to \eref{007}, we obtain 
\begin{equation}
  \langle \hat{s}_k^- \rangle_{\Phi(t)} = \frac{1}{2}e^{-\frac{\tau^2}{2N}}
  \label{83}
\end{equation}
for the expectation value of the spin lowering operator $\hat s_k^-$ (equal time anomalous Green's function) in the state $\Phi(t)$ in the thermodynamic limit.

The corresponding early-time classical motion is trivial. In the strong coupling limit $t\to0^+$, we neglect $\eps_j$ in \eref{ceom}. Spins in the classical ground state~\re{classini} with   $J_z=0$ are along the $x$-axis, $\bm S_j=\langle \hat {\bm s}_j\rangle_\mathrm{mf}=\frac{ \bm x }{2}$. Since the spins and the effective magnetic field $-2\bm\Delta$ are both parallel to the $x$-axis, the spins are stationary, i.e., $\langle \hat  s_k^-\rangle_\mathrm{mf}=\frac{1}{2}$. We see that \eref{83} is the early-time version of \eref{007}. At later
times the Gaussian stops decaying as a function of $t$ and saturates at $t=t_*$. Therefore, the conclusion below \eref{007} that the   thermodynamic ($N\to\infty$) and $t_0\to0^+$ limits  do not commute applies here as well.  In addition, we see that the $N\to\infty$ and $\eta\to+\infty$ (adiabatic) limits of the quantum solution do not commute as well, while taken in any order in the classical case they give
$\langle \hat  s_k^-\rangle_\mathrm{mf}=\frac{1}{2}$. 

\begin{figure}[t!]
\includegraphics[width=0.48\textwidth]{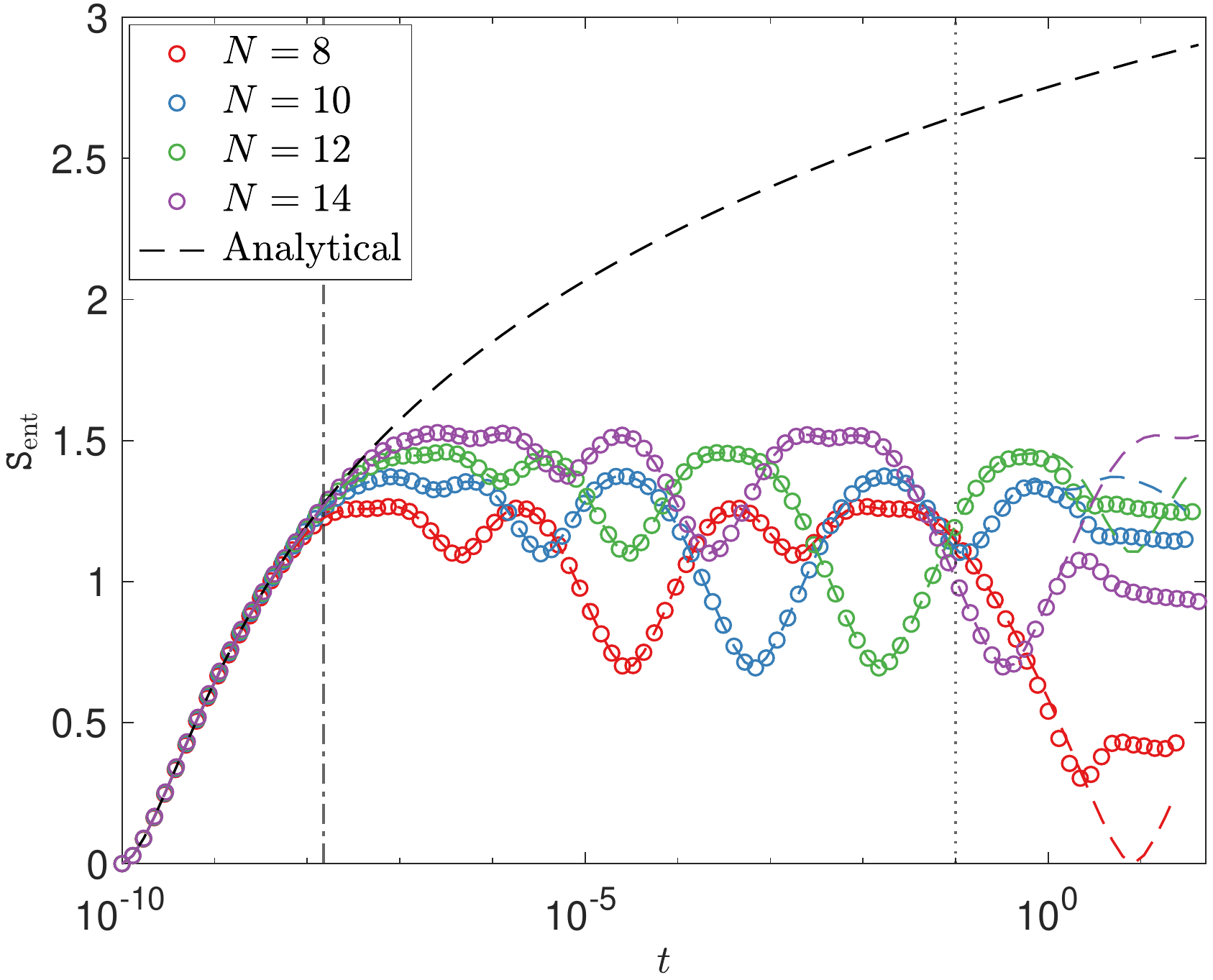}
\caption{\label{fig:s1}  Entanglement entropy $\mathtt{S}_\mathrm{ent}$ at short times  for the quantum BCS dynamics with time-dependent interaction strength $g(t)\propto \frac{\eta}{t}$. The initial state is the infinite coupling BCS ground state at $t=t_0$. System parameters are the same as in Fig.~\ref{fig:earlytimeOL}. Circles  represent  direct simulation of the dynamics with the full quantum BCS Hamiltonian, the corresponding colored 
  dashed curves  are $\mathtt{S}_\mathrm{ent}$   for the wave function~\re{bcsdecomp1} with a given $N$. Note that the early-time  approximation (neglecting the kinetic term in the  Hamiltonian)  accurately captures all the structures in $\mathtt{S}_\mathrm{ent}$  to the left of dotted vertical line  at $t=0.1$.    Black dashed curve  is the  analytic answer~\re{eq:coh_state_ent} obtained by taking the $N\to\infty$ limit on top of the early-time approximation.  It agrees with the numerically exact finite $N$ simulations   until $\mathtt{S}_\mathrm{ent}$ stops  growing at the Ehrenfest time $t_E$ [shown as a dash-dotted vertical line and given by \eref{ehrenfest} with $c_E=1.12$ and $N=10$] and finite size   oscillations (partial recurrences)  begin.  }
\end{figure}

Let us also provide a simple estimate of the characteristic time $t_*$ until which the early-time approximation, i.e.,  the neglect of $\hat H_0$ compared to $\hat H_\mathrm{int}$ is reliable. The estimate is based on the mean-field (classical) equations of motion~\re{ceom}.
We neglected $|\eps_j|$ compared to $|\bm\Delta|$ in these equations. At short times the classical spins  remain close to the $x$-axis
and therefore $|\bm\Delta|\approx \frac{\eta}{2t}$. We require  $|\eps_j|\ll \frac{\eta}{2t_*}$. Replacing $|\eps_j|$ with $\frac{W}{2}$, where $W$ is the bandwidth, we obtain $t_* \ll \frac{\eta}{W}$. Numerically, we find that 
 \beg
  t_*\approx\frac{0.1 \eta}{W}
  \en
is a reasonable estimate, see Figs.~\ref{fig:earlytimeOL} and \ref{fig:s1}. Note that in these figures $\eta=1$ and $W=1-\frac{1}{N}$, so $t_*$ is about 10\% larger than $t=0.1$, which is is hardly noticeable  on the logarithmic scale.

\subsection{Von Neumann entanglement entropy}
\label{renyi_sec}

We found in Sec.~\ref{ent_sec} that  $\ln N$ scaling of the von Neumann entanglement entropy $\mathtt{S}_\mathrm{ent}$ for large $N$ is generic for the quantum BCS model. In particular, it holds for the late-time asymptotic state $\Psi_\infty$,  projected BCS states, and exact  infinite coupling ground state.  We considered the case when the ratio of the number of fermion pairs $N_\up$ (equivalently, the number of up pseudospins) to the number $N$ of single-particle energy levels  $\frac{N_\up}{N}=\frac{1}{2}$, but we expect the $\ln N$ scaling to be valid in the thermodynamic limit for other finite ratios as well. 
Next, we determine the early-time $\mathtt{S}_\mathrm{ent}$    for the quantum BCS time evolution with coupling $g(t)\propto \frac{\eta}{t}$ starting from the BCS ground state  at $t=0^+$. We will find that $\mathtt{S}_\mathrm{ent}$  grows monotonically from zero with $\mathtt{S}_\mathrm{ent}\approx \ln \tau $ at large $\tau$. For finite $N$,  the growth saturates at $\mathtt{S}_\mathrm{ent}\propto \ln N$ as before. 

We postpone  derivations to Appendix~\ref{apdx:renyi_entropy} and  present only the main results and conclusions here. Let $N$ be even.
Our first step is to calculate the reduced density matrix $\rho_A$ for the subsystem $A$ consisting of $\frac{N}{2}$ spins for the early-time many-body wave function~\re{bcsdecomp1},    
\begin{equation}
    \rho_A =
    2^{-\frac{N}{2}}
    \sum_{K,K'=0}^{\frac{N}{2}}
    \Gamma_{K}\Gamma^{*}_{K'}
    e^{-\frac{\tau^{2}}{4N}(K-K')^2}
    \ket{K}\bra{K'}_A,
  \label{eq:coh_state_rdm}
\end{equation}
where 
\beg
\Gamma_{K} =  \mbinom{\frac{N}{2}}{K}^{\frac{1}{2}} e^{-\frac{i\tau}{N}\left[ K^2  - K \left(\frac{3N}{2}+1\right) \right]}
\en
 and $|K\rangle_A$ is the state of the subsystem $A$ with a definite number $K$ of up spins that is symmetric with respect to arbitrary permutations
of spins, i.e.,   with the maximum possible total spin $\frac{N}{4}$ and a definite value $K-\frac{N}{4}$ of its $z$-projection. Using  this result, we evaluate $ \mathrm{Tr}\left[\rho_A^n\right]$ in the limit $N\to\infty$  by replacing the sums with integrals  and employing the multi-dimensional saddle point method,
\begin{equation}
    \mathrm{Tr}\left[\rho_A^n\right] =
    \prod_{j=0}^{n-1}
        \left[1
          + \frac{\tau^{2}}{4} \sin^{2}\left(\frac{\pi j}{n}\right)\right]^{-\frac{1}{2}}.
\end{equation}

Now consider the R\'enyi entanglement entropies for integer $n$ defined as
\begin{equation}
  \mathtt{S}^\mathrm{R}_{n} = \frac{\ln\mathrm{Tr}\left[\rho_{A}^{n}\right]}{1-n}.
  \label{eq:renyi_ent}
\end{equation}
Treating $n$ as
a replica index that can be analytically continued~\cite{casini2009entanglement}, we obtain  the von Neumann entanglement entropy   in the $n\to 1$ limit as
 \begin{equation}
  \begin{gathered}
    \mathtt{S}_{\mathrm{ent}}
    = \sqrt{1 + \frac{\tau^{2}}{4}}
    \coth^{-1}\left[\sqrt{1 + \frac{\tau^{2}}{4}}\right] + \ln  \frac{\tau}{4}.
  \end{gathered}
  \label{eq:coh_state_ent}
\end{equation}
 We see that $ \mathtt{S}_{\mathrm{ent}}$ is intensive in the thermodynamic limit. It monotonically grows from  zero at $\tau=0$ $(t=t_0)$ behaving as $ \mathtt{S}_{\mathrm{ent}}\approx \ln \tau$ at large $\tau$.
  
 There is a simple  picture of the early-time dynamics that explains \eref{eq:coh_state_ent} and describes the saturation 
   and subsequent oscillations of the entanglement entropy for finite $N$ seen in Fig.~\ref{fig:s1}. First of all it is not difficult to see that  
   $ \mathtt{S}_{\mathrm{ent}}$ is bounded from above by $\ln \left(\frac{N}{2}+1\right)$. To show this, observe that the Hamiltonian with which the system
   evolves with respect to $\tau$ is    $\hat\jmath_+\hat\jmath_-$ up to a multiplicative constant, see \eref{evwrttau}. The collective spin  $\hat{\bm \jmath}=\hat{\bm \jmath}_A+\hat{\bm \jmath}_{\bar A}$, where $\hat{\bm \jmath}_A$ and
 $\hat{\bm \jmath}_{\bar A}$ are the total spins of subsystems $A$ and $\bar A$. Initially all spins-$\frac{1}{2}$ are along the $x$-axis and therefore the magnitudes of  $\hat{\bm \jmath}$, $\hat{\bm \jmath}_A$, and $\hat{\bm \jmath}_{\bar A}$ are $\jmath=\frac{N}{2}$, $\jmath_A=\frac{N}{4}$, and $\jmath_{\bar A}=\frac{N}{4}$. These magnitudes are conserved.  As a result the dynamics of subsystem $A$ is confined to a Hilbert space of dimension $\mathrm{dim} \mathcal{H}_A =2\jmath_A+1=  \frac{N}{2} + 1$. It is well-known~\cite{witten} that the entanglement entropy is bounded from above by $\ln(\mathrm{dim} \mathcal{H}_A)$, i.e., $\mathtt{S}_{\mathrm{ent}}\le \ln \left(\frac{N}{2}+1\right)$. 
 
 We see that the $\ln N$ scaling of the entanglement entropy in the BCS theory (see also Sec.~\ref{ent_sec}) is due to all to all interactions, which give rise to the collective spin $\hat{\bm \jmath}$.   Note also
 that the BCS order parameter is related to the collective spin as $\Delta = g\langle \hat \jmath_- \rangle$. The $\ln N$ behavior of
  $\mathtt{S}_{\mathrm{ent}}$ is a general feature of late-time asymptotic states of the BCS dynamics as well as the ground state and other low energy stationary states unrelated to the nonautonomous character of the Hamiltonian we study in this paper. We nevertheless discuss it for completeness.  
 
 A useful general result~\cite{lerose2020origin} for the von Neumann entanglement entropy in collective spin models, such as the early-time Hamiltonian~\re{eq:H_et}, is
 \beg
 \mathtt{S}_{\mathrm{ent}}
    = \sqrt{1 +  \langle \hat n_\mathrm{ex}\rangle }
    \coth^{-1}\!\left[\sqrt{1 + \langle\hat n_\mathrm{ex}\rangle }\,\right] + \frac{1}{2}\ln  \frac{ \langle \hat n_\mathrm{ex}\rangle }{4},
  \label{eq:coh_state_ent1}
  \en
  where $\langle \hat n_\mathrm{ex}\rangle=\langle \hat b^\dagger \hat b\rangle$ is the number of excitations -- the number of Holstein-Primakoff bosons for the collective spin $\hat{\bm\jmath}$ bosonized  via a Holstein-Primakoff transformation around the direction of $\langle\hat{\bm\jmath}(\tau)\rangle$.  To gain further insight into various features of $ \mathtt{S}_{\mathrm{ent}}$, consider the semiclassical  motion of this bosonic mode. This motion is one-dimensional and Hamiltonian. Nearby trajectories separate linearly in time, i.e., the growth of the  momentum and position with $\tau$ is linear. Since $\langle \hat n_\mathrm{ex}\rangle$ is quadratic in the momentum and position, we expect $\langle \hat n_\mathrm{ex}\rangle\sim\tau^2$. Indeed, comparing \esref{eq:coh_state_ent1} and \re{eq:coh_state_ent}, we see that  $\langle \hat n_\mathrm{ex}\rangle=\frac{\tau^2}{4}$. 
  
  Bounded one-dimensional Hamiltonian motion is periodic, which explains the post-saturation oscillations of 
  $\mathtt{S}_{\mathrm{ent}}$ (see Fig.~\ref{fig:s1})  approximately with the period of the underlying classical trajectory~\cite{lerose2020origin}.
  The time at which quantum fluctuations of the collective spin $\hat{\bm \jmath}$
  become comparable to its magnitude, i.e., $\langle \hat n_\mathrm{ex}\rangle\sim \frac{N}{2}$, is the Ehrenfest time $\tau_E$. Using $\langle \hat n_\mathrm{ex}\rangle=\frac{\tau^2}{4}$ and $\tau=\eta \ln\frac{t}{t_0}$, we obtain
 \beg
 \tau_E = c_E\sqrt{2N},\quad t_E=t_0 e^{c_E\frac{\sqrt{2N}}{\eta}},
 \label{ehrenfest}
 \en
where $c_E$ is a coefficient of order one.    Recall that  the entanglement entropy in the thermodynamic limit~\re{eq:coh_state_ent} grows as $\ln\tau$  for large $\tau$. This growth stops  and   finite size effects kick in at the Ehrenfest time scale    as at this point the number of bosonic excitations reaches its maximum possible value. Note also that $t_E$ is the time when the argument of the exponential function in \eref{83} becomes of order one, i.e., the quantum dephasing effect becomes appreciable.  Numerically, we observe that~\eref{ehrenfest}   provides a  reasonable estimate of the time when the finite $N$ entanglement entropy deviates from the $N\to\infty$ result~\re{eq:coh_state_ent}, see, e.g., the $t=t_E$ line for $N=10$ and  $c_E=1.12$ in Fig.~\ref{fig:s1} (dash-dotted vertical line).      We can also estimate  the saturation value of the von Neumann entanglement entropy for large $N$ as  $ \mathtt{S}_{\mathrm{ent}}\sim\ln\tau_E\sim\frac{1}{2}\ln N$, cf. \esref{bb} and \re{sentinf}.

\section{Conclusion}
\label{conclusion}
In this paper, we demonstrated that the far from equilibrium dynamics of BCS superconductors is classical    in the thermodynamic limit
  under certain  conditions. Specifically, we obtained   exact solutions for the quantum and classical (mean-field) dynamics of the BCS Hamiltonian with time-dependent coupling, $g(t)=\frac{1}{\nu t}$ launched from the ground state at $t=t_0\to 0^+$. We  explicitly determined    exact quantum   and mean-field   wave functions at long times, evaluated quantum averages of  a generic local observable in them, and proved that they coincide   in the thermodynamic limit.  It is clear that this must remain  true for a broad class of $g(t)$ and initial conditions. Nevertheless, it is worthwhile to verify that the mean field is similarly exact when $g(t)$ does not vanish at $t\to+\infty$. This can be done, for example,  by solving for the backward  time evolution from $t=+\infty$ to $t=0^+$ of    our model using the method we developed in this paper.
  
  On the other hand,   the classical  picture   breaks down dramatically for global quantities, such as the bipartite von Neumann entanglement entropy as we saw above and the Loschmidt echo as shown in Ref.~\onlinecite{echo}. Also noteworthy is the
  behavior of anomalous averages  -- expectation values of operators that do not  conserve the total fermion number, such as  the equal time anomalous Green's function. For these kinds of observables, the thermodynamic limit does not commute with the adiabatic ($\nu\to 0^+$) and $t_0\to 0^+$ limits due to quantum dephasing. Their quantum fluctuations are controlled by a new parameter that can be much larger than the inverse particle number -- the parameter controlling  the magnitude of equilibrium quantum fluctuations (finite size corrections). 
  
  These results provide a deeper understanding of the reasons behind the  success of mean-field theories in general, beyond our focus on  far from equilibrium superconductivity. In situations where they are believed  to be accurate (e.g., above a certain dimension),   we expect  the mean-field wave function  to capture the order parameter and other local observables, but not  the entanglement and other global properties of the true many-particle state.  A fascinating project is to investigate the interplay between quantumness and nonlocality in other ``mean-field'' models, such as  topological $p$-wave superconductors~\cite{read,victor1}, infinite-dimensional Hubbard model~\cite{gabi}, and magnon Bose-Einstein   condensates~\cite{ref1,ref2}.  In particular, an interesting question to ask in this connection is how well the mean-field approximation describes non-trivial topological properties, which are inherently non-local.

We saw that     the unitary time evolution  brings our system into a steady state similar to one of the nonequilibrium phases  in interaction quenched BCS superfluids~\cite{foster}. This state is a gapless superconductor with vanishing superfluid density, order parameter, and  pair-breaking excitation gap. As a consequence, its superfluid properties can only be revealed in energy resolved observables, such as the  spectral supercurrent density.
The steady state is nonthermal but is described by a nontrivial emergent generalized Gibbs ensemble, where the emergent integrals of motion are the single-particle level occupancy operators. We  only touched on the prospects of  realizing  our model  and testing our predictions in experiment in Introduction and Sec.~\ref{model_sec}, so this remains an important topic for future research.

We determined the dynamics of the von Neumann entanglement entropy $\mathtt{S}_{\mathrm{ent}}$ starting from the unentangled BCS (mean-field) ground state. Interestingly, the  growth of the entanglement  is entirely due to the interaction part of the quantum BCS Hamiltonian. The entanglement entropy is finite in the thermodynamic limit and grows monotonically as a function of time. Note that within the mean-field treatment $ \mathtt{S}_{\mathrm{ent}}=0$ at all times. In a finite system,   $ \mathtt{S}_{\mathrm{ent}}$ saturates at a value that scales as $\ln V$ with the system volume $V$.

This paper paves the way to a comprehensive theory of integrability of nonautonomous quantum Hamiltonians. We demonstrated  that   the off-shell Bethe Ansatz~\cite{babujian2,me,kitaev,volodya}, which provides an integral representation of solutions of the nonstationary Schr\"odinger equation, is a key ingredient of this theory.  Another key ingredient is the systematic method we developed here to obtain explicit physical results, such as the ones listed above in this section, from the integral representation. Our method should work equally well for other models that go through the off-shell Bethe Ansatz, e.g., the problem of molecular production in an atomic Fermi gas swept through a Feshbach resonance and
Demkov-Osherov, bow-tie, and generalized bow-tie  multi-level Landau-Zener models~\cite{me}. It is interesting  to apply our approach to these models and even more important to  generalize and broaden the  scope of  the off-shell Bethe Ansatz to include time-dependent models  unrelated to the Gaudin algebra.

\begin{acknowledgments}

We are grateful to V. Gurarie for helpful discussions. A.W. and J.H.P. are partially supported by NSF CAREER Grant No. DMR1941569 and the Alfred P. Sloan Foundation through a Sloan Research Fellowship.
We acknowledge the Beowulf cluster at the Department of
Physics and Astronomy of Rutgers University, and the
Rusty cluster from Flatiron Institute Research Computing, Simons Foundation that were used in obtaining the numerical results.

\end{acknowledgments}

 
 \onecolumngrid
\appendix

 \section{Solution of the classical equations of motion}
\label{classApp}

As discussed in the main text,  there are two ways to make Anderson pseudospins  classical. One option is to replace spin-$\frac{1}{2}$ with spin-$s$ and take the limit $s\to\infty$. The other option is to perform mean-field decoupling, $\langle \hat A_1 \hat A_2\rangle \to \langle \hat A_1\rangle \langle \hat A_2\rangle$, in the Heisenberg equations of motion for Anderson pseudospins. Both approaches result in the same  equations of motion~\re{ceom}. Here  we solve these classical equations of motion exactly and determine classical spins $\bm S_j$ explicitly at long times. In the main text, we  compare this answer with the thermodynamic limit of the exact quantum (spin-$\frac{1}{2}$) solution. We note that our result in this appendix also solves the generalized SU(2) Knizhnik-Zamolodchikov equations in the limit of classical spins and the boundary field $B\equiv\nu t\to+\infty$~\cite{me,kitaev,volodya}. Because we expect  classical (mean-field) dynamics to be able to reproduce  quantum dynamics only for $N\to\infty$, we primarily focus on this limit, even though
our method is suitable for finite $N$ as well.

Equation~\re{psi} is an exact integral representation for the solution of the nonstationary Schr\"odinger equation for  spins of arbitrary magnitudes. Let magnitudes of all spins be $s$. Our aim is to take the classical limit  $s\to\infty$ and $\hbar\to 0$ of the solution keeping $\hbar s=S$ fixed. The first step is to restore $\hbar$ in the  Schr\"odinger equation,
\beg
i\hbar \frac{ d \Psi}{dt}= \Biggl[  \sum_{j=1}^N 2\eps_j (\hbar \hat{s}_j^z)
  -\frac{1}{\nu t} \sum_{j,k=1}^N(\hbar \hat{s}_j^+) (\hbar \hat{s}_k^-)\Biggr] \Psi. 
  \label{nsse1}
\en
Note that $\hbar$ is dimensionless in our units. Canceling one factor of $\hbar$ on both sides and comparing to \eref{eq:H} with $g(t)=\frac{1}{\nu t}$, we see that the restoration of $\hbar$  amounts to the replacement $\nu\to \frac{\nu}{\hbar}$. Making this replacement in \eref{psi}, we further observe that 
the Yang-Yang action becomes,
\beg
{\cal S}(\bm\lam,\bm\eps, t)= 2\nu t\sum_\alpha \lam_\alpha +
2S\sum_j\sum_\alpha  \ln(\eps_j-\lam_\alpha)-
\hbar \sum_\alpha\sum_{\beta\ne\alpha}\ln(\lam_\beta-\lam_\alpha),
\label{S1}
\en
where 
\beg
S =\hbar s.
\en
 Expressions for $\Psi(t)$, $ \Xi(\bm\lam,\eps)$, and $\hat L^+(\lam)$ stay the same. 
 
Similar to the spin-$\frac{1}{2}$ case in Sec.~\ref{late_time_sec}, the integral in \eref{psi} localizes to the vicinities of the stationary points. The stationary point equations $\frac{\partial\mathcal{S}}{\partial{\lam_\alpha} }=0$ now are
\beg
\nu t+S\sum_j \frac{1}{\lam_\alpha-\eps_j}=\hbar \sum_{\beta\ne\alpha} \frac{1}{\lam_\alpha-\lam_\beta},\quad \alpha=1,\dots,N_\up.
\label{richi1}
\en
The difference with \eref{richi} is in the number $N_\up$ of rapidities $\lam_\alpha$. In \eref{richi},   $N_\up$  is the number of up spins which is smaller than the number $N$ of $\eps_j$. In \eref{richi1}, $N_\up$ is the amount by which the $z$-component $\jmath_z$ of the total spin is raised, $\jmath_z=N_\up-N s$. Since the magnitude $s$ of spins diverges, $N_\up$ also diverges and there are many more
$\lam_\alpha$ than $\eps_j$. 

As before, each $\lam_\alpha$ must tend to one of $\eps_j$ as $t\to+\infty$. Suppose $n_j$ of $\lam_\alpha$ tend to $\eps_j$. We denote the elements of this $j^\mathrm{th}$ degenerate subset of  $\lam_\alpha$ as $\lam_k^j$, where $k=1,\dots, n_j$. Note that $\sum_{j=1}^N n_j = N_\up$. Let
\beg
\lam_k^j=\eps_j+\chi_k^j,\quad \chi_k^j=\frac{\hbar z_k^j}{2\nu t}.
\en
We will see shortly that the new variables $z_k^j$ are independent of $t$ to the leading order in $t^{-1}$. All terms in \eref{richi1} that contain $\lam_\alpha-\eps_j$ with $\lam_\alpha$ not in the degenerate subset corresponding to $\eps_j$ or terms that contain  $\lam_\alpha-\lam_\beta$ with $\lam_\alpha$ and $\lam_\beta$ not in the same degenerate subset are negligible. Therefore, to the leading
order in $t^{-1}$, \eref{richi1} splits into $N$ decoupled sets of equations for $z_k^j$,
\beg
1+ \frac{2 s}{z_k^j}= \sum_{k'\ne k} \frac{2}{z_k^j -z_{k'}^j },\quad k=1,\dots,n_j.
\label{zkj}
\en
This is a set of $n_j$ equations for $n_j$ variables for each $j=1,\dots, N$. It is clear from \eref{zkj} that $z_k^j$ are $t$-independent. In fact, $z_k^j$ are  the roots of the associated Laguerre polynomial $L^{-2s-1}_{n_j}(z)$~\cite{sasha}. We will not need this property below as we will not evaluate integrals over $\chi_k^j$ by the steepest descent method.   

Let us see how the vector $ \Xi(\bm \lambda,\bm\eps)$ given by \eref{phi} behaves near stationary points. Observe that $\hat L^+(\lam_\alpha)\to \frac{\hat s_j^+}{\chi_k^r}$ when $\lam_\alpha\to\eps_r$. Therefore,
\beg
 \Xi(\bm \lambda,\bm\eps)\to \sum_{ \{ n_j\} }\left[ \frac{N_\up !}{ \prod_j n_j!} \prod_j \frac{ (\hat s_j^+)^{n_j} }{ \prod_k \chi_k^j} \right] |0\rangle.
\label{phitends}
\en
The summation here is over all sets $\{ n_j\}$ of $N$ positive integers such that
\beg
\sum_{j=1}^N n_j=N_\up.
\en
This corresponds to summing over all stationary points. The combinatorial factor in \eref{phitends} is the number of ways to choose
$N$ groups of variables out of $N_\up$ variables with $n_j$ elements in the $j^\mathrm{th}$ group. The action of    powers of spin-$s$ raising operators $s_r^+$ on the state $|0\rangle$
where all spins  point in the negative $z$-direction is
\beg
\prod_{r=1}^N (\hat s_r^+)^{n_r} |0\rangle=|n_1\dots n_N\rangle \prod_{r=1}^N \sqrt{ \frac{ (2s)! n_r!}{ (2s-n_r)!}  },
\en
where $|n_1\dots n_N\rangle$ is a normalized eigenstate of all $\hat s_r^z$ with eigenvalues $m_r=n_r-s$. \eref{phitends} becomes
\beg
\Xi(\bm \lambda,\bm\eps)\to  \sum_{ \{ n_j\} }  \prod_{j=1}^N \sqrt{ \frac{ (2s)! }{ (2s-n_j)! n_j! } }\,\, \frac{|n_1\dots n_N\rangle}{ \prod_k \chi_k^j}.
\label{phibin}
\en

We manipulate the Yang-Yang action~\re{S1} similarly to how we manipulated the stationary point equations~\re{richi1} neglecting $\chi_k^l$ in   $\eps_j-\lam_\alpha$ when $\lam_\alpha$ does not belong to the $j^\mathrm{th}$ degenerate subset and in $\lam_\alpha-\lam_\beta$ when $\lam_\alpha$ and $\lam_\beta$ are not in the same degenerate subset. We find
\beg
{\cal S}(\bm\chi,\bm\eps, t)= 2\nu t\sum_j n_j\eps_j +
2S\sum_j\sum_{k\ne j}  n_k \ln(\eps_j-\eps_k)-
\hbar \sum_j \sum_{k\ne j}n_j n_k \ln(\eps_j-\eps_k)+ \sum_j \mathcal{S}^j_\mathrm{deg}, 
\label{S2}
\en
where
\beg
\mathcal{S}^j_\mathrm{deg}= \hbar \sum_{k} z_k^j +2S \sum_{k} \ln(-\chi_k^j)- \hbar \sum_{k} \sum_{k'\ne k} \ln(\chi_k^j-\chi_{k'}^j).
\en
Recall that $\nu$ is of order $N$ and therefore the first three terms in ${\cal S}(\bm\chi,\bm\eps, t)$ are of order $N^2$ because each summation over the single-particle level index gives a factor of $N$. On the other hand, sums over the degenerate subspace
in $\mathcal{S}^j_\mathrm{deg}$ are of order $s$ and consequently $\sum_j \mathcal{S}^j_\mathrm{deg}$ is of order $N s$. Take the limits $N\to\infty$ and $s\to\infty$ so that $\frac{s}{N}\to0$. The last term in \eref{S2} 
is then negligible when $N\to\infty$.
As for the case of spin-$\frac{1}{2}$ in Sec.~\ref{late_time_sec}, we need to choose the branch of the logarithm so that $\ln(-1)=-i\pi$.  
It is not difficult to show that to order $N^2$ the imaginary part of the third term on the right hand side of \eref{S2} contributes only to the overall normalization constant. Together these observations allow
us to rewrite \eref{S2} as
\beg
\hbar {\cal S}(\bm\chi,\bm\eps, t)= 2\nu t\sum_j S_j^z\eps_j 
 -
2 \sum_{k> j}S_j^z S_k^z \ln|\eps_j-\eps_k|- 2i \pi S\sum_j j S_j^z, 
\label{S3}
\en
where $S_j^z=\hbar m_j$ are the $z$-components of the classical spins in the limit $\hbar\to 0$.  

We also need to express the square root of the binomial coefficient in \eref{phibin} in terms of $S_j^z=\hbar m_j= \hbar n_j -S$. Using Stirling's approximation, we obtain
\beg
\sqrt{ \frac{ (2s)! }{ (2s-n_j)! n_j! } }=\exp\left[ - \frac{(S+S_j^z)\ln(S+S_j^z)+ (S-S_j^z)\ln(S-S_j^z)}{2\hbar} \right].
\label{sterl}
\en
Let us summarize what we did so far.  We evaluated the ingredients in the general expression~\re{psi} for $\Psi(t)$ in the vicinity of the stationary points of the Yang-Yang action $\mathcal{S}(\bm \lam, \bm \eps, t)$ since at large $t$ the integral localizes to these vicinities. In \eref{sterl}, we took advantage of the fact that in the classical limit $\hbar\to0$ the magnitude of the quantum spin $\hat{\bm s}^z_j$ diverges, $s\to\infty$, while the magnitude $S=\hbar s$ and components, e.g., $S_j^z=\hbar m_j$, of the classical spin $\bm S_j$ remain finite. Substituting \esref{phibin} and \re{S3} into \eref{psi} and using \eref{sterl}, we find
\beg
\Psi(t) = \sum_{ \{m_j \} }  e^{-\frac{i E t}{\hbar}}e^{\frac{i \Omega}{\hbar \nu}} e^{-\frac{A}{2\hbar}}  |m_1\dots m_N\rangle,
\label{psi1}
\en
where the sum is over all $z$-projections $m_j$  such that $\sum_j m_j =\jmath_z$ and we integrated over all $\chi_k^j$ to derive \eref{psi1} . Since we neglected $\mathcal{S}^j_\mathrm{deg}$, these integrals are $\oint \frac{d \chi_k^j}{\chi_k^j}=2\pi i$ and contribute only to the overall multiplicative constant. The contours of integration over $\chi_k^j$ are guaranteed to enclose the origin as we start the time evolution in the instantaneous ground state at $t=0^+$, see the discussion below \eref{richi}. Quantities $A$, $E$, and $\Omega$ are
\begin{subequations}
\begin{align}
E &=\sum_j 2\eps_j S_j^z,\label{E}\\
\Omega &= 2 \sum_{k> j}S_j^z S_k^z \ln|\eps_j-\eps_k|,\label{omega} \\
A& = (S+S_j^z)\ln(S+S_j^z)+ (S-S_j^z)\ln(S-S_j^z) +\frac{4\pi j S  S_j^z}{\nu}. \label{A}
\end{align}
\end{subequations}
In the limit $\hbar\to 0$ only terms that minimize $A$ survive in \eref{psi1}. Minimizing \eref{A} with respect to $S_j^z$ at fixed $\sum_j S_j^z$, we obtain
\beg
S_j^z = -S \tanh(2Sa_j),\quad  a_j\equiv \frac{  \pi (j-\mu) }{\nu},
\label{sjz}
\en
where $\mu$ is the Lagrange multiplier (chemical potential) corresponding to $\sum_j S_j^z$ and we used $\tanh^{-1} z =\frac{1}{2} \ln\left( \frac{1+z}{1-z} \right)$.

  The state of the system~\re{psi1} is now a sum over only $m_j$ such that $\hbar m_j\to S_j^z$ with $S_j^z$ given
by \eref{sjz}. Note that this is more than one value of $m_j$, because, for example, $m_j$ and $m_j+1$ correspond to the same $S_j^z$ in the limit $\hbar\to0$. This allows us to also evaluate the $x$ and $y$ components of classical spins by taking the expectation value $\langle\hat s_j^-\rangle$ of the quantum spin in this state. We have
\beg
S_j^x -i S_j^y\equiv S_j^-=|S_j^-| e^{-\frac{i \Delta E t}{\hbar}}e^{\frac{i \Delta\Omega}{\hbar \nu}},
\en
where $\Delta E=E(m_j+1)-E(m_j)$ and $\Delta\Omega=\Omega(m_j+1)-\Omega(m_j)$ are the amounts by which $E$ and $\Omega$ change when we increase $m_j$ by 1 or, equivalently, increase $S_j^z$ by $\hbar$. We have
\beg
S_j^-=\frac{ S e^{-2i\eps_j t +i\varphi_j} }{\cosh(2S \zeta_j) },\quad \varphi_j=\frac{2\eta}{N}\sum_{k\ne j} S_k^z \ln|\eps_k-\eps_j|,\quad
S_j^z = -S \tanh(2S \zeta_j),\quad  \zeta_j\equiv \frac{  \pi\eta (j-\mu) }{N}.
\label{full}
\en
Here we traded $\nu$ for $\frac{N}{\eta}$.
This is the exact solutions of the Hamilton's equations of motion~\re{ceom} for the classical time-dependent BCS Hamiltonian~\re{eq:class_H} for $t\to+\infty$ and   $N\to\infty$ (these two limits commute as we discuss in Sec.~\ref{order_sec} for the quantum and at the end of Sec.~\ref{approach_sec} for the classical dynamics). As noted below \eref{psibcs}, to compare to mean-field
dynamics  starting from the BCS product state, we need to set the spin length $S=\frac{1}{2}$ in \eref{full}.

\section{\texorpdfstring{R\'enyi}{Renyi} entanglement entropy}
\label{apdx:renyi_entropy}
One of the key differences between quantum and classical systems is the presence of entanglement in the quantum case, i.e., of statistical correlations between subsystems that prevent the system from being in a state that is   a product of the states of the individual subsystems.
The degree of quantum entanglement between two subsystems $(A, \bar{A})$ is quantified by the von Neumann entanglement entropy defined as $\mathtt{S}_{\mathrm{ent}} = \mathrm{Tr}\rho_A \ln \rho_A$ where $\rho_A = \mathrm{Tr}_{\bar{A}}\rho$ is the reduced density matrix of $A$ and $\rho$ is the density matrix of the combined system.
Rather than computing $\mathtt{S}_{\mathrm{ent}}$ directly, it is more convenient to work with the R\'enyi entanglement entropy $\mathtt{S}^{\mathrm{R}}_n$ defined in Eq.~\eqref{eq:renyi_ent}.
Its usefulness   lies in the index $n$ which can be treated as a replica index to evaluate the von Neumann entropy in the limit $n\to 1$.
The R\'enyi entropy also encodes other entanglement measures such as the Hartley entropy for $n=0$ and the purity for $n=2$.
In what follows, we derive an expression for the R\'enyi entanglement entropy for arbitrary $n$ (and the von Neumann entanglement entropy by analytic continuation) for a bipartition of the  early-time wave function~\re{bcsdecomp1} for the dynamics with the quantum BCS Hamiltonian~\re{eq:bcs_model} [or equivalently with Hamiltonian~\re{eq:H} for $s=\frac{1}{2}$] with $g(t)=\frac{\eta}{Nt}$ starting from the infinite coupling BCS ground state~\re{alongx1} at $t=t_0\to 0^+$.

\subsection{Reduced density matrix}

In order to determine the R\'enyi entropy  $\mathtt{S}^{\mathrm{R}}_n$, we require the reduced density matrix $\rho_A$ for a bipartition of the system. As in Sec.~\ref{ent_sec}, we choose  $A$ to be the set of   spins $\hat{\bm s}_j$  that correspond the lower half of the single-particle energy spectrum $\eps_j$. We need to evaluate  $\rho_A$ for the state~\re{bcsdecomp1}. To do so, we first rewrite \eref{bcsdecomp1} in a more convenient form by substituting \eref{arb} into it and rearranging the coefficients by first splitting the sum into two terms ($N_\up\le \frac{N}{2}$ and $N_\up>\frac{N}{2}$) and then recombining it,
\begin{equation}
         \Phi(t) =
        2^{-\frac{N}{2}}
        \sum_{N_\up^A=0}^{  \frac{N}{2} }
        \binom{ \frac{N}{2} }{N_\up^A }^{\frac{1}{2}  }
        \left| N_\up^A \right\rangle_A
        \otimes
        \sum_{N_\up^{\bar A}=0}^{ \frac{N}{2}}
    e^{-i E_0(N_\up)\tau} 
        \binom{ \frac{N}{2}}{N_\up^{\bar A} }^{ \frac{1}{2}}
        \left| N_\up^{\bar A} \right\rangle_{\bar{A}},
\end{equation}
where  $N_\up=N_\up^{A}+N_\up^{\bar A}$ and $\tau=\eta\ln\frac{t}{t_0}$.
At $\tau = 0$ the system is in the product state~\re{alongx1}  [all spins point along $x$] and we see that the two sums decouple in this case as they should.
The density matrix of the system is then
\begin{equation}
        \rho(t) =
        2^{-N}
        \sum_{K,K'=0}^{ \frac{N}{2} }
         \binom{ \frac{N}{2}}{K}^{ \frac{1}{2}}
        \binom{ \frac{N}{2}}{K'}^{ \frac{1}{2}}
        \ket{K}\bra{K'}_A
        \otimes
        \sum_{Q,Q'=0}^{ \frac{N}{2}}
       e^{-i \left[ E_0(K+Q) - E_0(K'+Q')\right]  \tau} 
         \binom{ \frac{N}{2}}{Q}^{ \frac{1}{2}}
       \binom{ \frac{N}{2}}{Q'}^{ \frac{1}{2}}
        \ket{Q} \bra{Q'}_{\bar{A}},
\end{equation}
where we renamed the summation indices $N_\up^A\to K$ and $N_\up^{\bar A}\to Q$ for simplicity.
It remains to trace over $\bar{A}$ to find the reduced density matrix of $A$.
Performing the trace yields a factor of $\delta_{Q,Q'}$ which  consumes the sum over $Q'$.
In the large $N$ limit, we replace the remaining sum over $Q$ with an integral over $z = \frac{2Q}{N}$ and evaluate it using the saddle point method to arrive at
\begin{equation}
        \rho_A(t) =
        2^{-\frac{N}{2}}
        \sum_{K,K'=0}^{\frac{N}{2}}
        \sqrt{\binom{\frac{N}{2}}{K}}
        e^{-\frac{i\tau}{N}\left[ K^2  - K (\frac{3N}{2} + 1) \right]}
        \sqrt{\binom{\frac{N}{2}}{K'}}
        e^{ \frac{i\tau}{N}\left[ K'^2 - K'(\frac{3N}{2} + 1) \right]}
        e^{-\frac{\tau^{2}}{4N}(K-K')^2}
        \ket{K} \bra{K'}_A.
\end{equation}

\subsection{R\'enyi entropy}

Having determined $\rho_A(t)$, we now turn to calculating the R\'enyi entanglement entropy defined by \eref{eq:renyi_ent}.
When taking the trace of powers of $\rho_A$ the phases cancel so that
\begin{equation}
    \mathrm{Tr}\left[\rho_A^n\right] =
    2^{-n\frac{N}{2}}
    \sum_{x_1,\ldots,x_n}^\frac{N}{2}
    \binom{\frac{N}{2}}{x_1}\ldots\binom{\frac{N}{2}}{x_n}
    \exp\left[-\frac{\tau^{2}}{4N}\left[
            (x_1-x_2)^2 + \ldots + (x_n-x_1)^2
        \right]
    \right].
\end{equation}
As before, in the large $N$ limit we replace the sums with integrals and evaluate them using the multi-dimensional saddle point method.
The stationary point occurs at $x_1 = x_2 = \ldots = x_n = \frac{1}{2}$ and the matrix elements of the Hessian matrix  are
\begin{equation}
\mathrm{Hess}_{ij} = 
-\left(4 + \frac{\tau^2}{2}\right)\delta_{i,j} + \frac{\tau^2}{4}\delta_{i+1,j} + \frac{\tau^2}{4}\delta_{i-1,j}
\end{equation}
with eigenvalues given by~\cite{circulant}
\begin{equation}
    \kappa_j = -\left(4 + \frac{\tau^2}{2}\right) + \frac{\tau^2}{4} \omega^j + \frac{\tau^2}{4}\omega^{(n-1)j}
\end{equation}
where $\omega = e^{\frac{2\pi i}{n}}$ is a primitive $n^\text{th}$ root of unity.
Substituting $\prod_j \kappa_j$ for the determinant into the saddle point formula and simplifying we have
\begin{equation}
        \mathrm{Tr}\left[\rho_A^n\right] =
        \prod_{j=0}^{n-1}
            \left[1 + \frac{\tau^2}{16}\left(2 - e^{i\frac{2\pi j}{n} } - e^{-i\frac{2\pi j}{n} }\right)\right]^{-\frac{1}{2} }
\end{equation}
from which the R\'enyi entanglement entropy follows
\begin{equation}
    \mathtt{S}^{\mathrm{R}}_n(\tau) = \frac{-1}{2(1-n)}
    \sum_{j=0}^{n-1}
    \ln\left[1 + \frac{\tau^2}{16}\left(2 - e^{i\frac{2\pi j}{n} } - e^{-i\frac{2\pi j}{n} }\right)\right].
\end{equation}
To evaluate the von Neumann entanglement entropy we analytically continue to $n=1$ by writing the sum as a contour integral~\cite{casini2009entanglement}
\begin{equation}
    \mathtt{S}^{\mathrm{R}}_n(\tau)
    = \frac{-1}{2(1-n)}
    \sum_{j=0}^{n-1}
    \oint \frac{\mathrm{d}u}{2\pi i}
    \frac{\ln\left[1 + \frac{\tau^2}{16}\left(2 - u - u^{-1}\right)\right]}{u - e^{i2\pi j/n}}
    = \frac{-1}{2(1-n)}
    \oint \frac{\mathrm{d}u}{2\pi i}
    \left(\frac{n u^{n-1}}{u^n - 1}\right)
    \ln\left[1 + \frac{\tau^2}{16}\left(2 - u - u^{-1}\right)\right].
\end{equation}
In the limit $n\to1$~\cite{casini2009entanglement}
\begin{equation}
    \mathtt{S}^{\mathrm{R}}_{n\to 1}(\tau) \to
    -\frac{1}{2}\oint \frac{\mathrm{d}u}{2\pi i}
    \left(
        \frac{n}{n-1}\frac{1}{1-u} + \frac{\ln u}{(u-1)^2} + \mathcal{O}(n-1)
    \right)
    \ln\left[1 + \frac{\tau^2}{16}\left(2 - u - u^{-1}\right)\right],
\end{equation}
which becomes
\begin{equation}
        \mathtt{S}^{\mathrm{R}}_{n\to 1}(\tau) = \mathtt{S}_{\mathrm{ent}}(\tau) = \int_1^\infty \mathrm{d}\lambda
        \frac{\ln\left[1 + \frac{\tau^2}{16}\left(2 + \lambda + \lambda^{-1}\right)\right]}{(\lambda + 1)^2}
        =\sqrt{1 + \frac{\tau^2}{4}}\coth^{-1}\sqrt{1 + \frac{\tau^2}{4}} + \ln \frac{\tau}{4}.
        \label{eq257}
\end{equation}
 An immediate observation is that $\mathtt{S}_{\mathrm{ent}}(\tau)$ is  finite in the thermodynamic limit.  In the main text we provide a thorough qualitative explanation of \eref{eq257} and use it  together with other considerations to understand the behavior of $\mathtt{S}_{\mathrm{ent}}(\tau)$ for finite $N$ as well, namely,  its saturation at $\mathtt{S}_{\mathrm{ent}}\sim \frac{1}{2}\ln N$ at $\tau\sim\sqrt{N}$ and subsequent oscillations as a function of $\tau$.


\begin{thebibliography}{99}
\bibitem{clarify_finite} More precisely, $n$ must be such  that $\frac{n}{N_\mathrm{f}}\to0$ in the thermodynamic limit, where $N_\mathrm{f}$ is the total number of  fermions.
\bibitem{local} Since the BCS interaction is all to all, we can order $\eps_j$ on a chain (or arbitrary lattice) so that the $n$ levels $\eps_k$ involved are consecutive.  Then, the relative length $\frac{n}{N}$ of the $\eps_k$ segment  contracts to a point in the limit $N\to\infty$ justifying the term ``local''. This definition admits $n\sim N^a$  with $0\le a<1$.  
 In all our examples of local operators  $a=0$ and we therefore often assume $\frac{n}{N}\sim \frac{1}{N}$ without  loss of generality.   Let us also emphasize that our notion of locality   is unrelated to the property of being local in the real space.
\bibitem{note75} There is also a global phase proportional to $\ln t$ (see  note~\cite{factor12} below) and more generally to $\int^t dt'g(t')$, which    cancels in  $\left|C_{ \{s^z\}}\right|$.
\bibitem{verify22} More rigorously, we verified  that the term neglected in making the adiabatic approximation (see p. 330 of Ref.~\onlinecite{sakurai}) is of higher order in $t$ at small $t$.
\bibitem{factor12}  Let us mention for completeness   a nonessential  overall factor $\exp[-i\eta \frac{N_\up}{N} \ln t]$ in $\Psi_\infty(N_\up)$ that we omitted in \eref{psiinf} for simplicity.  We obtain this factor using \eref{roots}, which includes the $t^{-1}$ correction, as the saddle point value of $\lam_\alpha$ instead of $\lam_\alpha=\eps_\alpha$.  An alternative route is   the standard time-dependent perturbation theory in the limit $t\to+\infty$ treating $\hat H_\mathrm{int}$ as a perturbation to $\hat H_0$. The same term is present in the classical  solution. To incorporate    it into our answers for the late-time quantum and classical (mean-field) asymptotic states, simply replace $\varphi_k\to \varphi_k+\frac{\eta}{N} \ln t$ in the expressions for $S_k^\pm$, $u_k$, $v_k$, and $\langle \hat s_k^- \rangle_\mathrm{mf}$.
\bibitem{showed} We showed this for the quantum BCS time evolution  with a definite particle number in Sec.~\ref{order_sec}. Similar arguments
apply to the classical BCS dynamics considered in this section.
\end{thebibliography}
\end{document}